\documentclass[12pt]{iopart}
\usepackage{graphicx}
\usepackage{dcolumn}
\usepackage{bm}
\usepackage[mathscr]{eucal}
\usepackage{mathrsfs}
\usepackage{cite}

\usepackage{txfonts}

\begin{document}

\title[Total energy approach for the specific heat]{The total energy approach for calculating the specific heat of liquids and glasses}

\author{Koun Shirai$^{1,2}$}

\address{
$^1$ SANKEN, Osaka University, 8-1 Mihogaoka, Ibaraki, Osaka 567-0047, Japan \\
$^2$ Vietnam Japan University, Vietnam National University, Hanoi, \\
Luu Huu Phuoc Road, My Dinh 1 Ward, Nam Tu Liem District, Hanoi, Vietnam 
}

\begin{abstract}
The recent development of the calculation of specific heat ($C$) of liquids and glasses by first-principles molecular dynamics (MD) simulations is reviewed. Liquid and glass states have common properties in that there is no periodicity and the atom relaxation has an important role in their thermodynamic properties. These properties have, for a long time, hindered the construction of an appropriate theory of $C$ for these states. The total energy approach based on the density-functional theory (DFT) provides a universal method to calculate $C$, irrespective of the material states. However, aside from the convergence problem, even DFT-based MD simulations give different values for a thermodynamic property of liquids and glasses, depending on the setup of MD simulations. The essential problem is atom relaxation, which affects the relationship between the energy and temperature $T$. The temperature is determined by the equilibrium state, but there are many metastable states for glasses. Metastable states are stable within their relaxation times. We encounter the difficult problem of hysteresis, which is the most profound consequence of irreversibility. Irreversibility occurs even for quasistatic processes. This is the most difficult and confusing point in the thermodynamics literature. Here, a consistent treatment of both equilibrium properties and irreversibility in adiabatic MD simulations, which has no frictional term, is given by taking multi-timescales into account. A leading principle to determine the equilibrium is provided by the second law of thermodynamics. The basic ideas and the usefulness of the total energy approach in real calculations are presented.
\end{abstract}

\maketitle

\section{Introduction}
\label{sec:Introduction}
Specific heat ($C$) is a quantity of central importance in thermodynamics and material science. It is the change in the internal energy ($U$) against an infinitesimally small change in temperature $T$,
\begin{equation}
C_{X} = \left( \frac{\partial U}{\partial T} \right)_{X},
\label{eq:def-C}
\end{equation}
where $X$ denotes constraint, such as volume $V$. The enthalpy $H$ at a constant pressure ($P$) can be obtained by integrating the isobaric specific heat $C_{P}(T)$ with respect to $T$. Simultaneously, entropy $S$ can be obtained by the integration,
\begin{equation}
S(T) = \int_{0}^{T} \left( \frac{C(T')}{T'} \right)_{\rm rev} dT',
\label{eq:TDdef-S}
\end{equation}
provided that the integration path is reversible. Finally the Gibbs free energy $G$ can be obtained by $G=H-TS$. In this manner, the most fundamental quantity of the free energy, which determines all changes of material states, can be obtained solely by specific heat measurement. In spite of this importance, we do not known very much about the specific heat of liquids.

\subsection{Problems of liquids}
In elemental courses of thermodynamics and statistical mechanics, we learned the microscopic theory of specific heat for the gas and solid states. In the high-temperature limit, the isochoric specific heat $C_{V}$ per mole becomes insensitive to the material and is determined solely by the number of degrees of freedom $f$ of atom motions per atom:
\begin{equation}
C_{V} = \frac{f}{2} R,
\label{eq:classic-limit}
\end{equation}
where $R$ is the gas constant. For example, for monatomic gases, there are three degrees of translational motions, $f=3$. For monatomic crystals, there are additional three degrees of freedom for the potential energy, $f=6$, which is known as the Dulong--Petit law. However, no rule on such a high-temperature limit is known for liquids. Efforts were made to discover general rules for liquids but were unsuccessful. These previous attempts are described in a brief review by Granato \cite{Granato02}. Almost no description for liquids is found even in a monograph of specific heat \cite{Gopal66}, except the specific heat of liquid helium. Textbooks on liquid state itself are very rare compared with myriad literature on solid state physics. Even in those sparse textbooks of liquids \cite{Egelstaff-2ed,Hansen06,Beer72,Faber72,Shimoji77,Iida88,March-Tosi}, almost no description is found on the specific heat, except very formal matters. The only words presented therein are such that the construction of theory of liquid is very difficult. We may ask why it is so difficult.

There are fundamental difficulties for liquids.
\begin{enumerate}
\item{Degrees of freedom.} 
What is the meaning of the number of degrees of freedom appearing in specific heat? The answer to this question is obvious for gases. When a gas consists of $N$ atoms, the total number of degrees of freedom is $f_{\rm tot}= 3N$. For solids, $f_{\rm tot}= 6N$ because of additional degrees of freedom for the potential energy. The natural question is what $f_{\rm tot}$ is for liquids. Intuitively, we may expect that liquid is the intermediate case between gas and solid cases. However, the degree of freedom is a discrete quantity, namely, the presence or absence. It cannot be continuously changed.

\item{Anomalous specific heat.}
When we attempt to understand specific heat based on the number of degrees of freedom, a difficulty arises near phase transitions. $C$ often exhibits a quick increase near the transition temperature $T_{\rm tr}$ \cite{Pippard}. The $\alpha-\beta$ transition of quartz appears like a $\lambda$ transition: $C$ exhibits almost divergent behavior at $T_{\rm tr}$. The divergent behavior is also observed for liquids: for example, water near the critical point \cite{IAPWS96}. Large $C$ values are often found in supercooled liquids. Such large $C$ values are far from the range that can be explained by phonons.

\item{Relaxation effect.}
When the viscosity of a liquid becomes high, there is a time delay to reach equilibrium: the time to reach equilibrium is called the relaxation time $\tau$. The effects of relaxation appear in the time ($t$) and frequency ($\omega$) dependences of the $C$ at the glass transition. Because of this, supercooled liquids as well as glasses are traditionally treated as nonequilibrium states. However, there is no qualitative difference between the normal and supercooled liquids. The supercooled liquid state is certainly an extreme case in terms of viscosity but not an exceptional case: every liquid has a finite value of viscosity. A proper theory must treat both states on an equal footing.

\end{enumerate}
These issues are interrelated through atom relaxation of various kinds, as shown throughout this paper. 
Atom relaxation destroys eigenstates of the system. Traditionally, the statistical-mechanical treatment of an $N$-particles system starts from decomposing the total energy $E_{\rm tot}$ into the energies of the constituent particles, $\{ \epsilon_{j} \}$, as
\begin{equation}
E_{\rm tot} = \epsilon_{1} + \epsilon_{2} + \dots + \epsilon_{N}.
\label{eq:etot-sum-e}
\end{equation}
This decomposition in energy is obvious for ideal gases, because the motion of one atom is independent of the other atoms. For solids, all the atoms strongly interact with each other. But, by diagonalizing the Hamiltonian, independent quasi-particles called phonons are obtained \cite{Reif}. These phonons are eigenstates for the solid and the energy of $q$-th phonon, $\epsilon_{q}$, is the eigenenergy. The total excitation energy of a solid is obtained by summing as
\begin{equation}
\Delta E_{\rm tot} = \sum_{q} \bar{n}_{q} \epsilon_{q}
 = \bar{n}_{1} \epsilon_{1} + \bar{n}_{2} \epsilon_{2} + \dots ,
\label{eq:etot-sum-e1}
\end{equation}
where $\bar{n}_{q}$ is the Bose--Einstein occupation number. The specific heat of a solid is obtained by taking the derivative of equation (\ref{eq:etot-sum-e1}) with respect to $T$. This is a fairly standard procedure in statistical mechanics. Here, this manner of decomposition into quasi-particles is referred to as {\em the elemental excitation approach}. However, we should not forget that the summing property in equations (\ref{eq:etot-sum-e}) and (\ref{eq:etot-sum-e1}) is underpinned by an assumption of independent particle approximation, which is valid for eigenstates but not guaranteed for non-eigenstates.

For liquids, no elemental excitations are found in the sense that they form eigenstates. The consequence of this is discussed in more detail throughout this review. 
Despite the lack of firm grounds, recently many authors have adapted the phonon model also to liquids  \cite{Wallace98,Wallace02,Granato02,Trachenko08,Bolmatov11,Bolmatov12,Trachenko16, Proctor20,Baggioli21}. Wallace conducted a pioneering work on the liquid theory based on the phonon model \cite{Wallace94,Wallace97a,Wallace97b,Wallace98}. Recently, Trachenko advanced Wallace's theory one step further. His theory is gaining popularity through his textbook on the thermodynamics of liquid states \cite{Trachenko23-a}. The main idea of them is that the atom motions of a liquid can be considered a collection of instantaneous vibrations in a short timescale. The idea dates back to the defect model of liquids proposed by Frenkel \cite{Frenkel46}: there are random minima in the potential in a liquid; atoms are trapped at these minima in a short time period, creating instantaneous vibrations. Even though some successes of the phonon model in describing the thermodynamic properties of liquids are greatly acknowledged, there are fundamental reasons for doubting the phonon model for liquids: these are discussed in this review. 

\subsection{DFT plus thermodynamic method}
Up to now, no genuine elemental excitations of liquids that meet the independent-particle requirement are known. Such a ``liquion" probably does not exist. However, this does not mean that the calculation of specific heat is impossible. Instead of breaking to the eigenenergies of constituent particles, if we obtain directly the total energy, it is sufficient to calculate specific heat. Density functional theory (DFT) is the theory that enables this with high accuracy \cite{Callaway84,Parr-Yang89,QC21,Martin04}. Another quantity that is required to calculate $C$ is temperature, which is the most important state variable in thermodynamics. Temperature is normally well described by molecular dynamics (MD) simulation. Hence, the first-principles (FP) MD simulation is a suitable method for calculating the specific heat of liquids as well as solids. We here call this approach {\em the total energy approach}, in contrast to the elemental excitation approach.

Now, FP calculations are the most reliable method for describing material properties accurately. Many program codes of FP methods are available to material researchers in various fields, and anybody can use these codes without knowing the underlying principles. However, all of them do not necessarily give the same answers, aside from the convergence problem. There are many ways of performing FPMD simulations, but they often give different answers even though the same potentials are used. This is a big problem in calculating the melting temperature, which is described in section \ref{sec:meltingT-theory}. 
Readers may suspect that the determination of temperature in MD simulations is problematic. It is rather easy for most researchers of MD simulations. Temperature is the measure of equilibrium states (the zeroth law). Then, we need to know when equilibrium is reached in MD simulations. In general, there are numerous atom configurations. At low temperatures, the equilibrium state is uniquely determined by the ground-state configuration. The ground state can be determined unambiguously by energy minimization. However, for high temperatures, such as the melting temperature, the energy minimization method cannot be used: the equilibrium state is not the energy-minimum state. All configurations have finite lifetime (relaxation time). Accordingly, the final equilibrium state varies depending on the process parameters, such as cooling rate. Hysteresis is produced between heating and cooling processes. 
An extreme example that MD simulation creates spurious equilibrium states is demonstrated later (see figure \ref{fig:Est-lnD-si}). The question in this situation is thus how to determine the equilibrium state.

At a deep level, the above question is related to the fundamental problem of thermodynamics. There is a fundamental dilemma in thermodynamics theory. When we ask what temperature is, we can answer that temperature is a quantity that characterizes equilibrium. Then, we may ask what equilibrium is. Equilibrium is determined by temperature. Without referring to $T$, we cannot say equilibrium. The argument becomes a chicken-and-egg problem. This dilemma had not long been resolved.
Very fortunately, Gyftopoulos and Beretta (GB) have recently resolved this dilemma by describing equilibrium by its external effects \cite{Gyftopoulos}.

\vspace{2 mm}

\noindent 
{\bf Definition of equilibrium}.
{\it A material is said to be in equilibrium if work cannot be extracted from it without impacting its surroundings.} (\cite{Gyftopoulos}, p.~58)

\vspace{2 mm}

\noindent 
In the original form of BG, more precisely, the term of stable equilibrium is used for the equilibrium. Throughout this review, the term of equilibrium is understood to mean stable equilibrium. On defining equilibrium in this way, the second law of thermodynamics is stated as below.
\vspace{2 mm}

\noindent 
{\bf GB statement of the second law}.
{\it For a fixed $U$ and fixed constraints $\xi_{j}$, an isolated system has one and only one stable equilibrium state.} (\cite{Gyftopoulos}, p.~62)

\vspace{2 mm}
\noindent 
The cited expression is slightly modified from their original one in order to accommodate the present context. There are many ways of stating the second law. The GB statement is the latest of 21 statements \cite{Capek}. Some readers may not be familiar with this expression, and hence a brief explanation is attached in the Supplemental material of \cite{Shirai25-OrderParams}. In brief, the GB statement is equivalent to the maximum entropy principle. If the above statement is read as an isolated system changes its state in order to attain its maximum entropy, everyone will understand the statement. The state at the maximum entropy is the equilibrium state. The merit of using the GB statement is that no use of state variables $T$ and $S$ is needed for defining equilibrium. Even when we do not know how $T$ and $S$ can be defined, the spontaneous process automatically locates the equilibrium state \cite{note-axiomatic}. 

In the GB statement, the term constraint is used. Constraints are obstacles like the container's wall in which a gas is confined. This example is evident, but constraints are unclear inside materials. In short, the substance of constraint is an energy barrier of any kind. In solids, atom $j$ cannot move beyond the $j$-th unit cell. The object to inhibit the $j$-th atom from migration is the energy barrier $E_{b,j}$ built around the $j$-th atom. There is a one-to-one correspondence between constraint and state variable \cite{Reiss}. Time-dependent (dynamical) variables, such as atom positions, have thermal fluctuation. Many such variables will vanish when the time average is taken at equilibrium. However, some variables do not vanish after taking the average: that is, {\em the invariance against time average}. Such nonvanishing variables are qualified as the state variables. The constraint $E_{b,j}$ around the $j$-th atom makes the equilibrium position $\bar{\bf R}_{j}$ of solids as a nonvanishing quantity and thus a state variable. In this manner, all the equilibrium positions $\{ \bar{\bf R}_{j} \}$ become the state variables for solids \cite{Shirai18-StateVariable}. State $A$ of a solid is thus specified by a full set of the state variables,
\begin{equation}
A = A(T, \{ \bar{\bf R}_{j} \}).
\label{eq:stateAsolid}
\end{equation}
The significance of this relation will be clarified in the course of this review. 

\begin{figure}[ht!]
\centering
    \includegraphics[width=100mm, bb=0 0 780 650]{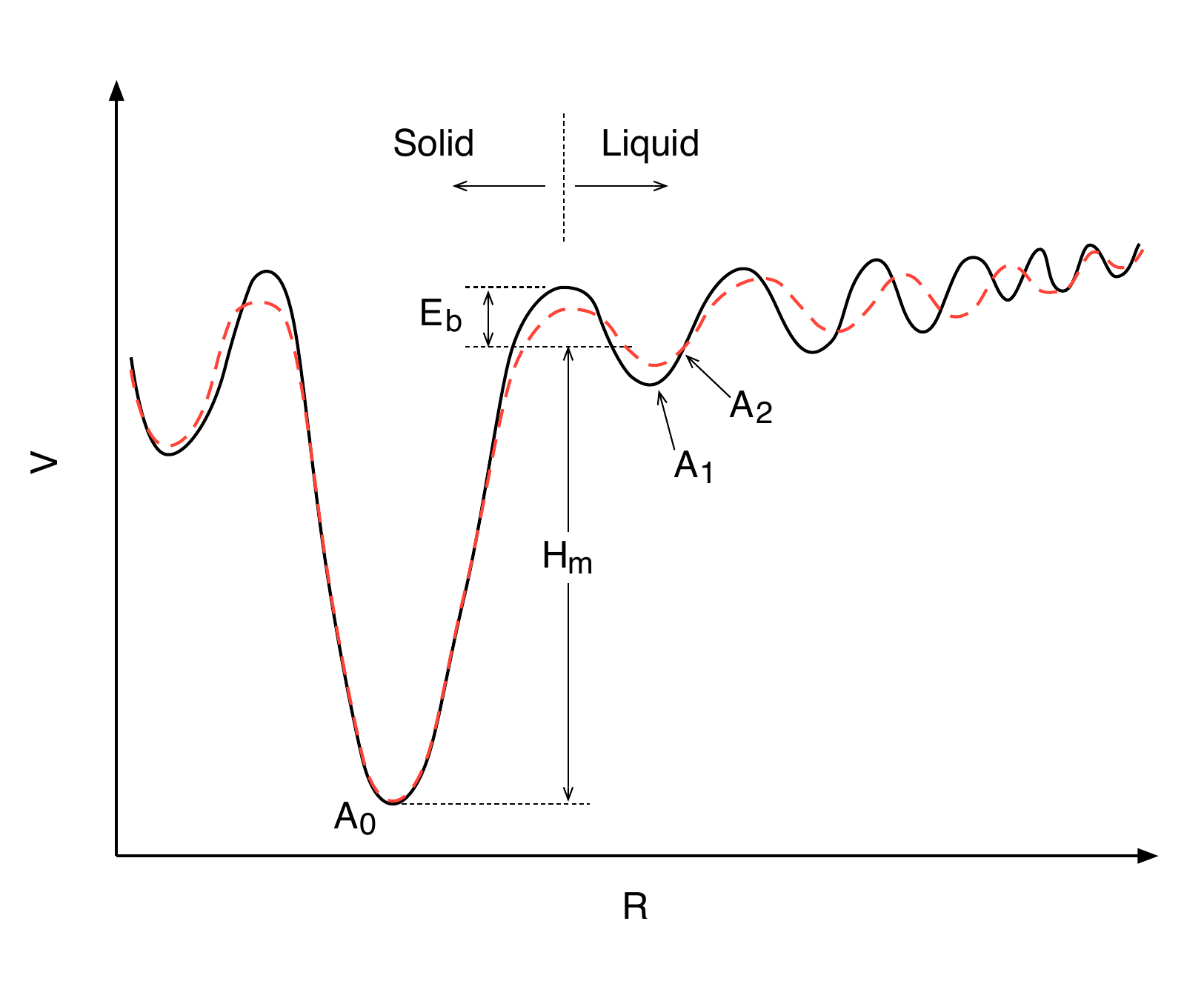} 
\caption{
Schematic of the adiabatic potential of a material that covers the solid and liquid states. When the environment of the system is changed, the adiabatic potential changes as indicated by the (red) dashed line.}
\label{fig:pot-relax} 
\end{figure}

The basic element of our approach consists of adiabatic MD simulation, which is here called {\em the adiabatic relaxation simulation}. This is essentially the same as what is called the $NVE$ simulation---constant $N$, $V$, and $E$---in usual textbooks \cite{Frenkel96,Allen87}. The method is an elemental one, and hence readers may ask what is new. Technically, nothing is new. The MD techniques described here are no more than those given in standard textbooks \cite{Frenkel96}. The novelty of this paper lies in the deep thermodynamic analysis on equilibrium and irreversibility, which are very difficult problems for in solids. Inside materials, various kinds of atom relaxation, energy dissipation, hysteresis, and memory effects, which are essentially the same phenomenon in terms of irreversibility, take place on different timescales. This makes consistent analysis very complicated.
Usually, calculations using either $NVE$ or $NPT$ ensembles are considered to give the same result, and hence the choice is made by considering which is convenient for the specific purpose. This is not true when relaxation processes are involved. This point is not well appreciated by many researchers of solid states, and hence, at this moment, the basic idea is quickly explained using a schematic of the adiabatic potential $V(\{ {\bf R}_{j} \})$ in figure \ref{fig:pot-relax}.
For solids, the unique equilibrium state $A_{0}$ is almost always reached, irrespective of the process in which the equilibrium state was reached. On the other hand, for liquids, there are many minima in the adiabatic potential.  Indeed, in adiabatic MD simulations, no thermodynamical significance of these minima is seen: the total energy is constant over various atom configurations, and there is no statistical difference between $A_{1}$ and $A_{2}$ in the figure. The adiabatic potential and thereby the energy barrier $E_{b}$ vary when the environment changes, as is demonstrated throughout this review.
When the relaxation processes are changed by some reason, the energy barrier is also changed. Then, the position of the local minimum is changed, as indicated by a red line in the figure. The equilibrium temperature is also changed. If an external heat bath is introduced in the MD simulation, the relaxation process is affected by the heat bath, and the adiabatic potential is changed accordingly. Consequently, the $U-T$ relation is changed from the original one. In effect, the specific heat and the melting temperature are altered. This is clearly undesirable, and this is why we avoid the use of a heat bath, even though it usually is innocuous. 
Further details of our approach are described in section \ref{sec:adiabatic-relaxation-MD}. 

\subsection{Generality of glass problems}
This review treats liquids and solids (specially glasses) on an equal footing. This is motivated by the utility of specific heat measurement.
Melting is the most familiar phase transition for solid-state physicists. Melting occurs at a discrete temperature $T_{m}$ called melting point. However, there are wider-class structural changes, such as polymerization and denaturation of proteins \cite{Haynie} for nonperiodic systems, whether or not they are called phase transitions. These phenomena share common properties with the glass case in that there is lack of the long-range order and thereby broadening of the transition temperature. The specific-heat measurement is widely and actively used for detecting such structural changes of nonperiodic systems beyond the realm of solid state physics \cite{Netsu-sokutei-HB-2}. We should know how the specific heat is changed without assuming the lattice periodicity. A good example is the glass transition. The glass transition is identified by the specific heat jump $\Delta C_{\rm gl}$. Although the glass transition has long been studied \cite{Jackle86,Gotze92,Angell99,Lubchenko07,  Heuer08,Berthier11,Wolynes12,Biroli13,Stillinger13,Berthier16}, the nature of this jump has not been understood well. Recently, by using the adiabatic relaxation simulation, the author's group has succeeded in reproducing the specific-heat jump at glass transitions without any empirical parameter \cite{Shirai22-SH,Shirai23-Silica}. The calculated values are comparable to the experimental values. Their studies clarify that the glass transition is essentially a structural transition but the transition is broadened owing to the finite values of relaxation times.
The effect of atom relaxation is reasonably expected for the glass transition, because the viscosities of glass-forming liquids are very high. On the other hand, its effect in normal liquids is not obvious. However, the effect appears in the temperature dependence of $C_{V}$ in normal liquids. This is demonstrated by a recent paper \cite{Shirai-EntropyLiquid}. Our approach is thus general to be applicable to any liquid. 
A constraint in the GB statement is a sort of discrete object of presence or absence. However, as we know the substance of constraint in a material is the energy barrier for atom motion, we recognize that the presence/absence of the constraint hinges upon the scope of a given problem. Any equilibrium is subjected to the constraint $E_{b,j}$. A finite value of the energy barrier renders the relaxation time $\tau_{j}$ finite through
\begin{equation}
\tau_{j} = \tau_{0} \exp\left( E_{b,j}/k_{\rm B}T \right), 
\label{eq:tau}
\end{equation}
where $k_{\rm B}$ is Boltzmann's constant and $\tau_{0}$ is the inverse of the attempt frequency of changes. There are many kinds of structural changes. Relaxation and growth are different sides of the same phenomenon. The characteristic time of the former is the relaxation time (or lifetime) $\tau_{\rm R}$, while the later is the growth time $\tau_{\rm G}$. To emphasize this nature of equilibrium, the timescale associated with a specific type of equilibrium is frequently cited throughout this review.

Traditionally, glass is treated separately from other materials due to its nonequilibrium character \cite{Jackle86,Gotze92,Angell99,Lubchenko07,Heuer08,Berthier11,Wolynes12,Biroli13,Stillinger13}. The commonly claimed reason for the nonequilibrium is that the current states of glasses are changing their structures toward the true equilibrium state, even though the changing rate is very slow. Here is the fundamental issue that the author casts questions, which are deeply discussed in section \ref{sec:discussion-glass}. In the thermodynamics context, the most difficult problem is the distinction between the static state and the almost static state with an infinitesimally slow rate of change. This problem was studied by Gibbs more than one century ago (\cite{Gibbs-TD}, p.~56), but unfortunately, the Gibbs solution is not widely shared. Nobody suspects that the gas in a container is in equilibrium once a uniform distribution is achieved. However, even a gas cylinder whose wall is made of robust metal has gas leakage, and it cannot retain the gas pressure beyond some years. A glacier is static in our daily timescale, but it moves in a geological timescale. Though the U$^{238}$ isotope is a stable nuclide, it will decay in $10^{10}$ years (Ref.~\cite{BerkleyPhysics-4}, in paragraph 37 of Chap.~7). Real materials always have defects to some extent and this causes degradation over a long time \cite{note-Eq-defect}. 
All these slow changes are relaxation processes towards a new equilibrium state. The relation between equilibrium and relaxation, which is the fundamental problem of glass, is a common problem in other materials, even though the timescale is different. The timescale of MD simulations is so short that, by cooling a liquid, we always obtain the amorphous phase for any material, unless a special technique is used.
Because MD simulation is the central method in this study, we should describe relaxation processes in a consistent manner, irrespective of the timescale or material used. We should not exclude glass as an exceptional case.

\subsection{Outline of this review}
In this review, we describe the foundations of the total energy approach for the specific heat of liquids (section \ref{sec:theory}) and demonstrate its usefulness in real calculations (section \ref{sec:Results}). The method of the adiabatic relaxation simulation is described at the last part of section \ref{sec:TheoryRelaxation}. As described above, the method itself is not new and not difficult to understand. The real difficulties come when the method is combined with thermodynamics. Equilibrium is the central issue for thermodynamics, while DFT does not identify which is the equilibrium state.
The author tried to write this review in a self-contained manner within a reasonable length. Fundamental problems of thermodynamics are, however, very deep and somewhat abstract. For a full understanding, lengthy explanations are required. These include the definition of the state variables of solids \cite{Shirai18-StateVariable} and glass \cite{Shirai22-SH}, the thermodynamic treatment of hysteresis \cite{Shirai24-hysteresis,Shirai23-Silica}, and the order parameter \cite{Shirai25-OrderParams}, which are explained with sufficient references in the respective papers. Many readers may be reluctant to read further lengthly references in order to understand this review. In this case, the author recommends those readers to judge the correctness of the theory on the basis of the results that the theory leads to. For doing so, it is vital to know experimental facts on the specific heats of liquids, which are reviewed in section \ref{sec:experiment}. Impatient readers may jump directly to section \ref{sec:Results} after reading section \ref{sec:experiment}.
If readers stick to the conventional view that hysteresis means nonequilibrium, reading first section \ref{sec:hysteresis-SH} is suggested. This study provides deep insights into the fundaments of thermodynamics. It impacts various areas, from the fundaments of statistical mechanics to the fronts of material research, and hence, it is worth suggesting these orientations of research, which are given in section \ref{sec:Development}.
%

\section{Facts of liquids}
\label{sec:experiment}
There is no difficulty in finding data on the specific heats of liquids: abundant data are available in handbooks and databases \cite{TPRC-6,Breitling72,nist-fluid-data}. New and updated data are published in J.~Phys.~Chem.~Ref.~Data. The difficulty of liquids lies in finding something meaningful behind these numerical values. Let us first review experimental facts about the specific heats of liquids in order to settle the targets of the theory.
First, the difference between isobaric $C_{P}$ and isochoric specific heat $C_{V}$ is noted.
For solids, the difference is given by 
\begin{equation}
C_{P} = C_{V} + C_{\rm te}.
\label{eq:Cp}
\end{equation}
The thermal expansion contribution $C_{\rm te}$ is given by
\begin{equation}
C_{\rm te} = TV \frac{(\alpha_{P})^{2}}{\kappa_{T}} ,
\label{eq:Cte}
\end{equation}
where $\alpha_{P}$ is the isobaric thermal expansion coefficient and $\kappa_{T}$ is the isothermal compressibility. It is useful to note that, in the limiting case of ideal gases, $\kappa_{T}$ becomes $\kappa_{T} = 1/P$ and $\alpha_{P}$ becomes $\alpha_{P} = 1/T$, which leads to the well-know formula $C_{\rm te}=R$. For solids, the contribution of the thermal expansion is normally small, i.e., $C_{\rm te} \ll R$, at room temperature. $C_{\rm te}$ increases with $T$, as indicated by equation~(\ref{eq:Cte}). Typical  $C_{\rm te}$ values of simple metal liquids are in the range of
\begin{equation}
C_{\rm te} = 0.4 \sim 0.5 R,
\label{eq:CllR}
\end{equation}
just above $T=T_{m}$. However, it sometimes exceeds $R$: rare-gas liquids are such examples \cite{Gladun66,Gladun71}. 
It is important to recognize that the value $C_{\rm te}=R$ in the case of ideal gas is not the upper bound of $C_{\rm te}$. The divergent property in $C_{P}$ near the critical point can be ascribed to the contribution of $C_{\rm te}$ in most cases \cite{Pippard}. However, this divergent property is debated: see \cite{Sengers15,Woodcock17,Woodcock18}. 
Bearing these facts about $C_{\rm te}$ in mind, we describe mainly $C_{V}$ of liquids hereafter.

\begin{figure}[ht!]
\centering
    \includegraphics[width=120mm, bb=0 0 900 768]{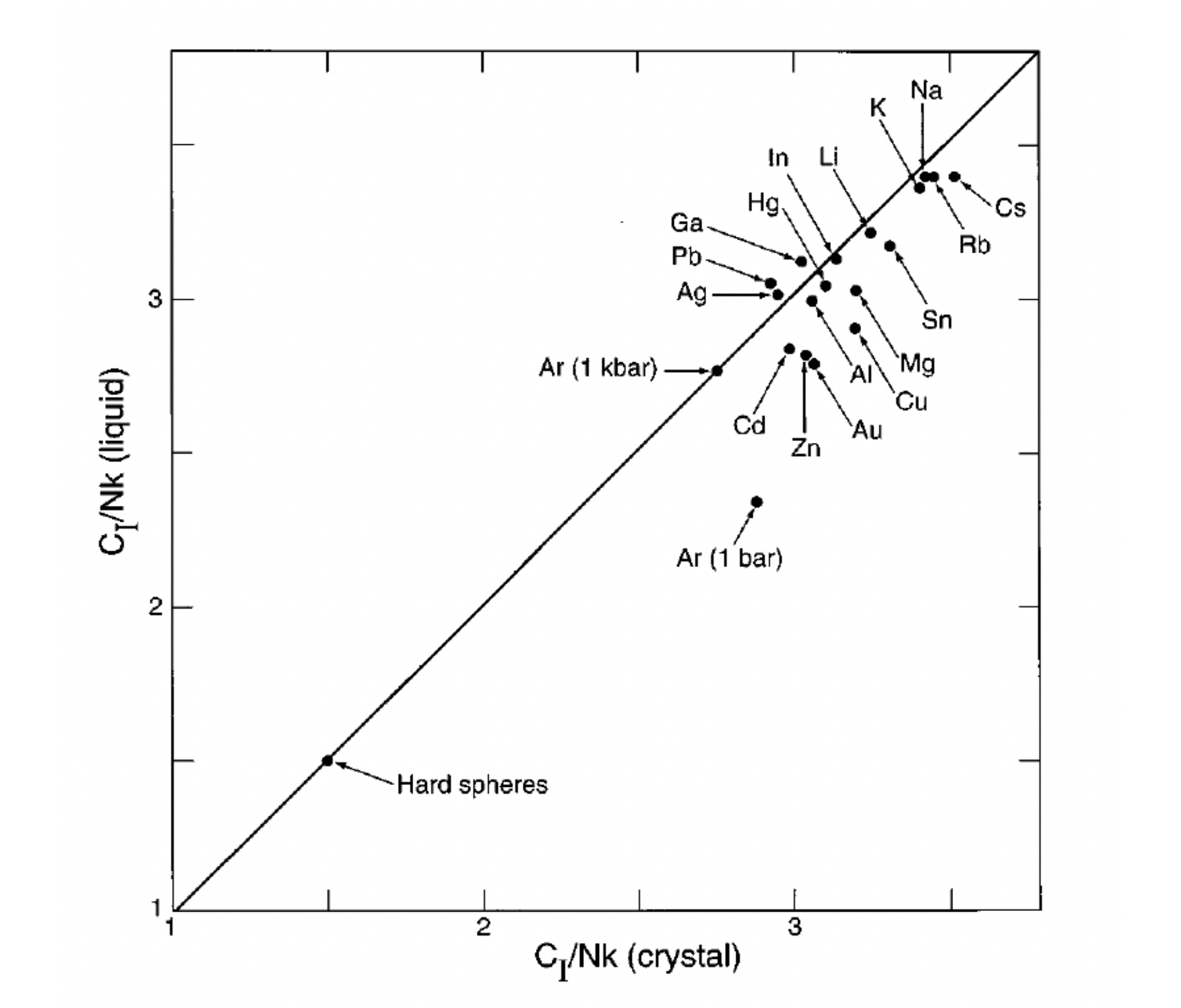} 
\caption{
Comparison of $C_{V}$'s of the crystal and liquid for simple metals near the melting temperature $T_{m}$. The figure is reproduced from the original paper of Wallace with permission \cite{Wallace98}. $C_{I}$ in the figure denotes the contribution of ions to $C_{V}$.}
\label{fig:Wallace98-1} 
\end{figure}

\subsection{$C_{V}$ near melting temperature}
\label{sec:CVatTm}
Although abundant experimental data of the specific heat of liquids are available in handbooks, most of the data are of isobaric specific heat $C_{P}$, in which the two contributions from $C_{V}$ and $C_{\rm te}$ are mixed. 
It is a great contribution of Wallace to compile the data on the isochoric specific heat $C_{V}$ of various single-component materials \cite{Wallace98,Wallace02}. 
In figure~\ref{fig:Wallace98-1}, his comparison of $C_{V}$'s of various monatomic metals near the melting temperature ($T_{m}$) is reproduced. $C_{I}$ in the figure means the contribution of ions to $C_{V}$ in order to distinguish the contribution of electrons ($C_{\rm el}$). The electronic contribution is usually very small, i.e., $C_{\rm el} < 0.1 R$, so that this contribution is ignored in this review.
Figure \ref{fig:Wallace98-1} clearly shows that $C_{V}$'s of both the solid and liquid are close to $3R$, i.e., the classical limit of the Dulong--Petit law. The fact of $C_{V} \approx 3R$ for solids is reasonably understood as $T_{m}$ is high enough to reach the classical regime such that the equipartition law holds. However, finding $C_{V} \approx 3R$ for liquid is somewhat surprising, because it is not uncommon to observe as large $C_{P}$ values as $C_{P} > 4 R$ for liquids. Finding $C_{V} \approx 3R$ thus convinced us to consider that the phonon model is applicable to liquids too. Of course, in the strict sense, the notion of phonons is not valid for liquids because of the lack of periodicity of crystals. However, there are sound waves in liquids too. Sound waves are density-fluctuating waves and correspond to the longitudinal acoustic mode in crystals. For the transverse acoustic mode, there seems to be no corresponding waves in liquids. However, Wallace pointed out that shear viscosity exists for liquids, and this somehow plays the role of restoring force for the transverse displacements of atoms. Accordingly, all the degrees of freedom $f=6$ are present for monatomic liquids.
The results of inelastic neutron and X-ray scattering experiments promote the idea of phonons for liquids. Phonon-like dispersions are observed from inelastic neutron scattering experiments \cite{Egelstaff-2ed,Copey74,Smith17,Khusnutdinoff20}, although normally they are referred to as collective modes. The observation of even a Brillouin-zone-like structure is surprising \cite{Suck05}. 

On closer observation, we notice that $C_{V}$ at $T_{m}$ is slightly larger than $3R$ for solids. Often, the amount exceeding $3R$ is called the excess specific heat. For solids, the excess specific heat is attributed to the contribution from anharmonicity \cite{Leibfried61}. For liquids, many $C_{V}$'s are also slightly larger than $3R$. In the case of liquids, the attribution of the excess specific heat to the anharmonic heat is suspicious, as explained later.

\subsection{Temperature dependence of $C_{V}$}
\label{sec:TdepCv}

\begin{figure}[ht!]
\centering
    \includegraphics[width=120mm, bb=0 0 420 330]{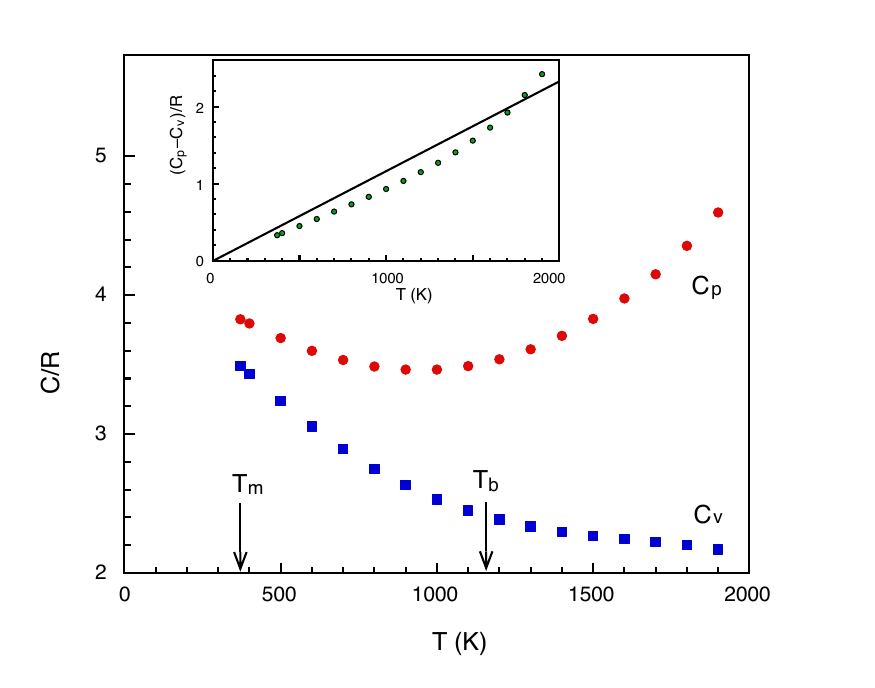} 
\caption{
Experimental values of $C_{P}$ and $C_{V}$ of liquid Na \cite{Fink95}.
The inset shows the difference $C_{P}-C_{V}$. The line in the inset indicates the $C_{\rm te}$ values calculated using equation~(\ref{eq:Cte}) with constant $\alpha_{P}$ and $\kappa_{T}$. The figure is reproduced from the original paper with permission \cite{Shirai-EntropyLiquid}.
}
\label{fig:Fink}
\end{figure}

For solids, the specific heat always increases with $T$. On the other hand, for liquids, it is not clear whether there is a general rule about the $T$ dependence of $C$. 
As an example, experimental data of liquid Na are shown in figure~\ref{fig:Fink}. $C_{P}$ initially decreases with increasing $T$ from $T_{m}$ and then it begins to decrease near the boiling temperature ($T_{b}$) \cite{Fink95}. In contrast, $C_{V}$ monotonically decreases with $T$ over the entire temperature range. 
The difference $C_{P}-C_{V}$ is shown in the inset of figure~\ref{fig:Fink}, with the $C_{\rm te}$ values calculated by equation~(\ref{eq:Cte}). Note that $C_{V}$ was obtained by a more involved method in the original study \cite{Fink95}.
From this comparison, we can understand that $C_{V}$ monotonically decreases with increasing $T$ and the initial decrease in $C_{P}$ near $T_{m}$ reflects this decrease in $C_{V}$. The contribution of $C_{\rm te}$ increases almost linearly in $T$, and it turns to overcompensate the decrease in $C_{V}$ near $T_{b}$. 

\begin{figure}[ht!]
\centering
    \includegraphics[width=120mm, bb=0 0 1024 768]{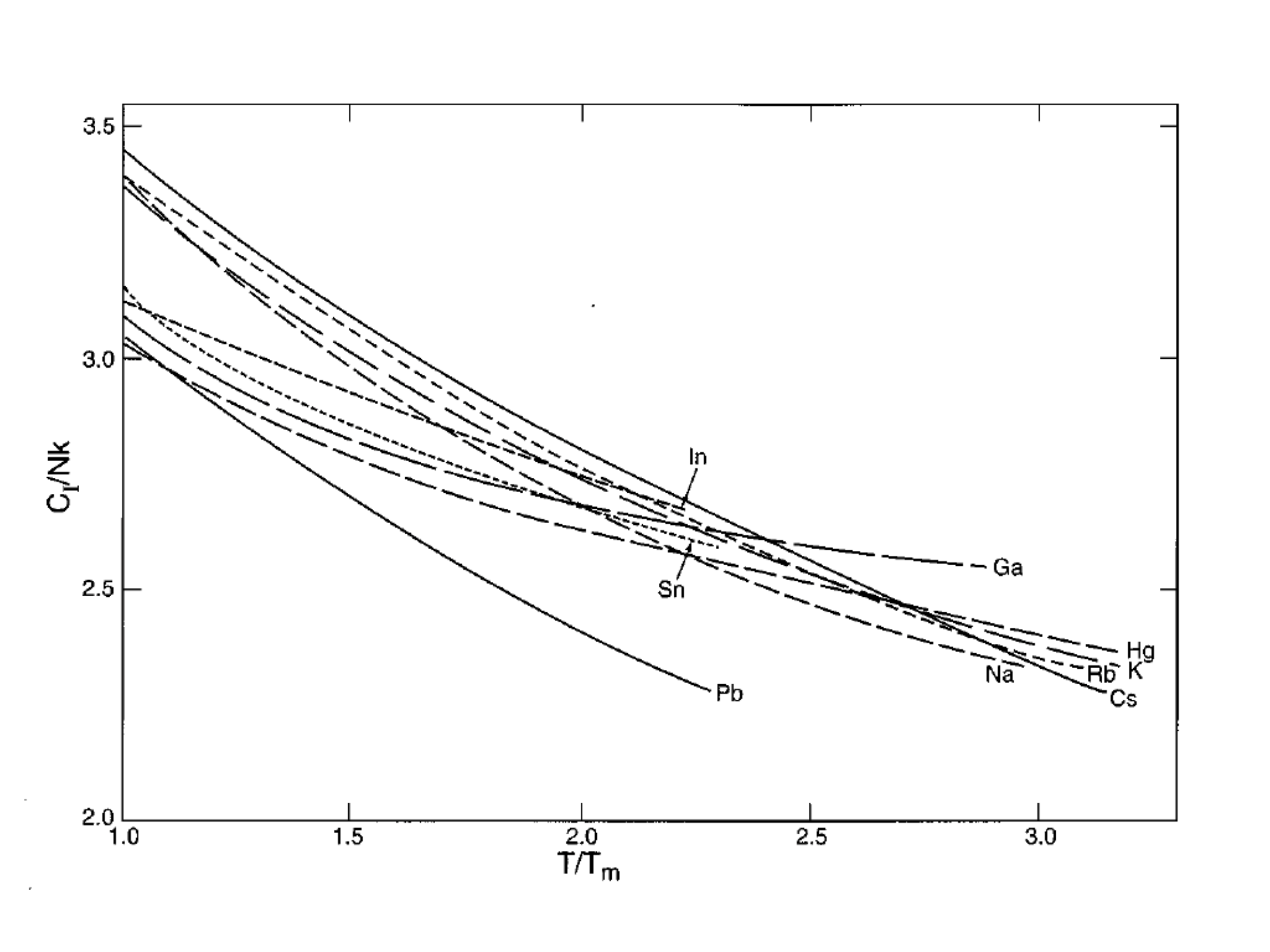} 
\caption{
Temperature dependence of $C_{V}$ of monatonic metal liquids. The figure is reproduced from the original paper of Wallace with permission \cite{Wallace98}.}
\label{fig:Wallace98-2} 
\end{figure}

It is also the contribution of Wallace that the decreasing behavior of $C_{V}$ is a common property of liquids, unless special mechanisms are involved. He examined the temperature dependence of $C_{V}$ for various liquids \cite{Wallace98}. His compilation is reproduced in figure~\ref{fig:Wallace98-2} from that paper. 
In all the cases, $C_{V}$ is approximately $3R$ just above $T_{m}$ and gradually decreases to approximately $2R$ at the boiling temperature $T_{b}$ \cite{Wallace98,Wallace02}.
This decreasing behavior is not linear in $T$, but it is useful to approximate the data as
\begin{equation}
\frac{dC_{V}}{dT} \sim -10^{-3} R {\rm /K}.
\label{eq:decreaseC}
\end{equation}
The chief subject of the theory on the specific heat of liquids is thus to give a quantitative account to this relationship.
At this moment, we infer that equation~(\ref{eq:decreaseC}) can be recast as
\begin{equation}
\frac{dC_{V}}{dT} \sim -\frac{C_{V}}{T}.
\label{eq:activationC}
\end{equation}
From this, an Arrhenius-type relation, $C_{V} \propto \exp( Q/k_{\rm B}T )$, is inspired. The suitability of this form is further elaborated in section~\ref{sec:TdepC-result}.

\subsection{$C_{V}$ of supercooled liquids}
\label{sec:CvofSL}

\subsubsection{Supercooled liquid state}
\label{sec:stateSL}

\begin{figure}[ht!]
\centering
    \includegraphics[width=120mm, bb=0 0 1024 768]{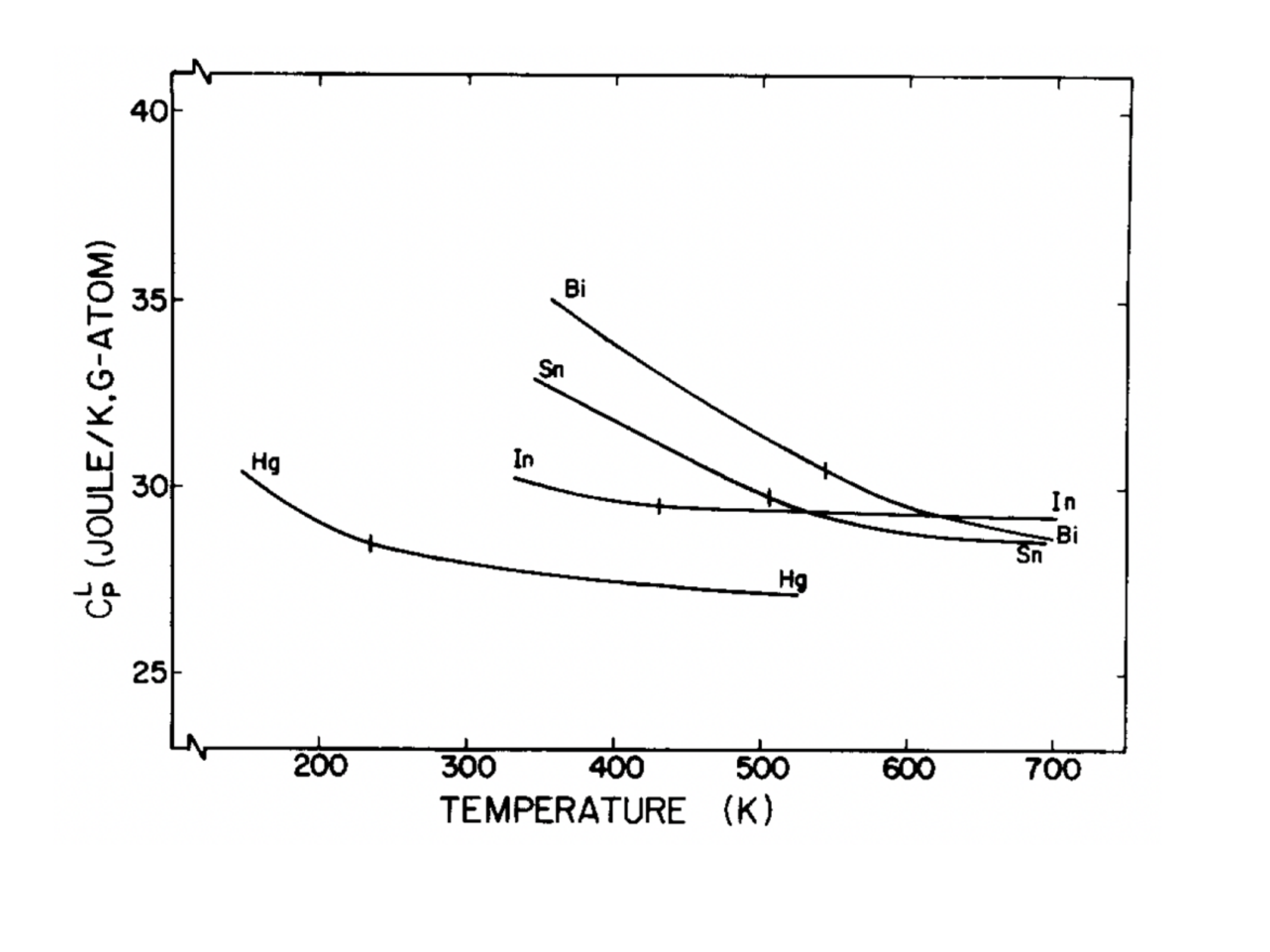} 
\caption{
Isobaric specific heat of various liquids. $T_{m}$ is marked by vertical bars. Note that $3R$ corresponds to 24.9 J/K$\cdot$gram-atom. The figure is reproduced from the original paper by Perepezko and Paik  with permission \cite{Perepezko84}.}
\label{fig:Perepezko84} 
\end{figure}

It is interesting to extend the temperature range of measurement of liquids below $T_{m}$, where the liquid is in a metastable state called the supercooled liquid. Although crystallization usually takes place immediately in the human timescale, a finite time is required to complete the crystallization process, which is called the crystal nucleation time $\tau_{\rm G,cr}$. Typically, $\tau_{\rm G,cr}$ ranges from ns to $\mu$s. Therefore, in normal experiments, the condition 
\begin{equation}
t_{\rm obs} \gg \tau_{\rm G,cr}(T_{m})
\label{eq:Time-nucleation}
\end{equation}
is easily satisfied, where $t_{\rm obs}$ is the time period of our observation. This condition can be regarded as the equilibrium condition for the obtained crystal.
For bulk liquids, it is difficult to cool at as faster a rate as $\tau_{\rm G,cr}^{-1}$. However, if fast cooling is,  by some means, realized to meet the condition
\begin{equation}
t_{\rm obs} < \tau_{\rm G,cr}(T_{m}),
\label{eq:quench}
\end{equation}
the crystallization is impeded and the supercooled liquid is obtained. The supercooled liquid state is stable in the time period $\tau_{\rm G,cr}$. Here, we have emphasized the $T$ dependence of $\tau_{\rm G,cr}$ by explicitly indicating it in the functional form; the $T$ dependence is very strong and is not even a monotonous function of $T$ \cite{Chandler10}. In fact, experimentalists have realized supercooled states for bulk materials by special techniques, such as flux method \cite{Kui84,Devaud85,Wilde94,Tachibana17}. For vast amounts of crystal growth, the growth rate is known to be controlled by heterogeneous nucleations at the surface/interface \cite{Shewmon69}. The main idea behind these techniques of creating supercooled liquids is that the heterogeneous nucleations are suppressed by immersing the target material into liquid media, rendering $\tau_{\rm G,cr}(T_{m})$ much longer. Such examples are shown in figure~\ref{fig:Perepezko84}, which is reproduced from the study by Perepezko and Paik \cite{Perepezko84}. They employed the droplet emulsion method to obtain the supercooled liquids. In this plot, the unit of gram-atom is used \cite{Brady80}. Although the plotted data are of $C_{P}$, by considering the general trend in equation (\ref{eq:CllR}), we see that $C_{V}$ must be close to $3R$ at $T_{m}$. Similar data are obtained for bulk metallic glasses \cite{Busch07}. 

It is observed that the specific heat increases quickly as $T$ decreases below $T_{m}$. 
We first note that there is no discontinuity in $C_{P}(T)$ at $T_{m}$. There is no qualitative distinction between the normal and supercooled liquids. For this reason, the distinction between them is not made in many notations in the followings, unless special need arises. They are distinguished only in relation to $T_{m}$. Recently, it has become possible to realize the supercooled liquid state for more difficult crystals, such as Ge \cite{Rhim00a,Li04} and Si \cite{Rhim97}, by the electromagnetic levitation method, which is a container-less method.
(Even though the specimen does not have any contact with a container, the droplet still has a surface. However, inhomogeneous nucleation is greatly suppressed by eliminating the strong inhomogeneities presented at the interface with a different material, which has many impurities and defects.)
The continuous increasing behavior at $T<T_{m}$ is observed also for these experiments.
Second, because $C_{V}$ already reaches $3R$ at $T_{m}$ and the contribution of $C_{\rm te}$ becomes smaller as $T$ decreases, the increase in  $C_{P}$ with decreasing $T$ means that $C_{V}$ becomes far larger than $3R$ below $T_{m}$.  As mentioned above, there is the view that the excess specific heat of liquids can be ascribed to anharmonic effects. The phonon contribution including the anharmonic effect is expected to decrease by lowering $T$ \cite{Leibfried61}. The continuous increase from $3R$ by lowering temperature thus cannot be explained by the phonon model. 
Third, even with the same sign of the negative temperature coefficient on $C_{V}$ at high temperatures, $T>T_{m}$, the mechanism of the $T$ dependence may be different in the magnitude. In contrast to the weak decrease in $C_{V}$ with $T$ near $T_{b}$, the increase in $C_{V}$ with decreasing $T$ at $T<T_{m}$ is very large and $C_{V}$ far exceeds $3R$. There is no good theory to explain this strong increase.

The condition (\ref{eq:quench}) suggests that the supercooled liquid will eventually crystallize if we wait for a long time. When a constant cooling rate $v_{c}$ is employed, after the supercooled liquid state is retained in a certain time period, it finally crystallizes at a certain temperature $T_{sc}$, which satisfies the condition
\begin{equation}
t_{\rm obs} = \tau_{\rm G,cr}(T_{sc}).
\label{eq:Tsc}
\end{equation}
$T_{sc}$ depends on $v_{c}$, so that $T_{sc}$ is a kinetic parameter. This crystallization process is associated with heat emission, which is called the recalescence heat $H_{\rm re}$, similar to the latent heat at $T_{m}$.

\subsubsection{Glass transition in the calorimetric measurement}
\label{sec:GTexp}

\begin{figure}[htbp]
    \centering
    \includegraphics[width=100 mm, bb=0 0 450 280]{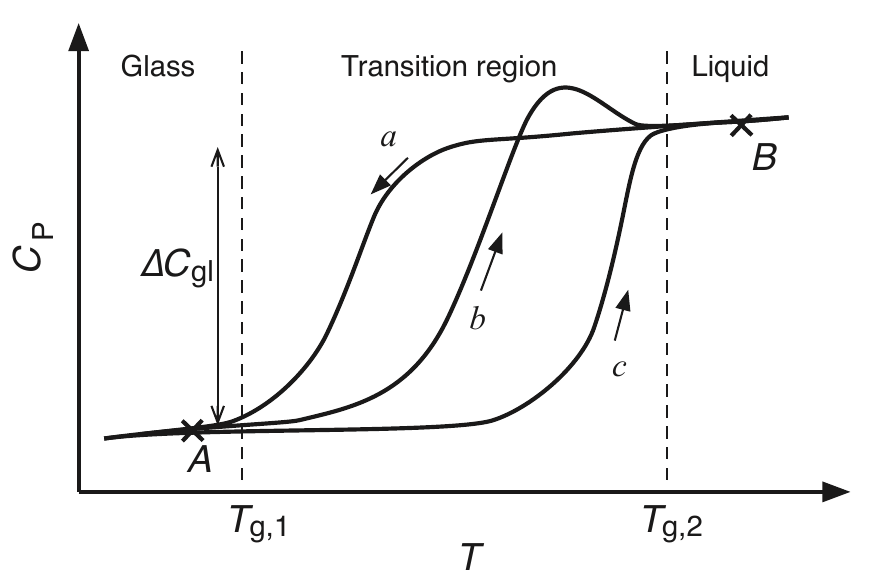} 
  \caption{Glass transition observed by the $C-T$ curve. Different cooling/heating rates are used. A faster rate is used in (b) than in (c).
  } 
  \label{fig:CT_Hysteresis}
\end{figure}
For the class of materials called glass, under the supercooled condition, condensation to a form of amorphous takes place before the crystal nucleation. This occurs when the condensation time $\tau_{\rm G,gl}$ in the glass is shorter than the nucleation time of the crystal \cite{note-GlassNucleat},
\begin{equation}
\tau_{\rm G,gl}(T_{g}) < \tau_{\rm G,cr}(T_{g}).
\label{eq:glass-trans}
\end{equation}
In equation~(\ref{eq:glass-trans}), $T_{g}$ is the glass transition temperature. The transition temperature is identified by the steep change in the $C-T$ curve, as shown in figure~\ref{fig:CT_Hysteresis}. Although the $T_{g}$ obtained in this manner is sometimes called by the specific name of calorimetric glass transition temperature, the steep changes are also observed in other properties, such as compressibility and thermal expansion, at the same temperature \cite{Kauzmann48}. Hence, it is reasonable to qualify the $T_{g}$ value as the true transition temperature. Unlike typical phase transitions, however, this transition occurs in a temperature range from $T_{g,1}$ to $T_{g,2}$: it has the width $\Delta T_{g} = T_{g,2}-T_{g,1}$. The temperature range within this width is referred to as {\it the transition region} \cite{Moynihan74}. Hence, there is no latent heat, and instead the jump in specific heat $\Delta C_{\rm gl}$ is observed in a narrow range $\Delta T_{g}$,
\begin{equation}
\Delta C_{\rm gl} = C_{P}^{\rm (sl)} - C_{P}^{\rm (gl)},
\label{eq:Cjump-glass}
\end{equation}
where $C_{P}^{\rm (sl)}$ and $C_{P}^{\rm (gl)}$ are isobaric specific heats of supercooled liquid and glass, respectively. Since Kauzmann's seminal paper on the glass transition \cite{Kauzmann48} was published, enormous numbers of studies have been accumulated on the specific heat of glass \cite{Wunderlich60,Moynihan74,Debolt76,Trachenko11,Ding12} (see reviews \cite{Hodge94,Nemilov-VitreousState}). Yet a consensus has not been reached regarding the meaning of the specific heat jump $\Delta C_{\rm gl}$. 
There is no useful relationship or empirical rule about the jump $\Delta C_{\rm gl}$ except the Prigogine--Defay (PD) ratio \cite{Nemilov-VitreousState}. The PD ratio (${\it \Pi}$) is given by the ratio of the specific heat jump to the other jumps in thermal expansion ($\Delta \alpha_{\rm gl}$) and compressibility ($\Delta \kappa_{\rm gl}$),
\begin{equation}
{\it \Pi} = \frac{\Delta C_{\rm gl} \Delta \kappa_{\rm gl}}{TV \Delta \alpha_{\rm gl}}.
\label{eq:PDratio}
\end{equation}
For glass materials, the ratio is always larger than unity \cite{Davies53a,O'Reilly62,Moynihan76,Moynihan81},
\begin{equation}
{\it \Pi} > 1.
\label{eq:PDratio}
\end{equation}
For the second-order phase transitions, this ratio must be equal to unity \cite{Pippard}. 
Since the pioneer work of Davies and Jones \cite{Davies53a}, it has been known that the deviation from unity is ascribed to the existence of plural order parameters. However, until today, it has not been known about what is the substance of this self-contradicting word, the order parameters for disordered systems, although there have been many proposals.

A difficulty in describing $\Delta C_{\rm gl}$ is that $\Delta C_{\rm gl}$ as well as $T_{g}$ has dependence on the cooling rate $v_{c}$ and on the past history in which the current state was obtained \cite{Moynihan71,Moynihan74,Debolt76}. Because of this history dependence, glasses are traditionally considered nonequilibrium materials. The hysteresis in the $C-T$ curve is illustrated in figure~\ref{fig:CT_Hysteresis}. When a glass-forming material is cooled from the liquid state (state $B$ at temperature $T_{B}$), a glass state $A$ is obtained along path $a$. Subsequently by heating $A$ to $T_{B}$, the original liquid state $B$ is recovered along path $b$. We have
\begin{equation}
\oint_{a \rightarrow b} dQ = \oint_{a \rightarrow b} C_{P} dT = 0.
\label{eq:circQzero}
\end{equation}
This equality must hold irrespective of paths, because the final state is the same as the initial liquid state $B$. In contrast, let us start from a glass state $A$ at $T_{A}$. The glass is heated to obtain a liquid state $B$, the heating path $c$ is generally different from $b$ because of the rate dependence. Then, when the liquid state $B$ is cooled down to $T_{A}$, the path integral (\ref{eq:circQzero}) does not vanish, even though the $C_{P}^{\rm (gl)}$ value is recovered:
\begin{equation}
\oint_{c \rightarrow a} C_{P} dT \neq 0.
\label{eq:circQNzero}
\end{equation}
This nonvanishing value (\ref{eq:circQNzero}) indicates that there is a difference in enthalpy, $\Delta H = H(A')-H(A) \neq 0$, meaning that the final state $A'$ is different from the initial state $A$. 
This fact tells us an important learning in thermodynamics that the specifications of the thermodynamic state must be made using a full set of state variables. If we take only $T$ and $C_{P}$ as the state variables, we would conclude that the final state is the same as the initial state. This is tantamount to saying that the same state has different energies, which contradicts the first law of thermodynamics that {\em the energy is a state function} (see \cite{Fermi}, p.~11). The significance that the energy is a state function plays the leading principle for investigating the thermodynamic description of hysteresis (see section \ref{sec:hysteresis-SH}). In either case, the Clausius integration gives negative values:
\begin{equation}
\oint \frac{dQ}{T} < 0.
\label{eq:Clausius-I}
\end{equation}
This says that the glass transition is an irreversible process, which yields many confusing arguments. That the glass transition is an irreversible process is, in itself, correct, but stating that the resulting state is a nonequilibrium state is erroneous.
Remember that the vapor in a turbine in an electric power generator undergoes an irreversible process because of strong turbulence and much heat loss. Yet the rejected water can be well described using only $T$ and $V$. See \cite{Shirai24-hysteresis}. The view that nonequilibrium processes result in nonequilibrium states contradicts the GB statement of the second law: there is always one unique equilibrium state for any spontaneous change of an isolated system.

Regarding the rate dependence of the glass transition, it is often claimed, in an exaggerated manner, that the transition temperature $T_{g}$ can be varied as low as desired if sufficiently slow cooling is realized. The fact is, however, that the calorimetric method determines $T_{g}$ within a few degrees K even though the cooling rate varies by several orders of magnitude \cite{Hodge94, Suga80}. The $T_{g}$ value converges within a reasonable range \cite{Mazurin07,Bruning92} if the experiment is performed with sufficiently slow cooling rates. The $T_{g}$ values are well established to be tabulated in standard references \cite{Kauzmann48,Rao02,Nemilov-VitreousState,Mysen-Richet05}, similar to melting temperatures. 
The same holds for the specific heat. By observing the hysteresis in the $C-T$ curve, some researchers claim that the specific heat of glass is not a material property but represents the process \cite{Hodge94}. Even though the specific heat $C_{P}(T)$ during the transition region changes as the cooling rate $v_{c}$ varies, the value of $C_{P}^{\rm (gl)}$ below $T<T_{g,1}$ is well reproduced \cite{Oblad37,Pohl81}, and accordingly investigating $\Delta C_{\rm gl}$ as the material property makes sense. The $\Delta C_{\rm gl}$ values are also available in the above references \cite{Kauzmann48,Rao02,Nemilov-VitreousState,Mysen-Richet05}. How to interpret $\Delta C_{\rm gl}$ is the current subject in this field. In a recent study,  Trachenko and Brazhkin presented an interesting approach to this problem by using a few parameters \cite{Trachenko11}.

In the temperature range $T_{g} < T<T_{m}$, the specific heat of supercooled liquid $C_{P}^{\rm (sl)}$ is larger than that of the corresponding crystal $C_{P}^{\rm (cr)}$. The difference is referred to as {\it the excess specific heat}, $C_{\rm ex}^{\rm (sl)}$, of the glass (or the supercooled liquid),
\begin{equation}
C_{\rm ex}^{\rm (sl)}(T) = C_{P}^{\rm (sl)}(T) - C_{P}^{\rm (cr)}(T).
\label{eq:excessC}
\end{equation}
Today, the excess specific heat of the supercooled liquid is generally interpreted as the contribution of {\it configuration}. 
When quantitative analysis is attempted to decompose the excess specific heat into the contributions of configuration and anharmonic effects, however, much confusion arises \cite{Goldstein76,Johari00,Smith17,Alvarez-Donado20}, and this issue is still unresolved today.

\section{Theory of specific heat}
\label{sec:theory}

\subsection{Total energy vs.~elemental excitation approaches}
\label{sec:Total-E-approach}

\subsubsection{General formulation}
\label{sec:GeneralForm}
DFT is a theory of ground states. The liquid state is not the ground state of material, and it exists only at finite temperatures, with the exception of liquid helium. An extension of DFT to finite-temperature systems was formulated by Mermin, who established that the thermodynamic potential of electrons is given by a functional of the equilibrium electron density \cite{Mermin65}. Recently, this notion of the electronic entropy has been recast within the general framework of the maximum-entropy principle by Yousefi and Caticha \cite{Yousefi24}. Thus, in principle, thermodynamic functions of a liquid can be expressed as functionals of the equilibrium electron density at a given temperature, $\bar{\rho}({\bf r}, T)$. Recently, efforts have been devoted to finding explicit expressions for the free energy of liquids as a functional of the atom density profile plus additional variables if necessary \cite{Oxtoby02,Wu-Li07}. 
However, knowing the concrete form of the functional appears unnecessary. Because of the large difference in the mass between electrons and ions, the electron subsystem is decoupled from the ion subsystem (the Born--Oppenheimer approximation) \cite{BornHuang}. The direct calculation of specific heat by MD simulations enables us to calculate all thermodynamic functions by equation~(\ref{eq:TDdef-S}).

Let us consider an adiabatic MD simulation of a system composed of $N$ atoms, irrespective of whether the system is a solid or a liquid. The $j$th atom having mass $M_{j}$ is at position ${\bf R}_{j}(t)$ with velocity ${\bf v}_{j}(t)$ at time $t$. In the Born--Oppenheimer approximation, the total energy $E_{\rm tot}$ of the system is given by the sum of the kinetic energy $E_{\rm K}$ and the potential energy $E_{\rm P}$ as
\begin{equation}
E_{\rm tot}(t) \equiv  E_{\rm P}(t) + E_{\rm K}(t) = 
    E_{\rm gs}( \{ {\bf R}_{j}(t) \}) + \frac{1}{2} \sum_{j} M_{j} v_{j}(t)^{2},
\label{eq:total-energy}
\end{equation}
where $E_{\rm gs}( \{ {\bf R}_{j}(t) \})$ is the ground state energy of the electrons, including the ion--ion potentials, for the instantaneous positions $\{ {\bf R}_{j}(t) \}$.
The internal energy in thermodynamics is defined at equilibrium and given by the time average of the microscopic total energy $E_{\rm tot}(t)$ of the system as
\begin{equation}
U = \overline{E_{\rm tot}(t)} = \overline{E_{\rm gs}( \{ {\bf R}_{j}(t) \})}
    + \frac{1}{2} \sum_{j} M_{j} \overline{ v_{j}(t)^{2} }.
\label{eq:internal-energy}
\end{equation}
When the volume $V$ of the system is fixed, the isochoric specific heat $C_{V}$ is obtained as $C_{V} = (\partial U/\partial T)_{V}$. This is the essence of the total energy approach.
Here, the electronic contribution to the specific heat ($C_{\rm el}$) is ignored. This treatment is appropriate for many materials. 

\subsubsection{Solid case}
\label{sec:Total-E-solid}
\begin{figure}[ht!]
\centering
    \includegraphics[width=100mm, bb=0 0 650 450]{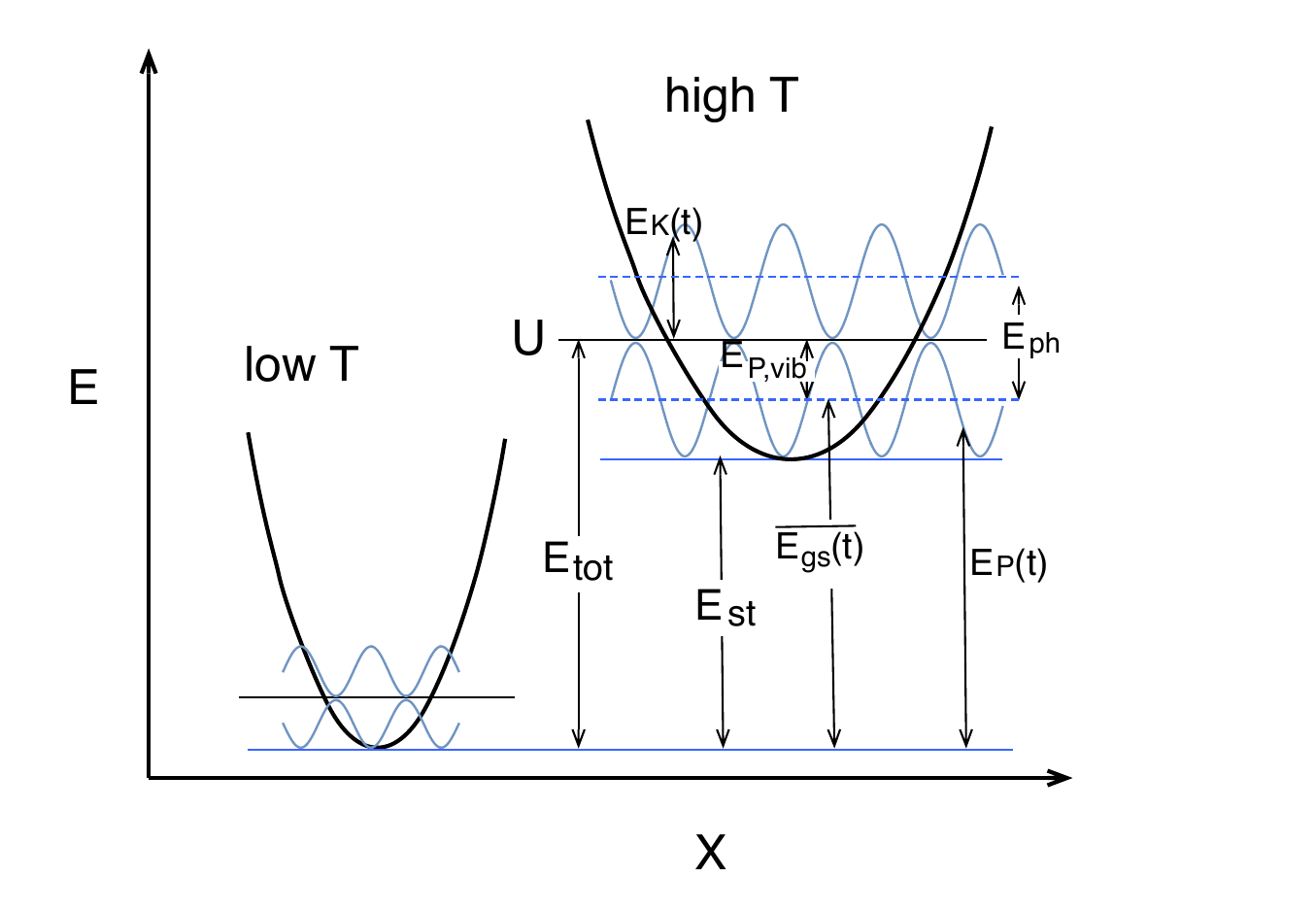} 
\caption{
Relationships between various energies in adiabatic MD simulations. As $T$ increases, the bottom of the adiabatic potential is raised by $E_{\rm st}$.
}
\label{fig:E-Xcurve} 
\end{figure}

For solids, we can go further. In this case, the constituent atoms fluctuate around their equilibrium positions. The instantaneous position of the $j$th atom can be expressed by the sum of the equilibrium position $\bar{\bf R}_{j}$ and a small displacement $\bar{\bf u}_{j}$ from this position as
\begin{equation}
{\bf R}_{j}(t) = \bar{\bf R}_{j} + {\bf u}_{j}(t).
\label{eq:Rexpansion}
\end{equation} 
The ground-state energy $E_{\rm gs}( \{ {\bf R}_{j}(t) \} )$ can be expanded in the Taylor series in terms of this displacement as
\begin{equation}
E_{\rm gs}( \{ {\bf R}_{j}(t) \} ) = E^{(0)} + E^{(2)} + E^{(3)} + \dots ,
\label{eq:gs-Taylor}
\end{equation}
where the linear term vanishes because of the equilibrium conditions. The zeroth and second terms are given by
\begin{equation}
E^{(0)} = E_{\rm st}( \{ \bar{\bf R}_{j} \} ), 
\quad E^{(2)} = \frac{1}{2} \sum_{i,j}  {\bf u}_{i}(t) \cdot {\bf D}_{ij} \cdot {\bf u}_{j}(t) \equiv E_{\rm P, vib},
\label{eq:gs-Taylor1}
\end{equation}
where ${\bf D}_{ij}$ is the dynamic matrix between the $i$th and $j$th atoms \cite{BornHuang}. $E_{\rm P, vib}$ together with $E_{\rm K}$ constitutes the phonon energy and its time average $E_{\rm ph}$ is expressed as
\begin{equation}
E_{\rm ph} = \frac{1}{2} \sum_{i,j} \overline{ {\bf u}_{i}(t)\cdot {\bf D}_{ij} \cdot {\bf u}_{j}(t)}
    + \frac{1}{2} \sum_{j} M_{j} \overline{ v_{j}(t)^{2} }.
\label{eq:phonon-energy}
\end{equation}
The zeroth term $E_{\rm st}( \{ \bar{\bf R}_{j} \} )$ is constant with respect to time and is here referred to as the {\em structural energy}. Thus, the internal energy $U^{\rm (so)}_{V}$ of a solid is decomposed of
\begin{equation}
U^{\rm (so)}_{V} \equiv U^{\rm (so)}_{V}(T,\{ \bar{\bf R}_{j} \} ) = E_{\rm st}( \{ \bar{\bf R}_{j} \} )  +  E_{\rm ph}(T),
\label{eq:U-solid}
\end{equation}
when $V$ is fixed. Correspondingly, the isochoric specific heat of solid becomes
\begin{equation}
C^{\rm (so)}_{V} = C^{\rm (so)}_{V}(T,\{ \bar{\bf R}_{j} \} ) = C_{\rm st}( \{ \bar{\bf R}_{j} \} )  +  C_{\rm ph}(T).
\label{eq:C-solid-R}
\end{equation}
Here, the most important variables determining each component are explicitly written. The existence of the structural part $C_{\rm st}$ was already known as the configurational specific heat by Prigogine and Defay in the early days (\cite{Prigogine54}, p.~293). However, at that time, the argument of this component was expressed by an internal variable or order parameter $Z$ whose substance was not known. Davies and Jones demonstrated that there must be plural order parameters $Z_{j}$ for the glass transition \cite{Davies53a} (see also \cite{Goldstein63,Goldstein73,DiMarzio74,Goldstein75,Gupta76,DiMarzio76,Lesikar80}). The nature of the order parameter has been argued from various viewpoints, such as four-point correlation function \cite{Berthier11}, polyhedron order \cite{Xia15}, orientational order \cite{Yang19}, and overlap function \cite{Parisi83,Franz97,Charbonneau14}. 
From the present context, it is seen that the substances of the order parameters of solids are the equilibrium positions of atoms, $\{ \bar{\bf R}_{j} \}$. In fact, it is demonstrated that the order parameter and state variables are equivalent in terms of the invariant property against thermal fluctuation \cite{Shirai25-OrderParams}.
Figure \ref{fig:E-Xcurve} indicates the summary of the interrelationships among various energies discussed up to here. At low temperatures, at which the harmonic approximation holds well, there is no temperature dependence on $E_{\rm st}$, and accordingly, $C_{V}$ is determined by $C_{\rm ph}$ alone. At high temperatures, $E_{\rm st}$ comes to appear in $C_{V}$. In this manner, $E_{\rm st}$ has strong $T$ dependence, which is reflected in the schematic change in the adiabatic potential in figure~\ref{fig:pot-relax}.

The functional relationship in equation~(\ref{eq:U-solid}) is what is called the fundamental equation in thermodynamics \cite{Callen}. As suggested by GB \cite{Gyftopoulos}, the name the fundamental relation of equilibrium (FRE) is more appropriate. 
\vspace{2 mm}

\noindent 
{\bf Fundamental relation of equilibrium for solids}.
\begin{equation}
U^{\rm (so)} = U^{\rm (so)}(T, V, \{ \bar{\bf R}_{j} \}).
\label{eq:FRE-solids}
\end{equation}

\vspace{2 mm}
\noindent 
The internal energy $U^{\rm (so)}$  depends on the equilibrium positions $\{ \bar{\bf R}_{j} \}$. This says that even displacing one atom to an interstitial position causes a change in $U$. 
Microscopically, this is obvious. However, when this fact is interpreted in the thermodynamics context, it turns out to be a great discovery for solving the historical conundrum in thermodynamics \cite{Shirai18-StateVariable}.
For a long time, it has been believed that the thermodynamic properties of one-component systems are determined only by two state variables, namely, $T$ and $V$; see for example, Callen (\cite{Callen},  Sec.~1.5). This is correct for gases but not for solids. The mechanical properties of steel have dependence on the past history of thermal and mechanical treatments. The $U$ of steel cannot be expressed solely as a function of $T$ and $V$. Researchers circumvented this difficulty merely by regarding these materials having hysteresis as nonequilibrium materials: they consider that nonequilibrium materials are exempted from thermodynamics laws. More than 80 years ago, Bridgman tackled this problem by taking plastic deformation as example \cite{Bridgman50}. He did not recommend rushing to conclude that the laws of thermodynamics do not apply to our system, if at first we are unsuccessful in finding a set of parameters that determine an energy function (Ref.~\cite{Bridgman61}, p.~59). Instead, he suggested suspecting that {\it we have not a complete list of parameters of state}.
Recognizing that the plastically deformed state cannot be uniquely specified using the state variables of $T$ and stress $\sigma$, he added strain $\varepsilon$ as an independent state variable (in elastic theory, $\varepsilon$ depends on $\sigma$). In this way, he was partly successful in describing plastic deformations. Unfortunately, real deformations are not so simple as to be describable by adding a single variable to a set of state variables. Additional {\it ad hoc} variables, called internal variables, were introduced by other researchers \cite{Kestin70,Rice71,Kestin92,Berdichevsky05}. However, nobody can say how many internal variables are needed or what the substance of the internal variables is. Now, we can answer the original question of the full list posed by Bridgman. The full list is given by a set of equilibrium atom position in addition to $T$, as indicated by the FRE of solids. 
Readers will find throughout this review how suitably the FRE (\ref{eq:FRE-solids}) describes the thermodynamic properties of solids.

Returning to the specific heat issue, we can calculate $E_{\rm ph}$ as
\begin{equation}
E_{\rm ph} = \int \hbar \omega \left( \bar{n}(\omega) + \frac{1}{2} \right) g(\omega) d\omega,
\label{eq:phonon-energy-integral}
\end{equation}
where $\hbar$ is the Planck constant and $g(\omega)$ is the phonon DOS. This is the the elemental excitation approach.
The phonon contribution to the specific heat can be obtained by the analytical form,
\begin{equation}
C_{\rm ph}(T) =
k_{\rm B} \int \left( \frac{\hbar \omega}{k_{\rm B} T} \right)^{2} 
\frac{e^{\hbar \omega/k_{\rm B}T}}{(e^{\hbar \omega/k_{\rm B}T}-1)^{2}}
g(\omega) d\omega,
\label{eq:C-phonon}
\end{equation}
as a function of $T$. In the MD simulations, the phonon DOS is obtained from the frequency spectrum $f_{v}(\omega)$ of atom velocity $v(t)$,
\begin{equation}
g(\omega) = 
 \frac{1}{3 N \frac{3}{2} k_{\rm B}T}  \sum_{j}^{3N} \frac{1}{2} M_{j} f_{v_{j}}^{\ast}(\omega) f_{v_{j}}(\omega).
\label{eq:phononDOS}
\end{equation}
Here, the subscript $j$ denotes a composite index referring to the atom index and Cartesian coordinates. The conservation of the number of degrees of freedom can be checked by
\begin{equation}
\int g(\omega) d\omega = 1.
\label{eq:sum-DOS}
\end{equation}
This check is important when the phonon model for liquids is investigated (in section \ref{sec:TdepC-result-phonon}). Because atom motions are treated classically in conventional MD simulations, the kinetic energy of atoms in the simulations does not obey the Bose--Einstein statistics, but instead, the classical equipartition law of energy,
\begin{equation}
E_{\rm P,vib} = E_{\rm K} = \frac{3}{2} N k_{\rm B}T,
\label{eq:equi-partition}
\end{equation}
holds. To fix this problem, first, this classical term in $\overline{E_{\rm gs}(t)}$ is removed from the total energy, giving the structural energy $E_{\rm st}$,
\begin{equation}
E_{\rm st} = \overline{E_{\rm gs}(t)}-E_{\rm K} = \overline{E_{\rm gs}(t)} - \frac{1}{2} E_{\rm ph}.
\label{eq:Est}
\end{equation}
Then, the phonon energy $E_{\rm ph}$ is evaluated using equation~(\ref{eq:phonon-energy-integral}). 
Finally, $E_{\rm tot}$ is recalculated by adding this $E_{\rm ph}$ to $E_{\rm st}$, thereby recovering the Bose--Einstein statistics. Practically, this modification of $E_{\rm tot}$ is important only at very low temperatures. 
There are various anharmonic effects, such as frequency shift and phonon--phonon interactions \cite{Leibfried61,Cowely64}, and correspondingly, various ways of treatment, such as quasi-harmonic and self-consistent anharmonic approximations, exist \cite{Bruesch82-1}. The problem of such perturbatic treatments is that it is uncertain whether summing up such perturbatic terms gives the correct total energy.
In contrast, our $E_{\rm st}$ contains all the contributions except the harmonic term, correctly giving the total energy,
\begin{equation}
E_{\rm st} = E^{(0)} + E^{(3)} + E^{(4)} + \dots \ .
\label{eq:st-expand}
\end{equation}
Only when the harmonic approximation holds, the result of the elemental excitation approach agree with that of the total energy approach: $U=E_{\rm ph}$.

\subsubsection{Liquid case}
\label{sec:Total-E-liquid}
For liquids, the atoms do not have unique equilibrium positions; thus expansion (\ref{eq:phonon-energy}) does not make sense \cite{note-INM}. We should stop at the general formula for the internal energy, equation~(\ref{eq:internal-energy}). Thereby, the internal energy of liquid, $U^{\rm (li)}$, becomes a simple form as a function of only $T$ and $V$,
\begin{equation}
U^{\rm (li)} = U^{\rm (li)}(T,V).
\label{eq:U-liquid}
\end{equation}
Atom positions do not have the invariance property against the time average; their values are indeterminant. Thus, they cannot be state variables. Equation (\ref{eq:U-liquid}) is the FRE of liquids, provided that the internal structure is absent or fixed even if present. The functional form of $U$ is the same as that of gases. Behind this simple functional form, however, the FRE of liquids is distinct from that of gases in its frequency dependence, which is described in the next section.
By taking the numerical derivative of $U(T)$ with respect to $T$, we can obtain $C^{\rm (li)}_{V}$, which has a similar functional dependence on $T$ and $V$ to that in equation (\ref{eq:U-liquid}). 

Despite that there are no equilibrium atom positions in liquids, recently, an increasing number of authors have adapted the phonon model to analyze the specific heat of liquids \cite{Wallace98, Bolmatov12,Trachenko16, Proctor20,Baggioli21}. The idea of phonons has been promoted by the observation of phonon-like dispersions in inelastic neutron scattering \cite{Egelstaff-2ed,Copey74,Smith17,Khusnutdinoff20}. Also, phonon-like DOS can be obtained by MD simulations by applying equation~(\ref{eq:phononDOS}). 
Let us investigate the applicability of the phonon model to liquids from the viewpoint of the foundation of statistical mechanics.
In textbooks of statistical mechanics, when the partition function of an $N$-particle system, $Z = \sum_{r} \exp( -E_{r}/k_{\rm B}T )$, is investigated, we tacitly assume that the system has a set of eigenstates $\{ r \}$. When the system comprises independent particles (or quasi-particles), the total energy of the system $E_{r}$ is given by the sum of the eigenenergies $\epsilon_{s}$ of individual particles, as in equation~(\ref{eq:etot-sum-e1}).
In quantum mechanics, the eigenstate means an infinite lifetime. The equilibrium distribution $\bar{n}_{s}$ is determined by the Bose--Einstein statistics for phonon systems, which is derived on the assumption of independent particles (\cite{Reif}, section~9.2). The total excitation energy of the phonon system is the sum of the phonon energies of the individual phonon modes, as in equation~(\ref{eq:phonon-energy-integral}).
In DFT, it is well known that the total energy is not the sum of the energies of the individual atoms. 
The additive property of equations (\ref{eq:etot-sum-e1}) and (\ref{eq:phonon-energy-integral}) results from the harmonic approximation for the atom excitations. The phonon modes diagonalize the Hamiltonian of the system and are thus the eigenstates for solids. The eigenstates are orthogonal to each other and thus there is no overlap between them, which guarantees the independent-particle description. The lifetime of a phonon can be regarded to be infinite. In this manner, the lifetime and independence of elemental excitations are intimately related.

The problem with liquids is that there are no eigenstates. Instantaneous motions of a liquid do not diagonalize the Hamiltonian. Thus, they overlap each other, and consequently, the summing property, equation~(\ref{eq:phonon-energy-integral}), does not hold. Therefore, equation (\ref{eq:phonon-energy-integral}) cannot be applied to liquids. A numerical example showing the consequence of the violation of the independent-particle assumption is illustrated in section~\ref{sec:TdepC-result}.

\subsection{Atom relaxations in glass transition}
\label{sec:Eq-Glass}
Atom relaxation is very important for understanding the thermodynamic properties of liquids. Without knowing the relation effects, we cannot establish full understanding of the thermodynamic properties of liquids. The effects of atom relaxation are observed most apparently at glass transitions. Hence, it is a good idea to study the effects through glass transitions, which is the subject of this subsection.
Traditionally, the topic of glass transition is evaded from thermodynamic description of materials because of the presence of hysteresis. However, now that we know a full set of state variables for solids, 
there is no reason to exclude the glass transition from thermodynamics.

\subsubsection{Equilibrium and relaxation}
\label{sec:Eq-Relax}

Specific heat $C$ is a thermodynamic property, which means that $C$ must be determined solely by the current state, independent of the previous process in which the current state was obtained. Equation (\ref{eq:def-C}) indicates that specific heat must be a state function like the energy is (see section \ref{sec:hysteresis-SH} for more detailed discussion). On the other hand, from the experimental viewpoint, $C$ is the response of a system to a change in temperature. The response is always associated with a time lag no matter how small it is. Therefore, the thermodynamic specific heat should be measured at a sufficiently long time after the relaxation process was ended. The time required to reach equilibrium is the relaxation time $\tau_{\rm R}$. When $\tau_{\rm R}$ becomes very long, the distinction between equilibrium and relaxation becomes ambiguous. This occurs for the glass transition. Careful treatment as to which are state variables is needed \cite{Shirai20-GlassState,Shirai21-GlassHysteresis}. We briefly describe this in the following. 

There are two criteria for thermodynamic equilibrium when the change in the internal energy, $\Delta U = \delta Q + \delta W$, is investigated. The one is {\em the thermal equilibrium}: there is no net heat exchange, $\delta Q = 0$. In this circumstance, the temperature, which is the most important state variable in thermodynamics, is uniquely determined. The time required to reach this condition is the thermal relaxation time $\tau_{T}$. The other is {\em the mechanical equilibrium}: there is no net work exchange, $\delta W = 0$. The time required to reach this condition is the mechanical relaxation time; here we call this more specifically {\it the structural relaxation time} $\tau_{\rm S}$, because we are interested in the properties of materials. Although the term is used collectively, different structural relaxation times exist for a solid. In fact, the structural relaxation time $\tau_{{\rm S}j}$ (or shortly $\tau_{j}$) can eventually be attached to an individual atom $j$ of the solid, so that there are as many $\tau_{j}$'s as the number of atoms, $N$.  $\tau_{j}$ is determined by the energy barrier $E_{b,j}$ surrounding the $j$th atom, as indicated by equation~(\ref{eq:tau}).
The energy barrier is a constraint in the thermodynamics context. Therefore, different atom configurations have different relaxation times: the $K$th structure has its own relaxation time $\tau_{K}$. A relaxation can be looked upon as the process in which the given configuration $K'$, which was previously an equilibrium state sustained by a set of constraints $\{ E_{b}^{K'} \}$, is transformed to a new equilibrium configuration $K$, when the constraints $\{ E_{b}^{K'} \}$ were removed. Therefore, the equilibrium of the state $K$ is established when the condition
\begin{equation}
\tau_{K} \ \gg \ t_{\rm obs} \ \gg \ \tau_{K'} \ {\rm and} \ \tau_{T},
\label{eq:transition-1}
\end{equation}
is met. For example, the two gases of hydrogen and nitrogen separated by a wall in a container are in equilibrium (state $K'$). When the wall is removed, the two gases immediately mix. The homogeneously mixed state ($K$) is a stable equilibrium state without causing further reactions at room temperature. However, over a long time, the mixed gases will react to produce ammonia gas (a new state $K''$); remember that the free energy of ammonia is lower than that of the mixed gas. Every material (state $K$) used for fuel is stable at normal conditions but becomes unstable immediately after ignition, leaving ash as the final product (state $K''$). The reaction $K \rightarrow K''$ occurs even at normal conditions but the reaction rate is infinitesimally so slow that we can preserve fuels with safety for a long time.
In this manner, the notion of equilibrium has meaning only in the relative sense (in the local scope). Relation (\ref{eq:transition-1}) says  that the end of an equilibrium state is the beginning of a new equilibrium state: Gujrati calls this the hierarchy of relaxation times \cite{Gujrati10,Gujrati18}.

\subsubsection{Thermodynamic description of glass transition}
\label{sec:GlassTr-relax}

The specific heat associated with equilibrium state $K$ can be defined when condition (\ref{eq:transition-1}) is met.
Normally, the relaxation time of thermal equilibrium, $\tau_{T}$, is so fast for both solids and liquids that there is no difficulty in determining the temperature. Experimentalists can easily measure the temperature of glasses even during the glass transition. On the other hand, the atom positions in a glass vary in the glass transition. The averaged positions, $\bar{\bf R}_{j}$ in equation~(\ref{eq:Rexpansion}), become indeterminate over the transition period. However, the viscosity of glass is so high that the rate of the structural change is very slow and a large difference appears in the response times between $\bar{\bf R}_{j}$ and ${\bf u}_{j}$. This large difference in the response times is called {\em the adiabatic separation of the second kind} in \cite{Shirai20-GlassState}. 
In this case, we can easily find such an experimental timescale that meets the condition
\begin{equation}
\tau_{K} \ \gg \ t_{\rm obs} \ \gg \ \tau_{T} \quad {\rm for \ any \ accessible \ }K.
\label{eq:transition-2}
\end{equation}
Under this condition, it has sense to measure $C$ as a function of $t$,
\begin{equation}
C(t) = C(T(t), \{ \bar{\bf R}_{j}(t) \}).
\label{eq:CfuncR-t}
\end{equation}
In equation~(\ref{eq:CfuncR-t}), the time average $\bar{\bf R}_{j}(t)$ is understood to be taken in a period much longer than $\tau_{T}$ but shorter than any $\tau_{K}$ in order to make the specification of structure meaningful. In this manner, the conflicting aspects, thermodynamic and kinetic, of the specific heat can be reconciled. On this understanding, we can speak of AC calorimetry. The functional dependence of $C$ on $\{ \bar{\bf R}_{j}(t) \}$ is the reason why hysteresis is observed in the $C-T$ curve for the glass transition. See section \ref{sec:hysteresis-SH} for further discussion on hysteresis.

After the completion of the glass transition from the liquid, the specific heat of glass $C^{\rm (gl)}$ is given as a function of equilibrium atom positions $\{ \bar{\bf R}_{j} \}$, 
\begin{equation}
C^{\rm (gl)} = C^{\rm (gl)}(T, \{ \bar{\bf R}_{j} \}),
\label{eq:CfuncR}
\end{equation}
as for solids in equation~(\ref{eq:FRE-solids}). For the liquid state, on the other hand, $C^{\rm (li)}$ has dependence solely on $T$ and $V$,
\begin{equation}
C^{\rm (li)} = C^{\rm (li)}(T, V),
\label{eq:Cliq}
\end{equation}
which is evident from equation~(\ref{eq:U-liquid}).
In the transition region, we reasonably suppose that the state is a mixture of solid and liquid parts. Although there are differences in nuance from author to author, there is a general consensus that one characteristic of the glass transition is dynamic heterogeneity: an inhomogeneous mixture of mobile and immobile parts \cite{Ediger00,Chandler10,Berthier11}. Hence, in the transition region, the specific heat $C^{\rm (inh)}$ can be expressed as
\begin{equation}
C^{\rm (inh)} = x C^{\rm (li)}(T, V_{l}) + (1-x) C^{\rm (gl)}(T, \{ \bar{\bf R}_{j}^{\rm (so)} \}),
\label{eq:CfuncR-inh}
\end{equation}
where $x$ is the mole fraction of the liquid part, $V_{l}$ is the volume of the liquid part, and $\{ \bar{\bf R}_{j}^{\rm (so)} \}$ is the set of the time average of atom positions in the solid part. In this manner, the state even during the transition region can be described by the thermodynamics method, as long as the adiabatic separation of the second kind is valid. If we focus on the difference $\Delta C_{\rm gl}$ between the glass and liquid, the knowledge about the detailed process during the transition region is unnecessary, because the specific heat is a state function.

\subsubsection{Effects of atom relaxation on specific heat}
\label{sec:EffectsOnC}

\paragraph{a. Rate dependence of glass transition temperature.}
The glass transition temperature is affected by the cooling rate $v_{c}$. This change is a direct consequence of the fact that the $C-T$ curve is affected by $v_{c}$. This $v_{c}$ dependence indicates that $C(T)$ is affected by the relaxation process. 
Detailed analyses the $v_{c}$-dependent $C-T$ curve from the relaxation viewpoint were made by Moynihan and others \cite{Moynihan74,Debolt76}. Many studies have followed since then; reviews of a large number of those studies are available in \cite{Kovacs79,Mathot84,Hodge94}. 

The main problem in the analysis is the nonexponential type of relaxation, which is nonexponential with respect to time $t$ and temperature $T$. The latter type is known as non-Arrhenius type behavior and constitutes the major challenge in the current glass research \cite{Angell88,Angell99,Angell00}. In particular, two kinds of {\em fragilities}, namely, kinetic and thermodynamic fragilities, have been seriously discussed up to now \cite{Angell95,Ngai99,Martinez01,Wang-LM02,Wang-LM03,Tanaka03}.
For the problem of nonexponentiality in time, further phenomenological parameters such as the stretched exponent, $\exp(-(t/\tau)^{\beta} )$, known as the KWW (Kohlrausch--Williams--Watt) function \cite{Williams70,Williams71} are commonly introduced. A similar manner of treating the response function in the frequency domain is known as the Cole--Cole plots for dielectric relaxation, which is an extension of a simple type of Debye formula in equation~(\ref{eq:kai}) \cite{Cole49,Havriliak66}. In either case, the attempts are to describe the deviation of the observed response from the exponential type, $\exp(-t/\tau)$, by introducing additional parameters while retaining a single relaxation time $\tau$. By increasing the number of assumed parameters, we will obtain a better fit. However, it is a common recognition that no set of parameters can cover a wide range of $T$ \cite{Hodge94}. Increasing the number of fitting parameters makes the physical meaning less clear. We note that there is no reason why only one relaxation time is involved. There are many equilibrium configurations $K$ for a given $T$, which have different energy barrier $E_{b}^{K}$. More importantly, the energy barrier $E_{b}^{K}$ is not constant but strongly varies near $T_{g}$. 
Today, it is almost agreed that the large deviation from the Arrhenius law cannot be understood without considering that the activation energy varies across $T_{g}$ \cite{Xia00,Lubchenko03,Dyre04,Keys13}. The major problem of these studies was that the evidence for the variation of $E_{b}$ with $T$ was indirect: their conclusions were derived from models. Direct showing the variation of $E_{b}$ with $T$ was awaited. A recent MD study by Han {\it et al}~revealed compelling evidence of the temperature dependence of the energy barrier by EAM (the embedded-atom method) \cite{Han20a}.
  
One difficulty remains that the apparent activation energy $Q_{\rm gl}^{\ast}$ obtained by specific heat and viscosity measurements is too large to understand based on chemical bonds: it is not rare to see more than 5 eV, which is a typical value of the cohesive energy \cite{Hodge94,Laughlin72}. %
It is unrealistic that $E_{b}$ largely exceeds the cohesive energy. For crystals, the magnitude of the melting enthalpy $H_{m}$ is of the order of $k_{\rm B} T_{m}$ (see Eq.~(\ref{eq:HmTm})), while $E_{b}$ is usually much smaller than $H_{m}$. See the energy scheme of figure \ref{fig:pot-relax}. This problem has been solved by addressing the manner of obtaining $Q_{\rm gl}^{\ast}$ by the conventional analysis of the Arrhenius plot: the conventional analysis represents the $E_{b}$ value with a considerable magnification \cite{Shirai21-ActEnergy}.

\paragraph{b. Frequency dependence of the glass specific heat.}
The time dependence of the specific heat in equation~(\ref{eq:CfuncR-t}) can be measured by AC calorimetry. Although the general formulae of the linear response theory was established in the 1950s \cite{Kubo-SP-II}, the application to specific heat was done only lately \cite{Birge85,Birge86}. Theoretical works were done further later \cite{Nielsen96,Hentschel08}.
In the simplest model of single relaxation time, the complex-number response function $\chi$ is given as
\begin{equation}
\chi(\omega) \propto \frac{1}{\omega(\omega -i \gamma)},
\label{eq:kai}
\end{equation}
which corresponds to the exponential type of time dependence $\chi(t) \propto e^{-\gamma t}$: see, for example, Ref.~\cite{Kittel-ISSP8} (p.~430). The real part of $\chi$ has the $\omega$ dependence as
\begin{equation}
\Re \{ \chi(\omega) \} \propto \frac{1}{\omega^{2}+\gamma^{2}}.
\label{eq:Re0kai}
\end{equation}
Experimental data of the specific heat of a supercooled liquid \cite{Birge85,Birge86} can be expressed by this simple model, equation~(\ref{eq:Re0kai}), where the corresponding decay rate is expressed as $\gamma_{C}$. In the limit $\omega \rightarrow 0$, the static specific heat becomes
\begin{equation}
C \propto \frac{1}{\gamma_{C}^{2}} \propto \exp\left( \frac{2 Q_{a}}{k_{\rm B}T} \right).
\label{eq:C0gamma2}
\end{equation}
This form is consistent with our early speculation, equation~(\ref{eq:activationC}), aside from the power index in $\gamma_{C}$. Equation (\ref{eq:Re0kai}) indicates that the specific heat becomes insensible to the high-frequency perturbations at $\omega > \gamma_{C}$. The characteristic relaxation time $\tau_{C} = 1/\gamma_{C}$ decreases as $T$ increases in the range $T>T_{g}$. 

Knowledge of the relaxation time of supercooled liquids can also be obtained from the temperature dependence of viscosity $\eta$. According to the transition-rate theory, $\eta$ is given by $\eta=\eta_{0} \exp (Q_{a}/k_{\rm B}T) $, where $Q_{a}$ is the activation energy for viscose motions \cite{Eyring36,Ewell37,Eyring64}. The pre-exponential factor $\eta_{0}$ is composed of several parameters but, after some mathematical manipulation, $\eta$ can be recast in a simple form $\eta=G_{\infty} \tau_{\eta}$ with the relaxation time $\tau_{\eta}$ and the high-frequency shear modulus $G_{\infty}$. Qualitatively, the temperature dependence of $\tau_{\eta}$ has the same behavior as that of $\tau_{C}$. Near the glass transition temperature, the $T$ dependence does not follow the Arrhenius type, but it is known to be described by the Vogel--Fulcher--Tammann (VFT) rule 
\begin{equation}
\eta = A  \exp \left( \frac{D}{T-T_{0}} \right).
\label{eq:VFT}
\end{equation} 
Here, $T_{0}$ is the characteristic temperature parameter. The value of $T_{0}$ is much lower than the caloric transition temperature $T_{g}$. In the glass literature, it is anticipated that the thermodynamic transition will occur at or near $T_{0}$.
However, note that, below $T_{g}$, the form of VFT is merely an extrapolation, because the viscosity measurement is difficult once the sample freezes in. Recent studies indicate deviations from the VFT rule at $T<T_{g}$ \cite{Stickel96,Hecksher08,Zhao13,Pogna15,Shirai21-ActEnergy}. Supporters of the VFT rule claim that their waiting times are insufficient. In this manner, the argument based on dynamics is endless, because there is no fundamental reason to inhibit future change. In thermodynamics, however, future behaviors are subjected to the restriction of the second law. See section \ref{sec:discussion-glass}.


\subsection{Microscopic theory of relaxation in liquids}
\label{sec:TheoryRelaxation}
Compared with the glass transition, the effects of atom relaxation seem not drastic for liquids. However, there are essential reasons why the relaxation effect must be considered for liquids. 
For solids, the relaxation locates the potential-minimum configuration, creating the eigenstates around the bottom of the potential, as illustrated in figure~\ref{fig:pot-relax}. On the other hand, for liquids, there is no thermodynamic significance in local minima in the adiabatic potential, as already explained in figure \ref{fig:pot-relax}. The role of relaxation is, rather, to destroy eigenstates.

\subsubsection{Statistical mechanics description}
\label{sec:SM-treat-relax}

The lack of eigenstates does not mean that the equilibrium notion is not applicable to liquids. Obviously, liquids have equilibrium states: we can speak of the equilibrium properties of a liquid by specifying $T$ and $V$, as in equation~(\ref{eq:U-liquid}).
Let us examine how the equilibrium is achieved when the system has no eigenstates.
In the MD simulations of a liquid, there is an instantaneous state $i$ with the total energy $E_{i}$. By collecting $i$ over simulations, we can define the probability (population) $P_{i}$ of the instantaneous state $i$. Here, indexes $r$ and $s$ are reserved for presenting eigenstates, while $i$ and $j$ are used to present instantaneous states. There is no guarantee that $P_{i}$ obeys the Bose--Einstein statistics, as explained in section~\ref{sec:Total-E-liquid}. Instead, $P_{i}$ is determined by the balance of transitions,
\begin{equation}
\frac{d}{dt} P_{i} = \sum_{j} W_{ji} P_{j} - \sum_{j} W_{ij} P_{i} ,
\label{eq:MasterEq}
\end{equation}
where $W_{ji}$ is the time-dependent transition probability from state $j$ to $i$. Equation (\ref{eq:MasterEq}) is known as the master equation for stochastic processes \cite{Reif}. The time-dependent transition probability $W_{ji}(t)$ is expressed as $W_{ji}(t) = \nu_{0} \exp( -\gamma_{ji} t)$, where $\gamma_{ji}$ is the transition rate of state $j$ to $i$ and $\nu_{0}$ is the attempt frequency. 
Here, $\gamma_{ji}$ is determined by the energy barrier $E_{b,ji}$ from state $j$ toward $i$ as $\gamma_{ji} = E_{b,ji}/\hbar$ \cite{Eyring64}. The equilibrium of the system is established by the detailed balance $(d/dt) P_{i}=0$ in equation~(\ref{eq:MasterEq}). When states $i$ and $j$ are eigenstates $r$ and $s$, their lifetimes become infinite, and therefore, it makes a sense to consider the limit of $t \rightarrow \infty$, namely, 
\begin{equation}
t \gg \tau_{rs},
\label{eq:Equilibrin-cond1}
\end{equation}
for any transition between $r$ and $s$, where $\tau_{rs} = 1/\gamma_{rs}$. 
Under this condition, the time dependence of the transition probability is replaced by the temperature dependence as, 
\begin{equation}
W_{rs} = \nu_{0} \exp(-E_{b,rs}/k_{\rm B}T).
\label{eq:Wrs-barrier}
\end{equation}
Then, the population ratio $P_{s}$ to $P_{r}$ is determined by the ratio of the forward to backward transition rates,
\begin{equation}
\frac{P_{s}}{P_{r}} =  \frac{W_{rs}}{W_{sr}} = \exp( -\beta (E_{b,rs} - E_{b,sr})) = \exp( -\beta (E_{r} - E_{s})).
\label{eq:detailed-balance}
\end{equation}
In the last step in equation~(\ref{eq:detailed-balance}), the barrier height between $s$ and $r$ is dropped by taking the difference between the forward and backward transitions; thus, the ratio $P_{s}/P_{r}$ is turned to be determined by only the energy difference $E_{r} - E_{s}$. The independence of the energy barrier on the populations is usually expected for the systems having eigenstates. 

However, for liquids, we cannot ignore the finite relaxation time; thus, condition (\ref{eq:Equilibrin-cond1}) is no longer valid. The equilibrium of a liquid is controlled by the dynamic balance between the incessant creation and destruction of phonon-like excitations. The balance $(d/dt) P_{i}=0$ holds only as the time average,
\begin{equation}
\left\langle \sum_{j} \nu_{0} \exp( -\gamma_{ji} t) P_{j} \right\rangle 
 = \left\langle \sum_{j} \nu_{0} \exp( -\gamma_{ij} t) P_{i} \right\rangle.
\label{eq:dynamic-balance}
\end{equation}
The brackets designate the time average.
Accordingly, the population $P_{i}$ is determined by the energy barriers in addition to the energy difference between the potential minima. 
The effect of the energy barriers on the population $P_{i}$ is demonstrated using extremely simple models of two- or three-level systems \cite{Peyrard20,Bisquert05}. 
$P_{i}$ has dependence on temperature, unlike phonon systems, where the phonon DOS, $g(\omega)$, is independent of temperature.

\subsubsection{Melting temperature}
\label{sec:meltingT-theory}
In model calculations, the issue of melting temperature can be treated separately from the theory of specific heat. In the total energy approach, to the contrary, everything is calculated self-consistently from the unique energy functional. The accuracy of the calculation of $T_{m}$ directly affects the $U-T$ relation near $T_{m}$, and hence, the accurate determination of $T_{m}$ is part of the study of the specific heat of liquids.

The melting temperature $T_{m}$ is a property specific to the material. Theoretically, it is determined by the temperature at which free energies of liquid ($G^{\rm (li)}$) and solid ($G^{\rm (so)}$) are coincident,
\begin{equation}
G^{\rm (so)}(T_{m}) = G^{\rm (li)}(T_{m}).
\label{eq:defTm}
\end{equation}
From this condition, the relationship between the latent heat $H_{m}$ and $T_{m}$ is deduced,
\begin{equation}
H_{m} = T_{m} \Delta S_{m},
\label{eq:HmTm}
\end{equation}
where $\Delta S_{m}$ is the entropy change at melting. For a wide range of materials, the order of $\Delta S_{m}$ is close to $k_{\rm B}$ (known as Richard's rule \cite{Iida88}), and we can estimate $H_{m}$ by $H_{m} \sim k_{\rm B} T_{m}$.
For specialists of MD simulations, the calculation of $T_{m}$ is known to be the most difficult subject. There are many cases that MD simulations show large deviations from experiment. For silica materials, even though the potentials were constructed so as to match the interatomic distance to the experimental value and correctly reproduced the negative thermal expansion of $\beta$-cristobalite and the $\alpha$-$\beta$ transition of cristobalite at approximately 500 K, the calculated $T_{m}$ value for $\beta$-cristobalite was greatly overestimated by 3000 K \cite{Yamahara01}. Similar overestimation was reported for silica glass by Takada {\it et al} \cite{Takada04}. Belonoshko wrote his surprise as ``it is puzzling why the interaction model of Matsui is successful in producing $PVT$ properties and solid--solid transformations but fail to give proper results for the pressure dependence of the melting temperature" \cite{Belonoshko94}.
The problem of large discrepancy in $T_{m}$ could be ascribed to the model potentials. DFT calculations should provide the correct value for $T_{m}$. Despite this expectation, it is surprising that even FPMD simulations often give overestimations for $T_{m}$ to such an extent \cite{Shirai25-MeltingT}. In order to fix the overestimation problem, several methods were proposed, depending on what cause is envisaged as the discrepancy in $T_{m}$. 
By observing that melting is initiated from the surface of a solid, the overestimation of $T_{m}$ in MD simulations is often ascribed to the lack of surface in the standard setup of MD simulation. This has motivated researchers to devise the solid--liquid coexistence method \cite{Morris94,Belonoshko94,Belonoshko01,Alfe05,Usui10,Hong13,Hong15,Geng24}.
Other investigations led thermodynamic integration \cite{Mei92,Sugino95,Frenkel96,Cheng19} and the Z method \cite{Belonoshko06,Alfe11}.

The overestimation problem is currently such a big problem that it cannot be furthermore described herein. Refer to \cite{Shirai25-MeltingT} for more details with sufficient references. Here, only the following comments are mentioned.
First, regarding the construction of model potentials, the importance of energy barrier for atom movements is suggested. When model potentials are constructed, attention is paid in order to match the bottom of the potential: the position and the depth of the potential bottom. However, little attention is paid to the saddle points of potential. The saddle points correspond to the energy barrier (see figure~\ref{fig:pot-relax}). An error in the energy barrier of 0.1 eV amounts to the error in $T_{m}$ of about 1000 K. This error would not, in principle, occur for DFT calculations, if the true energy functional were found. However, current DFT potentials are still approximations.
Second, even though there are many examples for the overestimation of $T_{m}$ in FPMD simulations, there are also cases that FPMD simulations of the standard setup give the correct value. Si and Na are such examples. This indicates that the missing surface in the standard setup cannot be a universal reason for the overestimation.
Third, it is reported that even when the same potentials are used, different $T_{m}$ values are obtained by different methods of MD simulations: for example, compare the thermodynamic integration method of Mei \cite{Mei92} to the coexistence method of Morris {\it et al}~\cite{Morris94}. This suggests that the difference in the relaxation process between these methods affects the calculated $T_{m}$ values. The last point has the most relevance to the present context (see sections \ref{sec:TdepC-result} and \ref{sec:melting-result}).

\subsubsection{Construction of adiabatic relaxation simulations}
\label{sec:adiabatic-relaxation-MD}

The GB statement of the second law guarantees that there is one unique equilibrium state for each of a given set of constraints, $(U, \{ \xi_{j} \})$. The actual connection between the initial conditions $(U, \{ \xi_{j} \})$ and the final equilibrium state is determined by the relaxation process. The relaxation processes simulated by the adiabatic MD method most faithfully represent the GB statement of the second law and hence provide the most reliable method of determining temperature.

\begin{figure}[ht!]
\centering
    \includegraphics[width=80mm, bb=0 0 720 500]{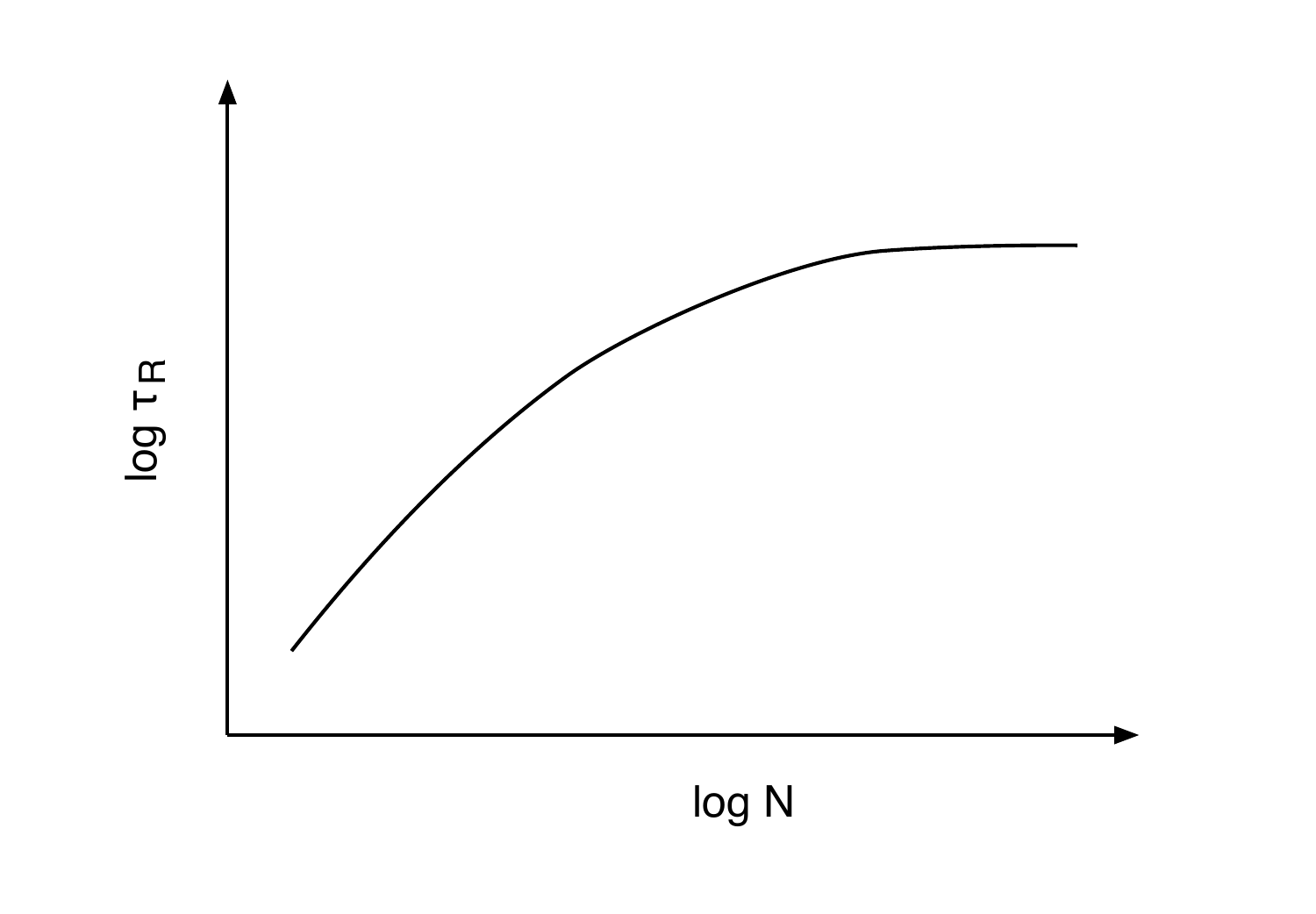} 
\caption{
Log--log plot of relaxation time $\tau_{\rm R}$ versus cell size $N$.}
\label{fig:TimeN} 
\end{figure}

Heat baths, such as the Nos\'{e}-Hoover thermostat are best avoided when thermodynamic properties need to be accurately calculated. In usual MD simulations, a thermostat is introduced in the equation of motion to control the system temperature through a frictional force,
\begin{equation}
M_{j} \left[ {\dot {\bf v}}_{j}(t) + \gamma {\bf v}_{j}(t) \right] = {\bf F}_{j}(t),
\label{eq:NewtonEq}
\end{equation}
where ${\bf F}_{j}$ is the force acting on the $j$-th atom and $\gamma$ is the frictional coefficient \cite{Frenkel96}. In quantum mechanics, there is no concept of frictional force; Schr\"{o}dinger equation involves no such frictional force.
In microscopic theory, {\em the friction represents the process by which the energy initially assigned to a specific motion is dissipated into the energy of the overwhelming number of other motions in an uncontrolled manner.} When the atom gets over an energy barrier, the energy previously stored in this atom is released to other atoms. This process is irreversible and entropy production occurs.
Adiabatic MD simulations automatically describe this energy dissipation process without introducing $\gamma$. 
The frictional coefficient is expressed by the relaxation time as $\gamma = 1/\tau_{\rm R}$. In MD simulations, $\tau_{\rm R}$ increases as the system size $N$ increases. Initially, $\tau_{\rm R}$ is approximately proportional to $N$, because the number of visiting configurations increases as $N$ increases (see figure \ref{fig:TimeN}). Eventually, $\tau_{\rm R}$ is saturated to the experimental value as $N \rightarrow \infty$. This is because the responses of far distant atoms are insensitive to the relaxation process started from a particular atom. In any case, the value of $\tau_{\rm R}$ in a MD simulation with a finite size $N$ is shorter than the experimental value. The adverse side of this fact is that the calculated value is different from the experimental value. The good side is that the experimentally observed phenomena can be reproduced in a much shorter timescale by using miniaturized systems. It is useless to prolong the simulation time as long as the experimental $\tau_{\rm R}$, even if it is possible. If there is no numerical error, $NEV$ simulations will recover the initial state, resulting in periodic behavior.

\begin{figure}[htbp]
    \centering
    \includegraphics[width=100 mm, bb=0 0 450 280]{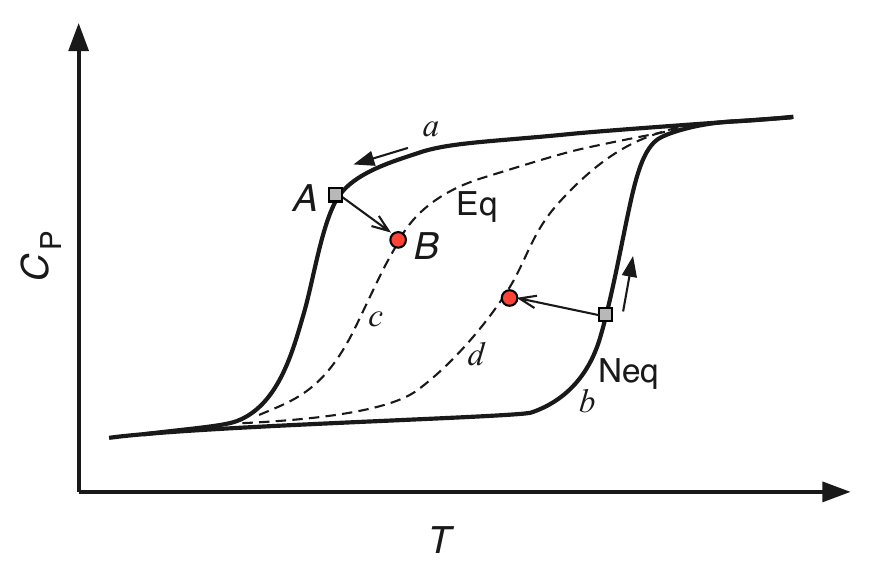} 
  \caption{$C_{P}-T$ curve of glass across the glass transition. Thick lines, $a$ and $b$, indicate real processes: each state in these lines is a nonequilibrium state. Dashed lines, $c$ and $d$ indicate quasistatic paths.} 
  \label{fig:CT_relax}
\end{figure}

With the above remarks in mind, let us describe how the adiabatic relaxation simulation is implemented. Let us apply the adiabatic relaxation method to an extreme case of glass transition in a sense of large viscosity. Once it is successfully implemented, it is easy to apply to other cases. In scanning calorimetry, a $C-T$ curve of the glass transition appears like the solid lines in figure~\ref{fig:CT_relax}. Because $T$ is scanned as a function of time, the relaxation processes are superimposed on the real $C-T$ curve, which makes the analysis complicated. To remove this complication, adiabatic calorimetry was devised \cite{Davies53,Davies53a}. Adiabatic calorimetry is established as a standard method of calorimetry \cite{Netsu-sokutei-HB-2}. At a point $A$ on the real path $a$, whose temperature is $T_{A}$, the sample is detached from the heat bath. Then, the sample is changed adiabatically towards the equilibrium state, whose temperature is $T_{B}$. By taking as many sampling points as possible, we obtain the continuous curve indicated by the dashed line in figure~\ref{fig:CT_relax}. The path so obtained represents the so-called {\em quasistatic} process in thermodynamics. Practically, when the rate of change in temperature, $v$, is so slow as to meet the condition
\begin{equation}
v  \ll \frac{\Delta T}{\tau_{R}},
\label{eq:quasistatic-condition}
\end{equation}
the process can be regarded as a quasistatic process, where $\Delta T$ is the step change in temperature. This quasistatic condition is equivalent to equilibrium condition (\ref{eq:Time-nucleation}).
Each state in a quasistatic process is an equilibrium state. However, quasistatic does not necessarily mean reversible (\cite{Zemansky}, sections 6.9--6.14). If the process involves friction, the process is irreversible.

Our method is very similar to this experimental situation of adiabatic calorimetry. 
In adiabatic MD simulations, the set of constraints $\{ \xi_{j} \}$ in the GB statement can be read as the initial atom positions $\{ {\bf R}_{j}(0) \}$. For a set of the initial conditions $(U, \{ {\bf R}_{j}(0) \} )$, a MD run is continued until reaching equilibrium, which is characterized by the state variables $(T, \{  \bar{\bf R}_{j} \})$. Symbolically, the GB statement can be read as a one-way correspondence,
\begin{equation}
(U, \{ {\bf R}_{j}(0) \} ) \rightarrow (T, \{  \bar{\bf R}_{j}^{K} \}),
\label{eq:adiabatic-correspondence}
\end{equation}
where $K$ indicates the particular atom configuration. The reversal correspondence does not hold because of the second law. This adiabatic MD run is repeated for as many values of $U$ as needed to obtain a continuous $C-T$ curve.
Again note that, even though each point of the so-obtained $C-T$ curve represents equilibrium, it is in a local equilibrium state and the whole of $C-T$ curve across the glass transition usually exhibits hysteresis. This is because there are a tremendous number of metastable configurations $\{ K' \}$ around $T_{g}$. Each $K'$ has a different energy barrier and hence a different relaxation time $\tau_{K'}$. The final state $K$ that the adiabatic MD run reaches depends on how the initial conditions on $U$ are changed. The step change in $\Delta U$ in our method corresponds to the change in cooling rate $v_{c}$ in the experiment. Let denote $\Delta T$ the resulting temperature step for each MD run.
Then, only those states that have shorter relaxation times as
\begin{equation}
\tau_{K'}  < \frac{\Delta T}{v_{c}},
\label{eq:eqcondition}
\end{equation}
can be accessible states. Within these accessible states, if the equilibrium conditions---the concrete conditions are given in the next paragraph---are met, the final state can be the equilibrium state. Therefore, usual thermodynamic analysis can be applicable to that state. See section \ref{sec:hysteresis-SH} for concrete manners of thermodynamic description of hysteresis. 
An important matter to notice is that amorphization is a common process for all materials, if sufficiently fast cooling is used, as indicated by the vapor deposition method and MD simulations.

For simulating slow cooling processes, the scanning $U$ is carried out by connecting many MD runs with a slight change in $U$; each run keeps a constant $U$ with succeeding the atom positions of the previous run. Each run is continued until the equilibrium state is reached. For the heating process from a solid, the succession of atom positions is not necessary, because the average positions in the previous run are always the same equilibrium positions of the solid. In adiabatic MD simulations, the control parameter is the initial temperature $T_{\rm in}$, at which the initial velocities of atoms are given through equation~(\ref{eq:equi-partition}). There are two criteria for equilibrium, as described in section~\ref{sec:Eq-Relax}. These criteria are implemented in the adiabatic relaxation simulations. The equilibrium temperature is determined by
\begin{equation}
\frac{3}{2} k_{\rm B}T = \frac{1}{2} \langle M_{j} \overline{ v_{j}(t)^{2} } \rangle_{j}.
\label{eq:equil-temp}
\end{equation}
Here, the brackets denote the particle average, whereas the bar indicates the time average.
Thermal equilibrium is reached when the time average of $E_{\rm K}(t)$ converges. For most of our MD simulations, $\tau_{T}$ is very short, on the order of 0.1 ps. For liquids, $\tau_{T}$ becomes longer but is still less than 1 ps. Thus, in most cases, the relaxation to equilibrium is controlled by the structural relaxation. The time evolution of the particle-averaged displacements, 
\begin{equation}
\langle \overline{\delta R_{j}(t)^{2} } \rangle_{j} = 
\frac{1}{N t_{\rm sm}} \sum_{j}^{N} \int \left( R_{j}(t-t_{0}) - R_{j}(t_{0}) \right)^{2} t_{0},
\label{eq:average-displace}
\end{equation}
is used to determine whether structural equilibrium is reached, where $t_{\rm sm}$ is the simulation time. For solids, it is considered the equilibrium state when $\langle \overline{\delta R_{j}(t)^{2} } \rangle_{j}$ is constant with respect to $t$. For liquids, it is considered the equilibrium state when $\langle \overline{\delta R_{j}(t)^{2} } \rangle_{j}$ shows a linear dependence on $t$ over the entire simulation time $t_{\rm sm}$. In the latter case, the slope, $D = (1/6) \lim_{t \rightarrow \infty} \langle \overline{\delta R_{j}(t)^{2} } \rangle_{j} /t$, gives the diffusion coefficient. 
For the liquids treated in this study, $\tau_{S}$ was on the order of a few ps. In this manner, we can manage the simulations of the glass transition in feasible simulation times.

At last in this section, we like to mention why we employ MD simulations for studying glass. The purpose, of course, depends on the researcher's intention, whether it is process simulation or material simulation. If he is interested in material properties like the author, details of the process are not a matter of central importance.
It is often claimed that the simulation time of current MD simulations is too short to compare with the experimental time so that the obtained results are unrealistic. In the experiment, the transition time of the glass transition, $\tau_{\rm gl}$, is of macroscopic timescale, i.e., one hour. It is impossible even for classical MD to access this timescale. 
However, we must ask why we need such a long simulation time as comparable to the experimental time.
We already know that the properties of a solid are uniquely determined by the structure alone, as indicated by the FRE (\ref{eq:FRE-solids}). If the same structure is obtained by some other method, the same properties must be obtained, irrespective of the processes used. 
Now, glasses are prepared also by the vapor-deposition method. The thin films are grown almost immediately in the human time scale. Usually, such thin films are less stable than the bulk form because of high concentrations of defects. Recently, however, it has been demonstrated that ultrastable glasses are obtained by the vapor-deposition method \cite{Swallen07,Singh13,Ediger17,Rodriguez-Tinoco22}. Surprisingly, the enthalpies of these thin-film glasses are lower than those of the corresponding glasses obtained from liquids. Aside from the interest in the mechanism, the important lesson from these experiments is that the stability of the obtained materials is not necessarily proportional to the time spent for the preparation. The important matter is whether the same structure as in the experiment is obtained in calculations. For assessing the suitability of the obtained material, powerful tools for the structural analysis are available for crystals. Unfortunately, for amorphous materials, there is no such powerful tool other than a gross method of the radial distribution function (RDF), which is not sensitive to the structural change in glasses. In this situation, the best way to assess the obtained glass is to compare the energy. By evaluating enthalpy rather than the simulation time, we can judge how good the simulated glass represents the real glass. An example is demonstrated in section~\ref{sec:Glass-result}.

\section{Simulation results}
\label{sec:Results}

\subsection{Specific heat of liquid sodium}
\label{sec:result-CofNa}

\subsubsection{Temperature dependence of specific heat of liquid}
\label{sec:TdepC-result}

The results obtained by the adiabatic relaxation method are illustrated by taking liquid sodium as example. Liquid Na is an important material for the application of nuclear power plants, and accordingly, rich experimental data on its thermodynamic properties are available \cite{Alcock94,Fink95}.

\begin{figure}[ht!]
\centering
    \includegraphics[width=100mm, bb=0 0 380 260]{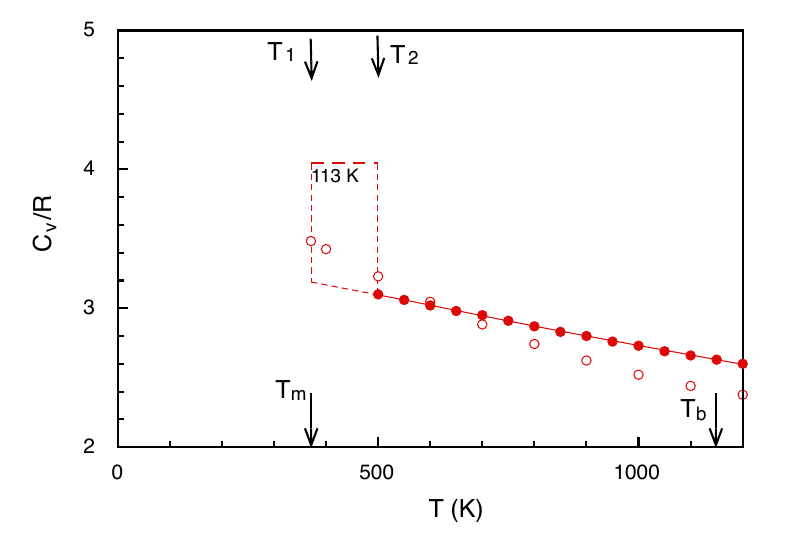} 
\caption{$C_{V}$ of liquid sodium in the range from the melting temperature, $T_{m}=371$ K, to the boiling temperature, $T_{b}=1156$ K. The calculated data are indicated by the red circles, and the experimental data in figure~\ref{fig:Fink} are indicated by the open circles. In the transition region from $T_{1}$ to $T_{2}$, a linear interpolation is used for the $U-T$ relation, leading to a constant $C_{V}$ indicated by the thick dashed line. The area enclosed by the dashed lines gives 113 K. The figure is reproduced from the original paper by Shirai {\it et al.}~with permission \cite{Shirai-EntropyLiquid}. 
}
\label{fig:Cv-liquid-calc-Na}
\end{figure}

In figure~\ref{fig:Cv-liquid-calc-Na}, the calculated values by the present method \cite{Shirai-EntropyLiquid} are compared with the experimental values \cite{Fink95}. 
The specific heat is directly obtained from the time-averaged internal energy $U(T)$ in MD simulations, using equation (\ref{eq:internal-energy}). As explained there, the total energy is the primal quantity in the total energy approach, while the phonon and structural energies are derived quantities. This view is shared with Han et al's study on the glass entropy \cite{Han20}.
There are technical problems when $C_{V}(T)$ is derived from $U(T)$. First, the fluctuation in $U(T)$ becomes large when the crystal melts, which is a common property of liquids. This fluctuation becomes particularly severe when a small-size cell is used. A direct calculation of numerical derivative of $U(T)$ magnifies the error in $C_{V}(T)$. The smoothing method using a polynomial was used for the raw data of  $U(T)$, resulting in the data in figure~\ref{fig:Cv-liquid-calc-Na}.
Second, the obtained melting curve $U-T$ in MD simulations has width $W_{m}$, which is discussed in section \ref{sec:melting-result}. The step-function-like behavior of the $U-T$ curve is rounded off. This rounded-off region, which is indicated by $T_{1}$ and $T_{2}$, is called {\em the transition region}. This is an artifact created owing to the finite size $N$. Hence, analyzing the details of the $T$ dependence in this region is physically nonsensical. For the present case of Na, the transition region encompasses the range $370 < T < 500$ K. 
The $C_{V}$ value obtained from the $U-T$ curve in this region contains some part of the latent heat, and thus becomes very large. We cannot distinguish the part of latent heat from that of sensible heat (measurable in specific heat experiment). 
Temporally, a linear interpolation was adopted for $U(T)$ in this region, giving the constant $C_{V} = 4.05 R$, as indicated by the thick dashed line. Thus, this $C_{V}$ value does not have physical meaning and should not be compared with experiment. Only the integrated value over the transition region has physical meaning. The integration $\int C_{V}(T) dT$ in this region (9.7 meV) plus the step $\Delta U_{m} = 20.7$ meV at $T_{m}$ gives the melting enthalpy, $H_{m}' = 30.4$ meV, which should be compared with the experimental value of 26 meV.

\begin{figure}[ht!]
\centering
    \includegraphics[width=120mm, bb=0 0 380 280]{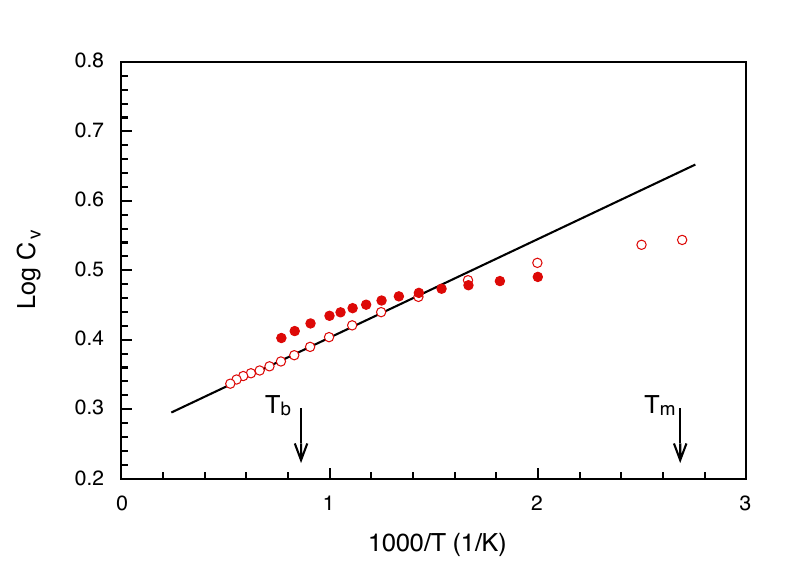} 
\caption{
Arrhenius plot of $C_{V}$ of liquid Na. The meaning of the symbols is the same as in figure \ref{fig:Cv-liquid-calc-Na}. The slope, $d \log C_{V}/d(1/T)$, is given to the experimental values \cite{Fink95} by the solid line, giving the activation energy $Q_{C} = 13.5$ meV.}
\label{fig:Arrhenius-Cv} 
\end{figure}

In figure \ref{fig:Cv-liquid-calc-Na}, the decreasing behavior of $C_{V}$ with increasing $T$ is clearly seen for $T>500$ K. This is the first time to demonstrate this decreasing behavior without using any empirical parameter.
The calculated value of the decreasing rate is $dC_{V}/dT = -0.94 \times 10^{-3} R$/K. This value is almost half the experimental value, $ -1.6 \times 10^{-3} R$/K. Presently, the origin of this discrepancy is unclear. However, reproducing the decreasing behavior without empirical parameters is very important for understanding the mechanism, which is described below. 

\begin{table}
\begin{tabular}{l  c r l}
  \hline \hline
  \multicolumn{2}{l}{Property}  &  $Q_{a}/k_{\rm B}T_{m}$ &  \\
   \hline
  Diffusion &  $Q_{D}$ & 3.3   &    \\ 
  Viscosity &  $Q_{\eta}$ & 2.2 &   \\ 
  Specific heat &  $Q_{C}$ & 0.42 &   \\ 
    \hline  
\end{tabular}
\caption{Values of $Q_{a}/k_{\rm B}T_{m}$ close to the melting point at 1 atm pressure. Data of $Q_{D}$ and $Q_{\eta}$ are cited from Egelstaff (\cite{Egelstaff-2ed}, Table 13.2)}
\label{tab:ActivationQ} 
\end{table}

In figure~\ref{fig:Arrhenius-Cv}, these data are replotted in the Arrhenius form. At high temperatures, the linearity of the Arrhenius plot is good, so that the Arrhenius analysis makes sense. By applying equation~(\ref{eq:C0gamma2}) to the experimental data, the activation energy $Q_{C}$ is obtained to be 13.5 meV. 
There are many experimental methods to obtain the activation energy. It is interesting to compare this value with the activation energies obtained by other measurements. In Table \ref{tab:ActivationQ}, these activation energies of liquid Na, including those obtained by diffusion ($Q_{D}$) and viscosity ($Q_{\eta}$) measurements, are compared. In the table, the activation energy $Q_{a}$ is normalized by the melting temperature $T_{m}$. As seen, $Q_{C}$ is by far smaller than the others. The difference in $Q_{a}$ is not surprising. All atoms have their own relaxation times $\tau_{j}$. They contribute different properties in different manners.
Such deviations from the proportionality are commonly observed in phenomenological laws on the transport coefficients, such as the Stokes--Einstein law \cite{Kumar06,Charbonneau14a} and the Wiedemann--Franz law. These laws give the ratios of different transport coefficients, which present the gross effects of atom (or electron) relaxations.
Ashcroft and Mermin point out that {\it the deviation is not due to the comparative rates at which electrons experience collisions, but to the comparative effectiveness of each single collision in degrading the two kinds of currents} (\cite{Ashcroft-Mermin}, p.~323).
In atom diffusion, the contribution of atom jumps most directly appears in the coefficient $D$. In viscosity, only the atom jumps as a group of atoms can contribute the viscosity. The involvement of atom jumps in the respective transport phenomena becomes weaker in the order of diffusion, viscosity, and specific heat. The least contribution is that of specific heat: there is no clear separation between atom jump and thermal motions. This trend can be clearly seen in Table \ref{tab:ActivationQ}.
In any way, it is rare to apply the Arrhenius analysis for the specific heat of liquids. This is because the $T$ dependence of $C_{V}$ is weak and is hidden by the larger $T$ dependence on $C_{\rm te}$ in the usual measurement of $C_{P}$. It is hoped to examine the Arrhenius analysis using as many liquids as possible.

\subsubsection{Phonon interpretation}
\label{sec:TdepC-result-phonon}

\begin{figure}[ht!]
\centering
    \includegraphics[width=120mm, bb=0 0 400 260]{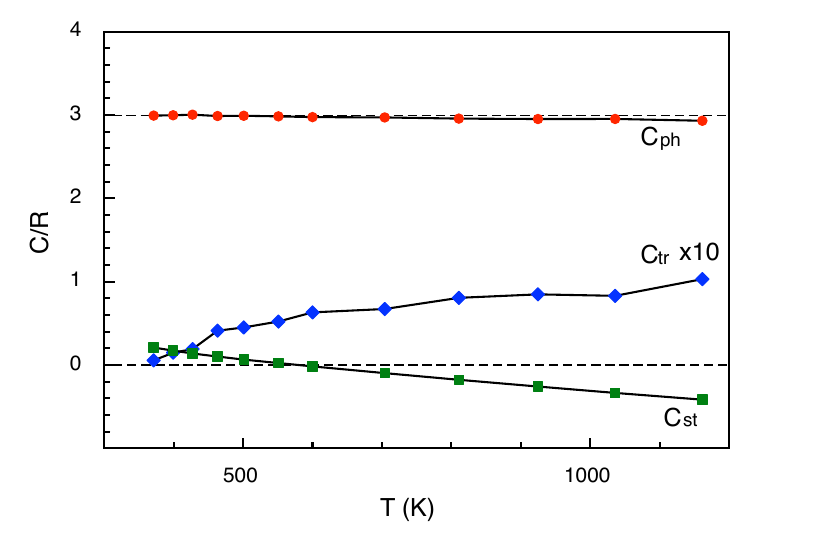} 
\caption{Three components of $C_{V}$ of liquid Na. Note that component $C_{\rm tr}$ is magnified by 10. The figure is reproduced from \cite{Shirai-EntropyLiquid} with permission.
}
\label{fig:C_comp_liq}
\end{figure}

Now, it is interesting to examine the phonon model for liquids.
According to the recent literature, the decreasing behavior of $C_{V}$ with $T$ could be explained within the phonon mechanism \cite{Wallace98,Trachenko08,Bolmatov11,Bolmatov12}. Because even for liquids the phonon-like DOS can be obtained by applying the Fourier-transform of velocities, via equation~(\ref{eq:phononDOS}), we can calculate the ``phonon contribution" to the specific heat, equation~(\ref{eq:C-phonon}). 
When a solid melts, the atoms begin to have diffusional motions and nonvanishing zero-frequency components $v_{j}(\omega=0)$ appear,
\begin{equation}
v_{j}(\omega=0) = \frac{1}{t_{\rm sm}} \int v_{j}(t) dt,
\label{eq:zero-freq-v}
\end{equation}
where $t_{\rm sm}$ is the simulation time. 
Let $x$ denote the fraction of the zero-frequency modes in the phonon DOS, and thus,
\begin{equation}
\int_{\omega>0} g(\omega) d\omega = 1-x.
\label{eq:zeroFreq-modes}
\end{equation}
This zero-frequency component creates the kinetic energy of pure translational motion, $E_{\rm tr} = \sum_{j} (1/2) M_{j} v_{j}(\omega=0)^{2}$, while there is no corresponding potential energy.
Thus, the total energy with constant $V$ comprises three components as
\begin{equation}
U_{V} = E_{\rm st}( \{ \bar{\bf R}_{j} \} )  +  E_{\rm tr}(T) + E_{\rm ph}(T).
\label{eq:internal-energy-4}
\end{equation}
The specific heat component corresponding to $E_{\rm tr}$ is denoted $C_{\rm tr}$. Under the assumption that the phonon model is valid for liquids, the fraction of the zero-frequency modes, $x$, controls the $T$ dependence of $C_{V}$.
When $T$ is such a high temperature that all phonons are thermally activated, the equipartition law of energy holds and classical limit becomes
\begin{equation}
C_{V} = x C_{\rm tr} + (1-x) C_{\rm ph} = \left[ \frac{3}{2} x +  \frac{6}{2} (1-x) \right] R.
\label{eq:Cliquid-expect}
\end{equation}
This could explain the $T$ dependence of $C_{V}$ of liquids. 
As described in section~\ref{sec:TdepCv}, $C_{V}$ is generally close to $3R$ just above $T_{m}$ and gradually decreases to approximately $2R$ at the boiling temperature $T_{b}$.
The longitudinal mode exists over the entire temperature range in the form of a sound wave. Wallace pointed out that the transverse modes also exist for liquids because the shear viscosity plays a role of the restoring force for transverse displacements. Therefore, it is reasonable to observe $C_{V} \approx 3R$ just at $T_{m}$ and the decrease in $C_{V}$ with further increasing $T$. As $T$ increases, the viscosity decreases, and the contribution of the transverse modes gradually diminishes, leaving the longitudinal mode as the sole active mode. Equation (\ref{eq:Cliquid-expect}) predicts $C_{V} = 2R$, when $x=2/3$ \cite{note-ext-supercritical}.
For the present calculation of liquid Na, the three components, $C_{\rm ph}$, $C_{\rm tr}$, and $C_{\rm st}$, are plotted in figure~\ref{fig:C_comp_liq}. The component $C_{\rm tr}$ is very small. Thus, the value of $C_{\rm ph}$ calculated using equation~(\ref{eq:phonon-energy-integral}) is virtually constant at the classical limit $3R$. Because of this constancy in $C_{\rm ph}$, $C_{\rm st}$ takes negative values. The negative specific heat suggests thermodynamic instability \cite{Lynden-Bell99,Landsberg87,Schmidt01,Gross02,Michaelian07,Michaelian09}. Obtaining the negative $C_{\rm st}$ in the present case is due to the fact that we overly subtracted $E_{\rm ph}$ from $E_{\rm gs}$ to obtain $E_{\rm st}$ in equation~(\ref{eq:Est}). This unphysical result is a consequence of adapting the independent assumption of phonon description to the liquid state. For liquids, the additive property (\ref{eq:etot-sum-e1}) (and also equation~(\ref{eq:phonon-energy-integral})) does not hold.

In the phonon model for liquid, it is a common idea to describe the decrease in $C_{V}$ with $T$ by introducing the cutoff frequency $\omega_{c}$ below which the corresponding potential does not contribute to the specific heat, in a similar manner to the case of pure translational motions. 
In the model by Trachenko {\it et al}~\cite{Bolmatov12,Trachenko16}, the transverse modes whose frequencies are lower than $\omega_{c}$---which they referred to as Frenkel frequency---are considered the diffusional motions, which have no potential part. As $T$ increases, $\omega_{c}$ increases and, accordingly, the number of transverse modes that contribute to $C_{V}$ decreases, leading to
\begin{equation}
C_{V} = R \left[ 3 - \left( \frac{\omega_{c}}{\omega_{D}} \right)^{3} \right],
\label{eq:Trachenko}
\end{equation}
where $\omega_{D}$ is the Debye frequency. In the Wallace's model, a slightly different interpretation is applied \cite{Wallace98}. He considered the upper bound for the phonon amplitudes to be that at which two atom vibrations do not overlap. For solids, this bound becomes infinity. Only those parts of phonons whose amplitudes do not exceed this bound can contribute to $C_{V}$.  Hence, the correction energy $E_{b}$ in his model has negative values. 
In either model, qualitative descriptions capture, to some extent, the atom dynamics of liquids. However, from the outset, the independent-particle assumption and thereby the application of the Bose--Einstein statistics are inappropriate. The degree of independence among the atom excitations decreases as $T$ increases. It has been shown that the overdamped motion of interacting particles obeys neither the Bose--Einstein nor Boltzmann distribution \cite{Andrade10}.

Trachenko and co-workers evaluated equation~(\ref{eq:Trachenko}) by replacing the cutoff frequency $\omega_{c}$ with $\tau_{c}^{-1} = G_{\infty}/\eta$ in a series of studies \cite{Trachenko08,Bolmatov11,Bolmatov12, Trachenko16}. This replacement causes the $T$ dependence of $C_{V}$ to be ascribed to the $T$ dependence of viscosity $\eta$. The origin of the $T$ dependence of $\eta$ is energy dissipation, and hence, their attribution to the relaxation effect turns to be physically reasonable, although the phonon model is mediated. They used the data of $T$-dependent $\eta$, which were obtained independently from the specific-heat measurement. They worked out on 21 liquids and compared the results with those of experiments, resulting in good agreements \cite{Bolmatov12}. The agreements for all the instances are particularly impressive. Fortuitous agreements could not occur for a  large number of examples. In spite of this, the present author would like to figure out an inconsistency in their model. 
As described above, the activation energies of $\eta$ and $C_{V}$ are different, so that parallelism between $\eta$ and $C_{V}$ over a wide range of $T$ cannot be expected. 
In paper \cite{Bolmatov11}, they found a systematic deviation from the experiments in the $T$ coefficient of $C_{V}$: the harmonic calculations always underestimate $dC_{V}/dT$. Then, they introduced an anharmonic effect by multiplying a factor $1+(1/2)\alpha_{P} T$ to the harmonic term of the internal energy, leading to the good agreement. The correction by $\alpha_{P}$ is so critical to this extent. However, $\alpha_{P}$ is thermal expansion coefficient, and the leading term of this effect already appears in the thermal expansion component $C_{\rm te}$, as in equation~(\ref{eq:Cte}). This term must be excluded from the calculation of $C_{V}$. The thermal expansion still has contribution to $C_{V}$ through the Gr\"{u}neisen parameters $\gamma_{G}$, which is proportional to $\alpha_{P}$. The treatment of these anharmonic terms is rather involved in the theory of anharmonicity for crystals \cite{Leibfried61}. In the perturbatic treatment, the effect of $\gamma_{G}$ appears in $C_{V}$ through $\gamma_{G}^{2}$, because of the second order perturbation. Therefore, the linear contribution of $\alpha_{P}$ in their model is questionable. 
The author does not claim that phonons are totally nonexistent in liquids. Not only longitudinal but also transverse modes are observed by neutron and X-ray scattering experiments: see chapters 14 and 15 of \cite{Egelstaff-2ed} and also \cite{Giordano10}. But they are observed as a result of external excitations by particles irradiations. The present issue is whether these externally excited modes obey the Bose--Einstein statistics.

\subsection{Specific heat across the melting temperature}
\label{sec:melting-result}

\begin{figure}[htbp]
  \centering
     \includegraphics[width=120 mm, bb=0 0 424 480]{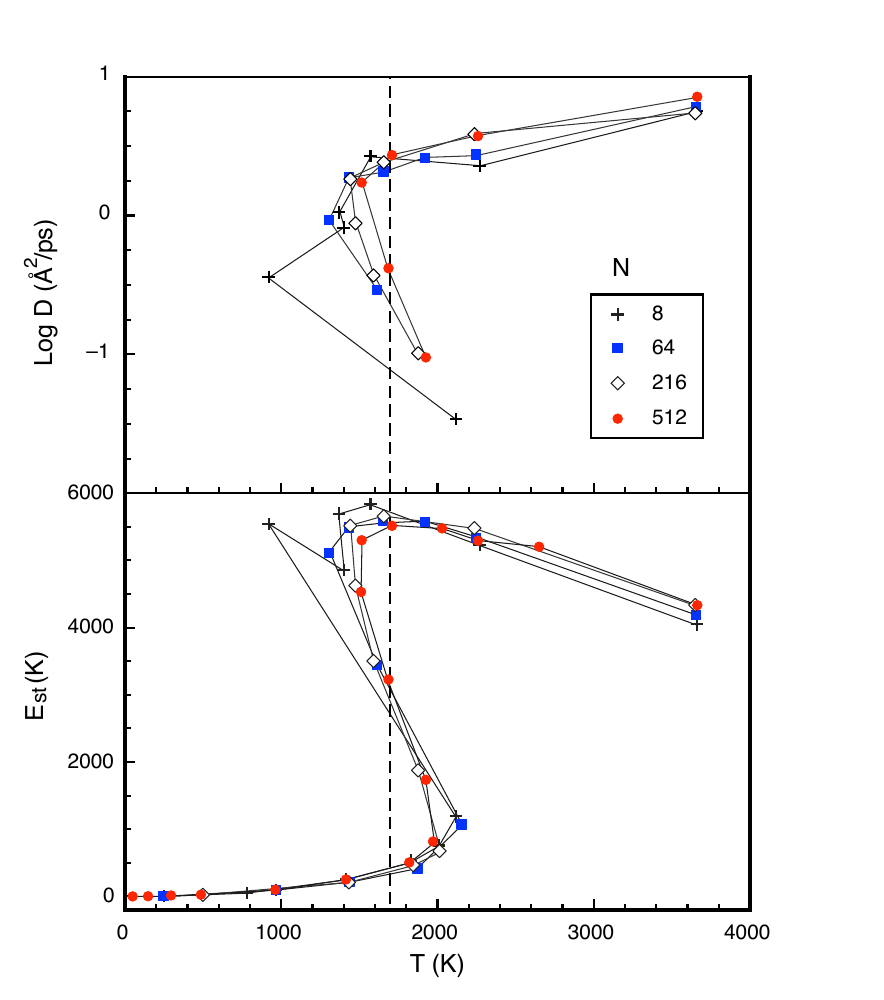} 
  \caption{Diffusion coefficient $D$ and structural energy $E_{\rm st}$ of Si. The cell sizes ($k$ meshes) are $N=8$ ($4^{3}$), 64 ($2^{3}$), 216 (1), and 512 (1). The experimental value of $T_{m}=1683$ K is indicated by the dashed line. For each $N$, the energy origin is set to the ground state energy of that cell. $E_{\rm st}$ is normalized by the number of atoms in the cell and is in units of K.  The figure is reproduced from \cite{Shirai25-MeltingT} with permission.
  } \label{fig:Est-lnD-si}
\end{figure}

The effect of atom relaxation appears most significantly at melting, for the obvious reason of structural change. Hence, it is interesting to examine what results the adiabatic relaxation method leads to. 
Figure~\ref{fig:Est-lnD-si} shows the result for silicon on heating the crystal \cite{Shirai25-MeltingT}. Plotted data are the structural energy $E_{\rm st}$ and diffusion constant $D$. The melting curves are examined by changing the cell size as $N=$8, 64, 216, and 512 atoms. 
In this figure, the data points are connected by lines to show the sequence of changing $T_{\rm in}$. 
As shown in figure~\ref{fig:Est-lnD-si}, $E_{\rm st}$ is virtually constant against $T$ up to $T=1000$ K, indicating that the crystal potential is well described by the harmonic approximation. Finite values of $D$ appear at around $T=2000$ K, confirming the onset of melting. In accordance with the variation in $D$, $E_{\rm st}$ exhibits an abrupt increase there.
If we define the $T_{m}$ by the midpoint in the transition region, we already have agreement with the experiment even using a small cell of $N=64$. In calculations of $T_{m}$ by MD simulations, the overestimation is often reported, as described in section \ref{sec:meltingT-theory}. The common explanation for this overestimation is the lack of surfaces in a standard MD setup. However, the present agreement in $T_{m}$ indicates that the surface effect cannot constitute the universal reason for the overestimation of $T_{m}$, unless the surface energy is very large.

In the obtained $E_{\rm st}-T$ curve, the melting temperature has the width $W_{m}$. In the experiment, melting occurs at a discrete temperature, and therefore, the $E_{\rm st}-T$ curve is expected to behave as a step function at $T_{m}$. The magnitude of the step should give the latent heat. As shown in figure \ref{fig:Est-lnD-si}, the width $W_{m}$ decreases as the size $N$ of the unit cell increases. Clearly, the obtained width is an artifact due to the use of finite-size cells. For finite-$N$ systems, there is an unavoidable uncertainty in temperature, $\Delta T$,
\begin{equation}
\frac{\Delta T}{T} = \frac{\Delta E_{\rm K}}{ \overline{E_{\rm K}} } = \frac{1}{\sqrt{N}}.
\label{eq:fluctuaion-T}
\end{equation}
Sometimes, it is referred to as the temperature fluctuation, but it may be better to call it the kinetic-energy fluctuation \cite{Freshbach87,Kittel88,Mandelbrot89}. Equation (\ref{eq:fluctuaion-T}) can be derived from a general relationship, 
\begin{equation}
C = k_{\rm B} \frac{ \overline{U^{2}}-\bar{U}^{2} }{\bar{U}^{2}}.
\label{eq:fluct-C}
\end{equation}
The abrupt transition is rounded off by this energy fluctuation. 
We have $W_{m} \rightarrow 0$ only at the thermodynamic limit, $N \rightarrow \infty$. It is noted that $W_{m}$ of 300 K remains even for the cell size of 512 atoms. The convergence with respect to $N$ is slow, as predicted by equation (\ref{eq:fluctuaion-T}).

Furthermore, we observe in figure~\ref{fig:Est-lnD-si} that the transition curves, both $E_{\rm st}$ and $D$, exhibit sigmoidal behaviors with negative slopes. The appearance of the negative sign depends on the material. The negative slope of $E_{\rm st}$ against $T$ means a negative specific heat, which is thermodynamically unstable \cite{Lynden-Bell99,Landsberg87,Michaelian07}. The negative slope of $D$ against $T$ also means dynamical instability. 
When the crystal melts, the energy distribution between the vibrational motions (in the solid portion) and translational motions (in the liquid portion) becomes unbalanced owing to this energy fluctuation. At temperatures slightly below $T_{m}$, those atoms with sufficiently high kinetic energy (higher than $k_{\rm B}T_{m}$) begin to convert their vibrational motions to diffusing motions, thereby decreasing the average kinetic energy $\langle E_{K} \rangle$ and increasing diffusion. In contrast, when this low-energy part of the distribution is overpopulated, the system temperature decreases. This makes these low-energy diffusing atoms being trapped in the crystal potential, leading to a release of the potential energy in the form of heat. This heat emission again increases the temperature of the system. In this manner, an oscillatory behavior appears around $T_{m}$. In the experiment, this sigmoidal behavior has been observed in clusters containing small numbers of atoms \cite{Schmidt01}.

The presence of this unstable region has some implications in the interpretation of MD simulations. 
Two extrema at $T_{1}$ and $T_{2}$ in this region are spurious equilibrium states. Accordingly, when the temperature is controlled by introducing a heat bath, there is a risk that the system happens to be trapped at either of the two extrema, depending on the process. 
There is a possibility that spurious equilibria happen in the Z method \cite{Belonoshko06,Alfe11}. Oscillatory behaviors between solid and liquid states are reported in \cite{Alfe11}.
In calculating $T_{m}$ of silicon by the thermodynamic integration, underestimations of $T_{m}$ by about 300 K are commonly reported \cite{Sugino95,Alfe03}. The spurious equilibria could be the cause of the underestimation of $T_{m}$.

The existence of the width $W_{m}$ for finite-size MD simulations has thus been resolved by the adiabatic relaxation method. Although this is an artifact, this finding reminds us of the following remarks regarding our understanding of heat and energy.

\begin{enumerate}
\item In MD simulations, the control of the average temperature does not guarantee that the simulation correctly presents the material properties at that temperature. The energy dispersion $\Delta E_{\rm K}$ is also important. This is particularly so when a thermostat is used. Even though the average temperature $\bar{T}$ is controlled at the intended temperature, the instantaneous temperature $T(t)$ oscillates around $\bar{T}$. For example, when an absorbed atom on a solid surface is studied with a thermostat, the desorption begins at a low average temperature $\bar{T}$. The instantaneous temperature of the absorbed atom can exceed the desorption temperature $T_{\rm ds}$, even though $\bar{T} < T_{\rm ds}$.

\item In the calorimetric measurement, the distinction between latent heat and sensible heat is clear. In finite-size MD simulations, this distinction becomes ambiguous. The finite width $W_{m}$ breaks the latent heat $H_{m}$ into the specific heat $C = H_{m}/W_{m}$. This suggests that the large specific heats often observed near phase transitions may be due to this effect: broadening of the transition due to the randomness of disorders. The part of latent heat is converted into the specific heat.
Some structural transitions exhibit so sharp feature in the $C-T$ curve that they are often interpreted as the $\lambda$ transition. The $\alpha-\beta$ transition of quartz is such an example. This transition is classified as higher-order transitions by Pippard \cite{Pippard,Pippard85}. It is clarified that the $\alpha-\beta$ transition of quartz has latent heat \cite{Richet82}, while the genuine $\lambda$ transition of liquid helium does not. Therefore, it is highly likely that the large increase in $C_{V}$ near the transition is ascribed to the broadening of the latent heat owing to disorders.
In a similar manner, the excess specific heat of supercooled liquids, $C_{\rm ex}^{\rm (sl)}$, can be understood to be a contribution of the structural energy: it has part of the latent heat. 
For the glass transition, it is reasonable to ascribe the excess specific heat of glass to the structural energy change $H_{m}- \Delta H_{gc}$ as
\begin{equation}
H_{m} - \Delta H_{gc} = \int_{T_{g}}^{T_{m}} C_{P}^{\rm (sl)} dT,
\label{eq:Hmg-Csl}
\end{equation}
where $\Delta H_{gc}$ is the enthalpy difference between the crystal and glass. 

\item The structural energy $E_{\rm st}$ changes mostly at $T_{m}$ but its change occurs also near $T_{m}$. 
Therefore, a change in the energy barrier $E_{b}$ can also be expected in this region.
Normally, the $T$ dependence of $E_{b}$ is not observed for crystals, because the temperature range in which $E_{\rm st}$ changes with $T$ is limited to only near $T_{m}$. However, for the glass transition, the range is extended from $T_{g}$ to $T_{m}$: thus, the change in $E_{g}$ can be easily observed.
In fact, deviations of the activation energy from the Arrhenius law are well known in this temperature ragion \cite{Angell91,Angell99,Xue22}.
\end{enumerate}

\subsection{Specific heat jump at the glass transition}
\label{sec:Glass-result}
The most challenging issue of specific heat may be the calculation of the specific heat jump at the glass transition, $\Delta C_{\rm gl}$. There has been no reliable method for predicting $\Delta C_{\rm gl}$, although a few models were proposed \cite{Wunderlich60,DiMarzio79,Trachenko11,Garden12}. It is indeed strange that DFT calculations had not been applied to this problem until the author did \cite{Shirai22-SH,Shirai23-Silica}. There are multiple reasons for this. The lack of a standard theory for liquids, the hysteresis of glass transition, and a big gap in the timescale between MD simulations and experiments hindered the application of DFT to the glass issue. These difficulties have been theoretically removed, as described in section \ref{sec:theory}. Now, we are in a position to apply the theory to a real example of glass. Below, a result for glycerol, which is a typical molecular glass, is described \cite{Shirai22-SH}.

\begin{figure}[htbp]
    \centering
    \includegraphics[width=120mm, bb=0 0 430 500]{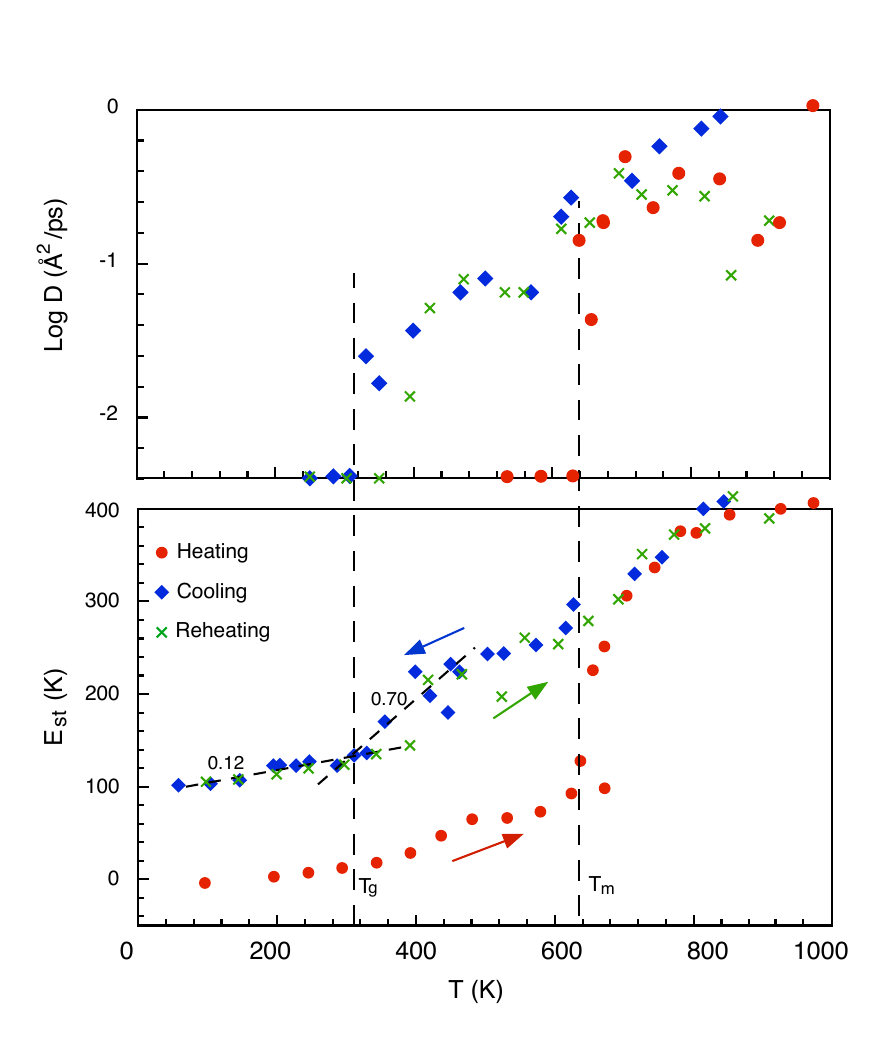} 
  \caption{Diffusion coefficient $D$ and structural energy $E_{\rm st}$ of glycerol as a function of $T$ in heating, cooling, and reheating processes.
Data points of $\log D =-2.4$ actually indicate $D=0$. These are plotted only because it clarifies where glass transition occurs. The figure is reproduced from \cite{Shirai22-SH} with permission.
  } \label{fig:D-Est}
\end{figure}

The changes in the thermodynamic properties of glycerol at the glass transition are studied. Figure \ref{fig:D-Est} shows the variations of the diffusion coefficient $D$ and the structural energy $E_{\rm st}$ over the temperature range covering the solid and liquid phases. By starting from crystalline glycerol at low temperatures, the sample was heated to melt and then the melt was cooled to obtain the glass state. Finally, the glass samples were reheated. The last step of reheating was added to check the hysteresis. We studied the glass transition during the cooling process. In the cooling process, the atom positions are succeeded between successive MD runs in order to simulate slow cooling. 

On heating the crystal, abrupt changes are observed in both $D$ and $E_{\rm st}$. The melting temperature can be identified as $T_{m}=635$ K. Compared with the experimental value of $T_{m}=291$ K, our result largely overestimates $T_{m}$. This overestimation of $T_{m}$ is already described in section \ref{sec:meltingT-theory}. The overestimation of $T_{m}$ seems to be a common property of oxide materials. There are several sources of error in calculating $T_{m}$, which were investigated in \cite{Shirai25-MeltingT}. Presently, the author considers the main reason to be the overbinding of the LDA/GGA functional. By considering that we presently do not have a better functional than GGA, we leave this error alone. Readers should read the following description while keeping this overestimation in mind.
On cooling the melt, $D$ vanishes at $T_{g}=310$ K. At this temperature, there is no jump in $E_{\rm st}$, which means no latent heat. Instead, there is a change in the slope $dE_{\rm st}/dT$. This corresponds to the jump in specific heat. Accordingly, the glass transition is unambiguously identified at $T_{g}=310$ K. The experimental value is $T_{g}=185$ K \cite{Rao02}. Again, an overestimation is observed, consistent with $T_{m}$.

\paragraph{a. Energetics}
The often-claimed criticism against the MD study of the glass transition is the large gap in the timescale between the experiment and the MD simulation; the difference in the cooling rate between them is more than a factor $10^{10}$. However, as already discussed in section~\ref{sec:adiabatic-relaxation-MD}, processes and properties are different things in the thermodynamic context. The quality of the obtained glass is not necessarily proportional to the time spent in its preparation. If the obtained structure is close to the real structure, our sample preparation should be considered successful. A suitable method for assessing the quality of the obtained glass is energy comparison.

For the melting enthalpy $H_{m}$ of the crystal glycerol, the calculated value has uncertainty ranging from 100 to 150 K/atom (11.5 to 17.4 kJ/mol) owing to a width $W_{m}$ of melting. Despite this uncertainty, the calculated $H_{m}$ is fair compared with the experimental value (18.3 kJ/mol) (\cite{CRC92}, p.~6-146). For glass, this uncertainty can be avoided by taking the energy of the crystal as the reference. 
The enthalpy difference between the glass and crystal, $\Delta H_{gc}$, is obtained from the extrapolation of $E_{\rm st}$ to $T=0$. From figure~\ref{fig:D-Est}, we obtain $\Delta H_{gc}=11.5$ kJ/mol (8.6 meV/atom). For the experimental data, $\Delta H_{\rm gc}$ was obtained to be 9.5 kJ/mol by analyzing the Gibson and Giauque data \cite{Gibson23}. The agreement in $\Delta H_{gc}$ between calculation and experiment is good. An energy difference of this order of magnitude is very common in materials research. This indicates that the usual method by DFT for assessing material properties is also valid in the glass research.

\paragraph{b. Specific heat jump}
A finite-size MD simulation intrinsically produces a finite width for the phase transition, which causes errors in the evaluation of $C_{V}$. For glass, fortunately or unfortunately, the transition has a width at the outset, and hence, the width caused by a finite size becomes not a serious problem. The practical problem may be the hysteresis. In fact, figure~\ref{fig:D-Est} indicates that a large hysteresis indeed occurs. However, because we are interested in the difference $\Delta C_{\rm gl}$ between the terminal states, i.e., the liquid and glass states, we can obtain a meaningful result.

As seen in figure \ref{fig:D-Est}, the obtained data are scattered, which makes it difficult to compute a numerical derivative. Hence, we evaluated $\Delta C_{\rm st}$ by linear fitting the $E_{\rm st}$ data in $T$ ranges above and below $T_{g}$ separately. These linear fittings are indicated by dashed lines in the figure. From this, we obtained $\Delta C_{\rm st} = 0.58 \ R$ for the contribution of structural energy to the specific heat jump.
In the cooling process, the phonon contribution $C_{\rm ph}$ almost continuously varies, so that the jump $\Delta C_{\rm ph}$ can be ignorable. The jump in the thermal expansion component, $\Delta C_{\rm te}$, was estimated to be at most $0.15R$. Therefore, the total jump $\Delta C_{\rm gl}$ is estimated to be $0.73 R$. This value is comparable to the experimental value, $0.70 \ R$ \cite{Kauzmann48}. Accordingly, we see that the jump $\Delta C_{p}$ is determined almost entirely by the contribution of structural energy. 

This is the first time to calculate the jump $\Delta C_{\rm gl}$ without any empirical parameter. Now we are able to study $\Delta C_{\rm gl}$ in a quantitative manner. It is quite a different situation from the previous status of study. Previously, only model calculations were available. A simulation using a variant of the Lennard--Jones type yielded divergent behavior in $C_{V}$ \cite{Pedersen10}, while a statistical method indicated a continuous change in $C$ to $T=0$ \cite{Yoshimori11}. The great merit of the total energy approach is clear in that the structural energy $E_{\rm st}( \{ \bar{\bf R}_{j} \} )$ can be calculated exactly, whereas, in model calculation, it is very difficult to model the complicated function $E_{\rm st}( \{ \bar{\bf R}_{j} \} )$ in terms of a few parameters.

\section{Developments and implications}
\label{sec:Development}

\subsection{Liquids with internal structure}
\label{sec:water}
In the analysis of experimental data discussed in section~\ref{sec:experiment}, we mainly focused on simple metals, that is, monatomic liquids. When a liquid has an internal structure, such as molecular liquids, the theory of specific heat can be extended by taking molecular vibrations into account. However, this extension is not so simple. A good example is water.
The specific heat of water is the most familiar to us as the definition of calorie. Yet we do not know the meaning of this value. By converting the unit cal/g to $R$, we obtain $C_{P} = 9.05 R$ for the specific heat of water. By applying equation (\ref{eq:classic-limit}), we see $f_{m} =18$ for the molecular freedom.
This is simply the number of degrees of freedom expected to water, $f_{m} =18 = 6 \times 3$. It appears as if the Dulong--Petit law holds: all phonon modes would be fully activated. This is impossible, however. The frequencies of the molecular vibrations of water molecule are so high, $\omega = 1595,  3657$, and $3756$ cm$^{-1}$, that these modes cannot be thermally activated at around room temperature \cite{note-CofWater}. See also a study on entropy \cite{Zhang11}.

The large value of specific heat of water cannot be explained by the phonon model. The structural contribution $E_{\rm st}$ must be considered. But a simple model of the energy barrier against atom movements has difficulty in explaining the large specific heat. Water is known as the least viscose liquid. 
Similarly, large values are found in ammonia liquid. Water and ammonia share a common property of polar liquids, namely, their large dielectric constants. Long-range interactions through polarizability must play the crucial role in their specific heats, though the details are unclear. 
Many studies are devoted to treating the dielectric effect on $E_{\rm st}$ \cite{Sceats80,Berens83, Shiga05,Shinoda05,Vega10,Cheng19}. The results of those studies are in rather good agreement with experiment; the errors are within about 10 \%, while all values are underestimations. In those studies, various potentials with special code names such as TIP4P were used. The author does not know which is the most suitable one; this is a subject in the realm of quantum chemistry. The long-range interactions are, in principle, included in the total energy approach through equation~(\ref{eq:internal-energy}). However, in the current status, DFT functionals have difficulty in treating special types of long-ranged interactions such as van der Waals interaction \cite{Berland15}. In any case, it is truly difficult to understand why such a large energy is required to raise the temperature of water by one degree, even though hydrogen bonding is weak and the viscosity of water is the lowest among the existing liquids; all these suggest the least relaxation effect.

Liquid selenium has a large $C_{p}$ of more than $4 R$ \cite{Shu80}. A similar large value is observed for liquid sulfur \cite{West59}. These are homopolar liquids, and accordingly, a different mechanism from the water case must be involved. For these materials, polymerization occurs in the liquid phase. An extra heat similar to the latent heat of melting is required. However, the treatment of polymerization is not as simple as this analogy is valid. These polymerizations are discussed in connection with liquid-liquid phase transitions (see \cite{Greer98,McMillan04,Holten12,Henry20}). All these issues remain as challenges for the total energy approach.

\subsection{Hysteresis in specific heat}
\label{sec:hysteresis-SH}

At the beginning of this review, the author wrote that, for solids, the equilibrium state can be uniquely determined by MD simulations, irrespective of relaxation processes. However, this is not true when there are multiple minima in the potentials. This occurs when defect states are considered. There are various kinds of defects, from point defects, dislocations, to macroscopic defects like stacking faults, in real materials. Combining various defect configurations $K$ to their relaxation times $\tau_{K}$ produces the process dependence on the final equilibrium state, which is known as hysteresis. 
Traditionally, hysteresis phenomena are considered nonequilibrium states, because a thermodynamic state must be specified solely by the current values of state variables and must be independent of the process. Mathematically, hysteresis is defined by multivalued functions \cite{Brokate96,Morris11,Muschik93}. 
However, in the thermodynamics context, this manner of definition must be reappraised because we now have a full list of state variables for solids because of FRE (\ref{eq:FRE-solids}). In the following, we explain this by taking the hysteresis in specific heat as an example. A full exposition is given elsewhere \cite{Shirai24-hysteresis}.

\begin{figure}[ht!]
\centering
    \includegraphics[width=80mm, bb=0 0 600 400]{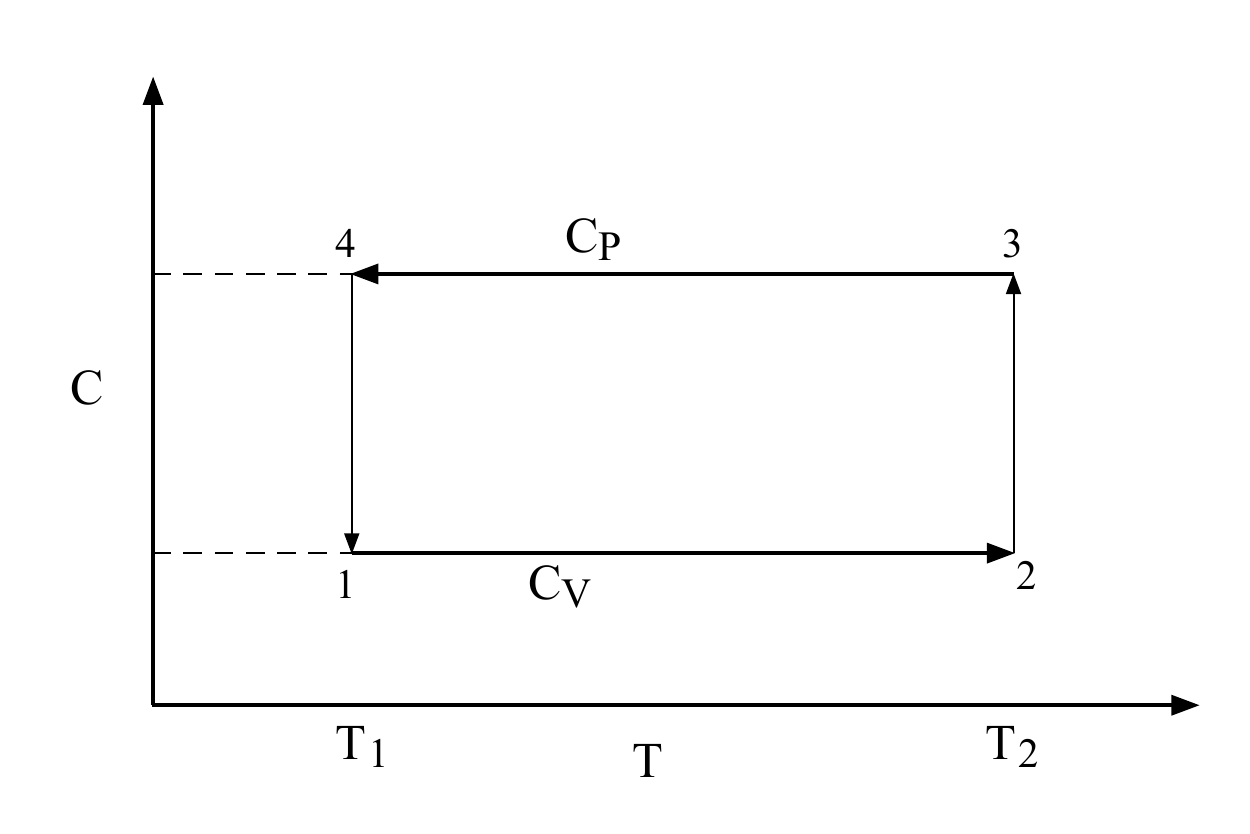} 
\caption{Specific heat of an ideal gas in a heat cycle through isochoric heating ($1 \rightarrow 2$) and isobaric cooling ($3 \rightarrow 4$). The two paths are connected by isothermal volume changes. The whole process is reversible but yet exhibits a hysteresis loop.}
\label{fig:CVhysteresis-gas} 
\end{figure}

Let us consider a measurement of the specific heat of an ideal gas in a cylinder with a piston. The measurement is performed in a heat cycle as shown in figure \ref{fig:CVhysteresis-gas}. The heat cycle starts from isochoric heating ($1 \rightarrow 2$) with a constant $V_{1}$ by fixing the piston. A constant $C_{V}$ value is obtained throughout this process. The pressure is increased from $P_{1}$ to $P_{2}$. At state 2, the piston is made free to retain a constant temperature at $T_{2}$. The gas system undergoes isothermal expansion ($2 \rightarrow 3$) and pressure is reduced to $P_{3}$ from $P_{2}$. The measurement is continued in isobaric cooling ($3 \rightarrow 4$) until reaching the initial temperature $T_{1}$. Throughout this process, the measurement gives a constant $C_{P}$ value. At state 4, the volume is reduced to $V_{4}$. Finally, by performing isothermal compression from $V_{4}$ to $V_{1}$, the gas restores the initial state 1. After the completion of this heat cycle, we find forming a hysteresis loop in the $C-T$ measurement, as shown in the figure. Every stage of the cycle can be performed in a quasistatic manner, and thus, the whole process can be reversible. This example indicates that the traditional way of defining hysteresis by a multivalued function must be seriously reappraised. 

\begin{figure}[ht!]
\centering
    \includegraphics[width=100mm, bb=0 0 700 400]{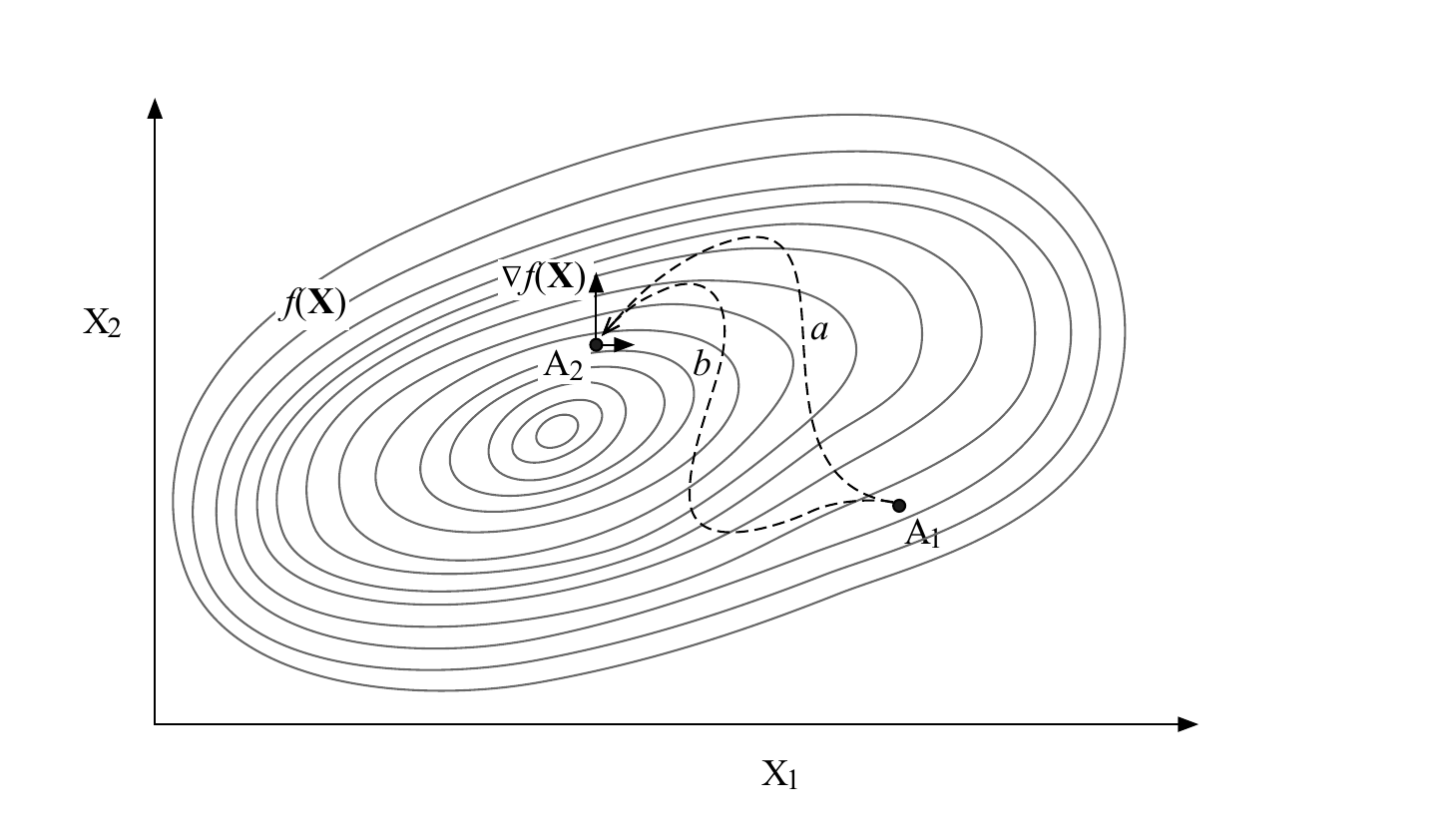} 
\caption{
State function $f(X_{1}, X_{2})$ in a two-dimensional state space $(X_{1}, X_{2})$. There are various paths ($a$ and $b$) connecting two states $A_{1}$ and $A_{2}$. The gradient $\nabla f$ is also a state function, which does not depend on the paths.}
\label{fig:two-dim-contour} 
\end{figure}

The essential problem is how to specify the thermodynamic state of solids. 
Let us consider a state function $f(X_{1}, X_{2})$ in a two-dimensional state space, as indicated in figure \ref{fig:two-dim-contour}. It may be useful to imagine $U(P, V)$ of a gas as example. $f=U$ is plotted as a contour map in the two-dimensional space $(P, V)$. The energy difference $\Delta U$ between two states $A_{1}$ and $A_{2}$ does not depend on the paths connecting these terminal states. This is the meaning that $U$ is a state function. Specific heat is the gradient of $U$ in this space, $\nabla U$, and thus is a vector quantity, whose components are given by
\begin{equation}
C_{X_{i}} = \left( \frac{\partial U}{\partial T} \right)_{\{X_{j}\}},
\label{eq:gradient}
\end{equation}
where the subscript $\{X_{j}\}$ means a set of all the {\em independent} variables. Thus, all $X_{j}$'s are held fixed. Only the variable $Y_{i}$ conjugate to $X_{i}$ can vary.
We see that $C$ is a state function, and does not depend on the paths connecting the terminal states \cite{note-CasStateV}.
The appearance of a multivalued function of $C$ originates from the way of plotting $C(P, V)$ on the one-dimensional space $T$. If we use the full space, we find that specific heat is a single-valued function in the multidimensional space. The multivalued character disappears! Now it is clear how to treat hysteresis. According to the advice of Bridgman (section \ref{sec:Total-E-solid}), we must use a full set of state variables $\{ X_{j} \}$. According to FRE (\ref{eq:FRE-solids}) for solids, we have to use the equilibrium atom positions $\{ \bar{\bf R}_{j} \}$.
In this manner, specific heat {\em does} describe the state of materials, even if a hysteresis loop is observed, provided that the measurement is performed quasistatically.

Once this mechanism of the appearance of hysteresis is understood, we can give an accurate definition of hysteresis. Hysteresis is observed as a response in a property $Y$ of the material, when the external field $X$ is changed in a cyclic manner. Now, hysteresis is unambiguously defined as follows.

\vspace{2 mm}

\noindent 
{\bf Definition of hysteresis}.
{\it Hysteresis is a process where the initial state $A_{i}$ of the system cannot be recovered after one cycle of the external field $X$ without causing any change in the environment.} \cite{Shirai24-hysteresis}

\vspace{2 mm}

\noindent 
We have found that hysteresis is no more than an irreversible process, in which the system does not restore the original state when the input field is returned to the initial value. Even though the property $Y$ of the material under investigation is returned to the initial value, the remaining state variables $\{ X_{j} \}$ are not returned to the initial values. To restore all the $\{ X_{j} \}$ to the initial values, extra work is required, which leaves impacts on the environment.

The important lesson here is that even though the current state $A$ was obtained by an irreversible process, if the current state is static or quasistatic in the sense of Eq.~(\ref{eq:transition-1}), the state $A$ is equilibrium and thus can be described by state variables. By using the full set of state variables, usual thermodynamic methods can be applied to this state.
In a hysteresis loop, there are many such final states that are static states. Whenever the final states are static, these final states are equilibrium states. Their equilibrium properties are completely described by FRE with state variables $\{ \bar{\bf R}_{j} \}$. There are as many equilibrium states as the number of variables $\{ \bar{\bf R}_{j} \}$ for a fixed $T$ and $P$. This is not the special case of glass but is general for any solid. A crystal has many defect configurations, each being equilibrium state specified by $\{ \bar{\bf R}_{j}^{K} \}$. These are equilibrium states within $\tau_{R}^{K}$. This view is fully compatible with the GB definition of equilibrium: it is impossible to extract net work from an equilibrium state without impacting the environment. In fact, our experience shows that we cannot obtain net work from defective crystals without impacting the environment.

\subsection{Revisiting the thermodynamic characterization of glass}
\label{sec:discussion-glass}
Although glass physics has been extensively studied for over a century, many fundamental problems remain. Many models and different interpretations for glass properties have been proposed from time to time, which are written in many review papers \cite{Kauzmann48,Jackle86,Gotze92,Angell99,Nemilov-VitreousState,Lubchenko07,  Heuer08,Berthier11,Wolynes12,Biroli13,Stillinger13,Berthier16}.
Discussing correctness of these models is not the purpose of this review. Here, we are concerned with the thermodynamic foundations of glass properties. The common understanding of glass is that the current states of glass are nonequilibrium states. The fundamental reason for this nonequilibrium characterization is the observation of the history dependence of glass properties; they cannot be uniquely described solely by $T$ and $P$. However, as already seen in the preceding subsection, hysteresis alone does not imply nonequilibrium. Since the traditional interpretation of nonequilibrium loses the fundamental reasoning, we need to revisit the thermodynamic foundations of glass physics from the viewpoint of the present theory.

\subsubsection{Transition temperature}
\label{sec:TransitionTemp}
As already described in section \ref{sec:GTexp}, by observing the rate dependence of the glass transition, an exaggerated view that the transition temperature $T_{g}$ can be varied as low as desired is claimed. This endows authority to the view that the glass transition is not a thermodynamic transition. However, the fact is different, as described there. The calorimetric transition temperature $T_{g}$ converges within a few degrees K, if sufficiently slow cooling is employed. The $T_{g}$ values are well tabulated in handbooks as a material property, as $T_{m}$ is.

For oxide glasses such as silica, the crystal nucleation time $\tau_{\rm G,cr}$ at $T_{g}$ is known to be much longer than the experimental timescale $t_{\rm obs}$, so nucleation cannot be observed \cite{Fokin03,Nascimento11,Zanotto17,Abyzov23}. However, since the crystal state is already an inaccessible state owing to the high energy barrier, why do not we consider the equilibrium problem within the scope of accessible states? Equilibrium is defined only within a given timescale, that is, equation (\ref{eq:transition-1}).
The formation of the glass is the best solution when crystal nucleation is inhibited. We know that the $T_{g}$ is saturated if sufficiently slow cooling rates are used. Thus, we observe the glass formation in the timescale,
\begin{equation}
\tau_{\rm G,cr} \gg t_{\rm obs} \gg  \tau_{\rm G,gl},
\label{eq:R-G-gl}
\end{equation}
which is a specific case of equation~(\ref{eq:transition-1}). 
In fact, even glass researchers concede that glasses do not change their structures for long times: the Egyptian jugs were created 2500 years ago, amber glasses are $10^{7}$ years old, volcanic glasses are $10^{8}$ years old, and the moon beads are more than $10^{9}$ years old (see the column ``the paradox of old glasses" in \cite{Berthier16}). Thus, the stability of glass is not the paradox but indeed indicates its equilibrium character.

An interesting issue about relation (\ref{eq:R-G-gl}) is whether the glass transition can be regarded as a quasistatic process if the cooling is performed at sufficiently slow rates.
As pointed out in section~\ref{sec:GTexp}, experiments show that the negative values are observed in Clausius integration~(\ref{eq:Clausius-I}) for glasses, indicating the irreversibility of the glass transition. From this observation, it was seriously suspected, by Kivelson and Reiss, the validity of the thermodynamics analysis obtained by equation~(\ref{eq:TDdef-S}) \cite{Kivelson99}, because the equation presumes the reversibility of the integration path. Many debates were invoked in connection to the violation of the third law: the residual entropy $S_{\rm res}$ violates the third law \cite{Speedy99,Mauro07,Gupta07,Goldstein08,Reiss09,Goldstein11,Takada13,Moller06,Aji10}. After serious debates, the consensus has been shared that the effect of irreversibility is not so serious to invalidate the measured values of $S_{\rm res}$. Careful calorimetry experiments have accuracy to such a degree that the measured value of the order of 0.1 cal/(K$\cdot$mol) is physically meaningful. This means that the irreversibility of the local path is ignorable and usual thermodynamic analyses are valid for the glass transition, provided the measurement is performed quasistatically in the sense of equation (\ref{eq:R-G-gl}).
As already described in section \ref{sec:GTexp}, even if the same $T$ and $C_{P}$ are recovered by quasistatic path, the initial state has not been recovered and hence the the whole process is an irreversible process. 
A successive small changes in atom configuration $K' \rightarrow K''$ can be made quasistatically by using slow rates of change. The equilibrium relation (\ref{eq:detailed-balance}) holds there. However, the change in $K$ is accumulated in a heat cycle in $T$, so that the final configuration $K_{f}$ reaches far from the initial configuration $K_{i}$.
Further analysis of the residual entropy is discussed in \cite{Shirai22-res}.

\subsubsection{Kinetic transition}
\label{sec:kinetic-trans}
A traditional view of the glass transition is that the glass transition does not accord with Eherenfest's standard classification of phase transitions \cite{Pippard,Jaeger98}. It is not a first-order transition because of the lack of latent heat. The entropy $S$ is continuous. Neither is it a second-order one because the PD ratio ${\it \Pi}$ is deviated from unity. Since Kauzmann characterized the glass transition as a new class of transition, {\em kinetic transition} \cite{Kauzmann48}, this characterization is widely accepted. The term kinetic transition is also used in statistical models, such as mode-coupling theory, with a slightly different nuance \cite{Gotze92,Gotze99,Miyazaki07-8}. Here, we reexamine this characterization of the kinetic transition from a purely thermodynamic viewpoint. Thermodynamic consideration must precede kinetics, as DiMarzio stated \cite{DiMarzio95}.

The thermodynamic transition is described by the presence of the transition temperature at which the free energies of two phases cross each other. This condition is expressed by equation~(\ref{eq:defTm}). Crystallization occurs at a discrete temperature $T_{m}$ in association with the latent heat $H_{m}$. On the other hand, glass is said to be a frozen liquid. Because of the high viscosity of the glass-forming liquid, the atoms are arrested before reaching their equilibrium positions. However, every crystal is a ``frozen liquid" if the aspect of the forming process is focussed on. The substance that arrests the glass atoms is no other than the potential. In section \ref{sec:Glass-result}, it was shown that the jump $\Delta C_{\rm gl}$ is brought about by the potential change $E_{\rm st}$; clearly bonding is formed among atoms. At this point, there is no difference from the usual solidification. 

Certainly, the glass transition temperature $T_{g}$ is affected by the process, as repeatedly discussed. However, if we violate the quasistatic condition (\ref{eq:transition-1}) in the preparation, $T_{m}$ of all crystals is turned out to have process dependence. Easy examples are thin-film processes. There is no idea of ``the crystallization temperature" when a film is prepared by a vapor deposition method. Operationally, there is the preparation temperature $T_{\rm prep}$, at which experimentalists obtain the thin-film form. It is normally lower than the $T_{m}$ of a bulk crystal \cite{Nishinaga97}:
\begin{equation}
T_{\rm prep} < T_{m}.
\label{eq:TprepTm}
\end{equation}
Around $T_{\rm prep}$, the growth rate ($v_{\rm gr}$) has temperature dependence, $v_{\rm gr}(T)$, so that the crystallization temperature cannot be defined.
Surprisingly, by molecular-beam-epitaxy (MBE) method, experimentalists obtain thin films of higher quality than the bulk: for GaAs, the $T_{\rm prep}$ by MBE is lower than half its bulk $T_{m}$ \cite{Newman-Vahidi15}. 
The temperature-dependent $v_{\rm gr}(T)$ implies the activation energy. Relation (\ref{eq:TprepTm}) indicates that vapor deposition processes can be categorized as supercooled processes: solid nucleation is retarded until reaching temperatures significantly lower than the bulk $T_{m}$ because of the energy barrier. The film growth is controlled by the energy barrier $E_{b}$, as seen in figure~\ref{fig:pot-relax}. We can use the term kinetic transition to this case.

By restricting the growth freedom into a two-dimensional space, a much longer time of nucleation ($\tau_{\rm G,tf}$) is required than in the bulk growth. The growth mechanisms of thin films are very complicated and thus $v_{\rm gr}(T)$ is determined by many factors \cite{Nishinaga97}. For the present purpose, it is enough to estimate it by the order of magnitude. The nucleation time in the vertical direction $\tau_{\rm G,tf}$ can be estimated by $\tau_{\rm G,tf} \sim a_{0}/v_{\rm gr}$, where $a_{0}$ is the lattice spacing. The typical deposition rate $v_{\rm gr}$ in thin-film processes is on the order of 1 \AA/s. This order-of-magnitude estimation shows that
\begin{equation}
t_{\rm obs} \sim \tau_{\rm G,tf}.
\label{eq:2d-growth-rate}
\end{equation}
This clearly violates the quasistatic condition of bulk growth, equation (\ref{eq:Time-nucleation}), and hence, it is reasonable to see no crystallization temperature. The associated activation energy $Q_{\rm G,tf}$ of the thin-film growth typically ranges from 0.1 to 2 eV depending on the material and growth mode \cite{Lewis-Anderson78}. The relation of the energy barrier $E_{b}$ to the latent heat $H_{m}$ is shown in the energy landscape of figure \ref{fig:pot-relax}. 
In thin-film processes, the inequality (\ref{eq:TprepTm}) holds, and there seems no lower bound in $T_{\rm prep}$. On the other hand, for the glass transition of a bulk glass from the liquid, there is the lower bound in $T_{g}$, as explained in section \ref{sec:GTexp}. More importantly, there is no concept of the temperature dependence of the growth rate for bulk glasses. Hence, it is inappropriate to categorize the glass transition from liquid as a kinetic transition. In view of equation (\ref{eq:R-G-gl}), the glass transition can be regarded as a thermodynamic transition. A quantitative analysis of the free energy is given next.

\subsubsection{Free-energy analysis of the glass transition}
\label{sec:Free-energy-glass}

\begin{figure}[htbp]
    \centering
    \includegraphics[width=80mm, bb=0 0 550 650]{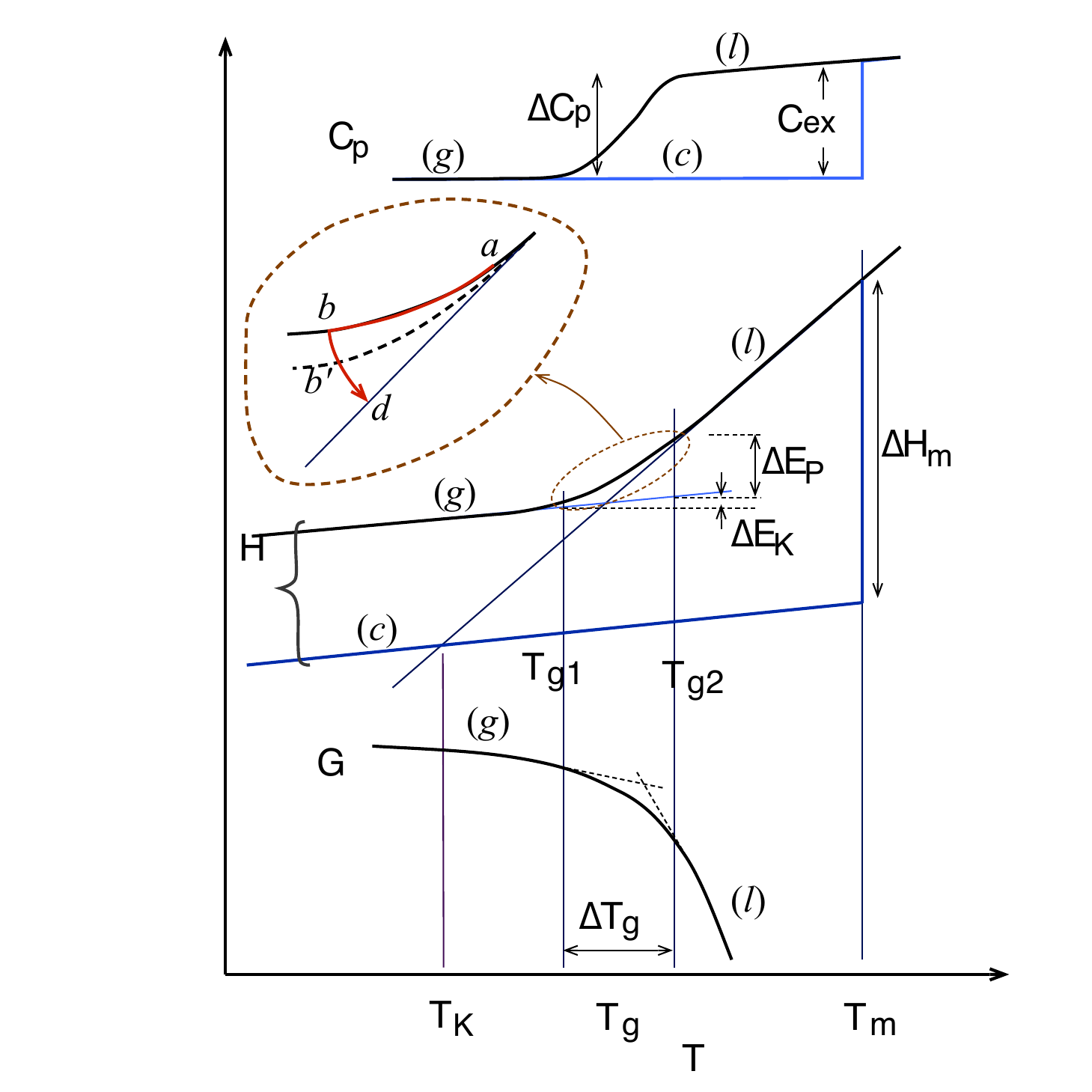} 
  \caption{Schematics of the temperature dependence of specific heat $C_{P}$, enthalpy $H$, and Gibbs free energy $G$ across the glass transition. Caloric measurement is performed from crystal ($c$) through liquid ($l$) to glass ($g$) states.
  } \label{fig:CHG-Tcurve}
\end{figure}

Although glasses were previously considered nonequilibrium materials, conventional thermodynamic methods were routinely applied to glasses. By integrating specific heat over a relevant temperature range, thermodynamic functions, such as free energies, were obtained and tabulated in data handbooks, even though the integration is allowed only for reversible processes, as indicated in equation (\ref{eq:TDdef-S}). For a long time, we have been content with this logical inconsistency between the measurement and theory. Indeed this logical inconsistency is the starting point of Kivelson and Reiss's argument on the residual entropy of glass \cite{Kivelson99}.
Now, this inconsistency has been resolved. We can use the usual thermodynamic methods even for hysteresis, provided it is obtained quasistatically. Let us analyze the details of the $C-T$ curve in the transition region. 

In figure~\ref{fig:CHG-Tcurve}, a typical $C_{P}-T$ curve is shown together with the enthalpy $H$ and the Gibbs free energy $G$ curves that are derived from the $C_{P}-T$ curve. 
Recognizing the importance of the relaxation effect, Davies and Jones studied the $C_{P}-T$ curve by adiabatic calorimetry \cite{Davies53a}. On cooling the liquid, the $H-T$ curve begins to deviate from the $H-T$ curve of the liquid at state $a$, as shown in the inset. Note that the thin line of the $H-T$ curve below $T_{a}$ is merely an extrapolation of the liquid enthalpy $H^{\rm (li)}(T)$ at $T>T_{a}$. If the temperature scanning is stopped and detached from the heat source at some moment (state $b$), then the enthalpy of glass $H^{\rm (gl)}$ is adiabatically changed toward the equilibrium state. They supposed that the final state must be the liquid state $d$ on the extrapolated line. Actually, there is no evidence that the final state is the liquid state, as described in section \ref{sec:TransitionTemp}. The fact of not detecting the liquid state is commonly explained by the very slow motions of atoms: the relaxation time is so long that we cannot observe the final equilibrium state within a reasonable time of the experiment. The occurrence of relaxation is true but they missed the possibility that the equilibrium curve is another line of the glass state $b'$, which is a more stable glass state than the currently obtained state $b$. 

The supposition of Davies and Jones is incompatible with the principle of thermodynamics. The thermodynamic stability condition under the constraint of a constant $T$ is the minimum of the Gibbs free energy $G$ but not of the enthalpy $H$. At the bottom of figure~\ref{fig:CHG-Tcurve}, the $G-T$ curve obtained from the $C_{P}-T$ curve is plotted. For $T<T_{g}$, the inequality
\begin{equation}
G^{\rm (gl)}(T) < G^{\rm (li)}(T),
\label{eq:Gcompare}
\end{equation}
holds because of the upward concavity of the free energy $G(T)$. Although the figure is drawn schematically, relation (\ref{eq:Gcompare}) is general and is independent of material. A real calculation of the $G-T$ curve for a concrete material is shown in \cite{Shirai20-GlassState} (see figure~A1 in that paper). In the transition region, the $C_{P}-T$ curve significantly changes depending on the cooling/heating rate, and often an overshoot feature is observed \cite{Hodge83,Hodge94}. Despite this, relation (\ref{eq:Gcompare}) always holds, no matter how complicated the $C_{P}-T$ curve is.
The inequality (\ref{eq:Gcompare}) is deduced from the general relation (\cite{Zemansky}, p.~372),
\begin{equation}
\frac{\partial^{2} G}{\partial T^{2}} = -\frac{C_{P}}{T} < 0.
\label{eq:Gcurvature}
\end{equation}
The positive definiteness of specific heat guarantees the upward concave property of the Gibbs free energy. The opposite relation $G^{\rm (gl)} > G^{\rm (li)}$ would be possible only when the specific heat has a negative sign. But negative specific heat is thermodynamically unstable, as already discussed \cite{Lynden-Bell99,Landsberg87,Michaelian07}. Experimental facts indicate that the specific heat of glass is always positive over the entire temperature range. Therefore, the glass state must be more stable than the liquid state at $T<T_{g}$. The liquid state cannot be recovered by the relaxation process after the glass transition.
The inequality (\ref{eq:Gcompare}) indicates that the glass transition is a thermodynamic transition, contrary to the traditional belief. 
The simple reason why glass is formed is that the free energy of glass becomes lower than that of the liquid. 

\subsubsection{The Prigogine--Defay ratio and phase transition}
\label{sec:discussion-PDratio}
The PD ratio ${\it \Pi}$ is an important parameter when discussing phase transitions. However, there is confusion, when we discuss the inequality of ${\it \Pi}$. The equality is derived when the entropy and volume are continuous at a discrete transition temperature $T_{\rm tr}$: $\Delta S=0$ and $\Delta V=0$ (\cite{Zemansky}, p.~385). These two conditions, respectively, lead to
\begin{equation}
\frac{dP}{dT} = \frac{1}{TV} \frac{\Delta C_{P}}{\Delta \alpha_{P}}
\label{eq:Eherenfest-1}
\end{equation}
and
\begin{equation}
\frac{dP}{dT} = \frac{\Delta \alpha_{P}}{\Delta \kappa_{T}},
\label{eq:Eherenfest-2}
\end{equation}
from which the equality ${\it \Pi}=1$ follows. When Davies and Jones applied the PD ratio to glasses, they ignored the width: $\Delta T_{g} \rightarrow 0 $. Otherwise, it is meaningless to speak of the jump $\Delta C_{\rm gl}$: $\Delta C_{P}$ is defined by the difference in $C_{P}$ between the terminal states at $T_{g,1}$ and $T_{g,2}$. Recent discussions on the PD ratio \cite{Tropin12,Schmelzer12,Garden13} argue the kinetics of $C_{P}$ within the width. Therefore, these arguments are missing the original target.

The argument of Davies and Jones that ${\it \Pi}>1$ is a consequence of the plurality of the order parameters is correct. The number of order parameters of a glass, $N_{\rm op}$, is greater than one \cite{note-DiMarzio}. 
The only problem that has been left as an open question is what is the substance of the order parameters of a glass. FRE (\ref{eq:FRE-solids}) of solids clearly indicates that the equilibrium positions of atoms are the order parameters as well as state variables. In this manner, the order parameters and state variables are considered to be the same quantities in the thermodynamics context. Refer to \cite{Shirai25-OrderParams} for a detailed discussion on the nature of order parameters.

Now that the substance of the order parameters has been clarified, we can state the physical meaning of the PD ratio \cite{Shirai23-Silica}, which was previously unknown.
With some reasonable approximations about $\alpha$ and $\kappa$, the PD ratio can be recast as
\begin{equation}
{\it \Pi} = \frac{ \Delta C_{P}}{\Delta C_{\rm te}}
    = \frac{ 
    \mathrm{(Change\ in\ the\ total\ energy)} }{ \mathrm{(Contribution\ of\ isotropic\ volume\ change)} }.
\label{eq:PDratio3}
\end{equation}
Because ${\it \Pi} > 1$, the glass transition occurs mainly as a result of the change in the internal structure that determines $E_{\rm st}$. The contributions of thermal expansion and phonons are negligible.
In this respect, there is a merit to use the PD ratio for evaluating how the structural change contributes to the phase transition. The specific heat jump itself indicates the structural change $\Delta E_{\rm st}$. However, the specific heat also reflects the energy fluctuation, as indicated by equation (\ref{eq:fluct-C}).
This means that it becomes insensitive to the change in $T$ at high temperatures; the change in $E_{\rm st}$ appearing in $C$ is hidden by the uninterested change $3R\Delta T$ due to the classical equipartition rule. Thus, ${\it \Pi}$ is better than $\Delta C_{\rm st}$ for assessing how much the structural energy changes for a unit change in $T$, because the insensitivity to temperature change is eliminated by taking the ratio $\Delta C_{P}/\Delta C_{\rm te}$. In calorimetric measurement, silica appears to have a little change between the glass and liquid states: the jump $\Delta C_{\rm gl}$ of silica is the smallest among various glasses. The distinction between these two phases is not clear. However, the difference becomes clearer by looking at the ${\it \Pi}$ value of silica, which is the largest, indicating a significant change in the internal structure.

Lastly, a comment about the Eherenfest's classification of phase transitions is mentioned \cite{Pippard, Zemansky}. From the historical perspective, the original manner of classification was often a subject of criticism \cite{Jaeger98}.
The traditional classification of phase transitions is based on the analytic behavior of free energy, namely, continuity in the derivatives of function. The analytic behavior can be discussed only when the transition process is reversible. When the transition process is irreversible for some reason, hysteresis appears and hence the analytic behavior is destroyed. This happens for glass transition. Therefore, it is inappropriate to adapt the Eherenfest's classification to the glass transition in a mathematical manner. However, if the physical meaning of transition is considered, the glass transition is essentially a structural transition, and in this sense it can be classified as a first-order transition with broadening transition temperature.


\subsection{Implications on statistical mechanics}
\label{sec:implicationSM}
The author is afraid of writing this subsection: the subjects discussed here are clearly beyond the author's competence. However, thermodynamics is a universal theory of physics, whose validity goes beyond material science.  A famous maxim by Eddington is cited here: {\em But if your theory is found to be against the Second Law of Thermodynamics I can give you no hope} \cite{Eddington35}.
This holds even for the extreme material of a black hole \cite{Bekenstein80,Davies78,Jacobson96}. If discrepancies are found in any field, the present theory must be wrong. It is with this believe that the author presents this subsection.

Extensivity and intensivity are important notions in thermodynamics but are still in debate at the forefront of research \cite{Jaynes92,Gross02,NonextEntropy04,Saadatmand20}. The distinction is clear in the elemental course of thermodynamics. Energy and entropy are extensive variables, because the additive property holds. If one gram of iron is divided in half, the sum of the energies of the divided parts, $U_{1}$ and $U_{2}$, gives exactly the energy of the original piece, $U_{0}$:
\begin{equation}
U_{0} = U_{1} + U_{2}.
\label{eqE0sum}
\end{equation}
On the other hand, for temperature and pressure, this additive property does not hold. These are intensive variables. However, under more careful investigation, we will find that the distinction becomes ambiguous. Pressure is an intensive variable, but when the total pressure of a gas mixture is decomposed to the partial pressures, we find pressure to be extensive. In contrast, the extensivity does not hold for mixing entropy.  Further examples are shown by Grandy \cite{Grandy} (chapter 5). In this manner, the notions of extensive and intensive are only approximations. The total energy and total entropy are significant, which are in accordance with the spirit of DFT.

For the above dividing process, equation~(\ref{eqE0sum}), the question is to what extent we can finely divide the system. For gas systems, the additive property is valid up to the atomic level. For solids, the additivity may be valid at micronmeters or similar scales. However, at the atomic-size scale, it is not. DFT clearly states that the total energy is not the sum of the individual atom energies. Rather, the energy of an individual atom is not well defined there. In spite of this, because thermal excitations are such gentle excitations, the corresponding Hamiltonian can be diagonalized for small displacements and can be expressed by the sum of their eigenstates, equation~(\ref{eq:etot-sum-e1}). The phonon theory is so often used that it is easily forgotten that the summation property holds only approximately. A similar situation is seen for electronic excitations. In the elemental course of solid state physics, we learned that the Sommerfeld formula holds well for describing the specific heat of the electron system, which is based on the free-electron approximation \cite{Ashcroft-Mermin}. However, the experimental values of the specific heat coefficient $\gamma_{e}$ differ from neither those of free electrons nor those calculated on the basis of band theory. The deviation is significant for heavy-electron systems \cite{Ueda-Onuki98}. To explain the large deviation, the Landau theory of Fermi liquids was devised, showing that the independent description is valid only under restricted conditions \cite{Nozieres64}. Outside the validity regime, the summing property does not hold, or more appropriately, no suitable quasi-particles exist for describing such strong excitations. 
An extreme case is the entropy of a black hole. The entropy of a black hole, $S_{\rm bh}$, is proportional to its surface area $A$ but not to the volume, i.e., $S_{\rm bh} \propto A$ \cite{Bekenstein80}. The interaction inside the black hole is so strong that the additive property does not hold even at the macroscopic scale.

In information theory, the fundamental problem is how to determine the probability distribution $p_{i}$ for a given set of events (states) $x_{i} \in X$ from an incomplete set of information ($X$ is the set of events). The maximum entropy principle (MEP) by Jaynes offers the leading principle \cite{Jaynes57,Jaynes57a}. The probability distribution is determined by maximizing the entropy functional $H = -\sum_{i} p_{i} \ln p_{i}$ with given constraints,
\begin{equation}
\bar{f}_{k} = -\sum_{i} p_{i} f_{k}(x_{i}).
\label{eq:Jaynes-constr}
\end{equation}
Here, $\bar{f}_{k}$ is the expectation value of a property $f(x)$ and is assumed to be known beforehand. For independent-particle systems, the prediction of MEP agrees with the Maxwell--Boltzmann distribution.
The big problem is how to generalize the MEP to strongly interacting systems. For interacting systems, the extensivity and additivity of energy do not hold. Intensive effort has been recently devoted to this fundamental problem \cite{Presse13r,Presse13,Saadatmand20,Jizba19}.
Interaction means correlation among its components. One way to incorporate the correlation is to introduce a relative sense to the probability: i.e., conditional probabilities. The probability is considered to be always updated for each observation. The MEP is extended using the notion of relative entropy \cite{Shore80,Caticha21,Cover-Thomas}.
Another way of generalization is to devise an appropriate form of entropy functions to produce nonextensivity and nonadditivity \cite{Tsallis11,Tsallis15,Jizba19,Tsallis23}. Presently, no satisfactory theories exist. This problem arises because we wish to assign eigenenergies to constituent particles of the system. The problem does not occur in the total energy approach. The same is also true for entropy. Only the total entropy has physical meaning when interactions between the constituent particles are strong. The specific heat of liquids offers a good example for studying the extensivity and additive properties. A greate amount of experimental data are already available for liquids, while measurement for astronomic objects and nuclear matters is difficult.

\section{Summary}
\label{sec:Summary}
At the beginning of this review, the author posed a question why the construction of the theory of the specific heat of liquids is difficult. 
If we see a snapshot of atom positions of a liquid, we cannot distinguish it from that of a glass. The worse is taking the time average: this makes the density uniform, whose value is almost the same between the liquid and crystal. The essential difference of liquids consists in the atom relaxation. This dynamical aspect is easily missed if only these static quantities, such as DOS and RDF, are inspected.

For glass materials, there are a tremendous number of metastable structures $\{ K \}$, with each having a different relaxation time $\tau_{R,K}$. Within the transition region, the specific heat has various values in accordance with equation (\ref{eq:CfuncR-inh}), depending on the cooling conditions. 
Finally one of them, say $K_{1}$, is selected and becomes the stable equilibrium state at $T$ much lower than $T_{g}$. Once the structure is uniquely determined, the FRE and thereby the specific heat of this particular sample contain the structural information by $K_{1}$ only.
For liquids, on the other hand, there is no unique structure. The system visits all the accessible configurations $\{ K \}$. Accordingly, the FRE is determined solely by $T$ and $V$, 
and the information of involved configurations is averaged over in $U^{\rm (li)}$ and $C^{\rm (li)}$.
However, the visiting rates to individual configurations are affected by the atom relaxations, and thus the FRE has dependence on the frequency $\omega$ at which the measurement is done. This reflects the $\omega$ dependence and the $T$ dependence on $C^{\rm (li)}$.

The characteristic behavior of the glass transition and the negative $T$ coefficient in $C_{V}^{\rm (li)}$ of liquid have been well reproduced by the total energy approach. Neither can be understood without taking the effect of relaxations into account. In statistical mechanics, energy relaxations are treated by the master equation with a complete set of transition rates $\{ \gamma_{ij} \}$. On the other hand, the master equation does not answer the question of how the transition rates are determined. FPMD simulations describe the relaxation processes and therefore automatically determine the transition rates without recourse to experimental data. However, this automatic calculation of the correct answer is not guaranteed without appropriate treatment of the relaxation simulation. The adiabatic relaxation is recommended, because this method most faithfully simulates spontaneous changes while retaining the correct correspondence between the constraints and the equilibrium state.
This correspondence is one of the statements of the second law of thermodynamics. Throughout this review, the multiple views with respect to relaxation time are emphasized through equation (\ref{eq:transition-1}). 
This is because an equilibrium is established by balancing the many competitive configurations $K$.

The dynamical properties of liquids are often studied by neutron and $X$-ray scattering methods. If we compare the collective modes of a liquid with the phonon spectrum of corresponding solid, these are very similar for simple metals, except broadening. Hence, it may be natural to apply the phonon model to the specific heat of liquids. However, the fundamental difficulties in the phonon model have been illustrated in section \ref{sec:TdepC-result-phonon}. The essential problem of the specific heat of liquids is that the degrees of freedom ($f$) are not conserved quantities: $f$ is continuously reduced from 6 at the solid end to 3 at the gas end. Destroying the degrees of freedom is the consequence of atom relaxations. This invalidities the independence assumption in the elemental excitation approach.

The total energy approach brings the energetic investigation to glass research in a quantitative manner. This enables us to assess how the simulated glass structure is realistic.
Calculations of the specific heat offer a basic element of material characterization for glasses. 
From the fundaments of thermodynamics, a significance of the total energy approach for nonperiodic systems is that it clarifies which are state variables through the evaluation of the internal energy. The correct form of FRE resolves many conceptual difficulties in thermodynamics, such as the hysteresis and irreversibility of solids, which are the reasons why thermodynamic investigations were evaded for glasses. Lastly, it is emphasized that all the important thermodynamic properties described here are derived from only one principle, namely, the GB statement of the second law with appropriate understanding of constraints.

\section*{Acknowledgment}
The author thanks my colleagues for useful discussions throughout this study. He also thanks Prof.~Trachenko (Queen Mary Univ.~London) and Prof.~Zaccone (Univ.~Milan) for discussing the specific heats of glasses and liquids, and Prof.~Sugawara (Akita Univ.) for instruction on the calorimetric methods.
He thanks Myu Research (myu-group.co.jp) for the English language review.

\section*{References}


\begin{thebibliography}{100}
\bibitem{note-axiomatic}
In the axiomatic approach of thermodynamics, the adiabatic processes are used
  to classify equilibrium states before defining temperature. However, we note
  that various restrictions about the systems and processes, such as
  single-component systems and equilibrium states, are posed, depending on the
  author \cite{Buchdahl,Landsberg70,Giles70,Lieb99,Lieb13}. This causes
  confusion. The GB treatment is the most general one. Such unnecessary
  restrictions have been removed there.
\bibitem{note-Eq-defect}
Someone says that defect states are not in equilibrium. However, it is a
  learning from the elemental thermodynamics that a crystal with an equilibrium
  concentration of defects has a lower free energy than the perfect crystal
  does. In this respect, the perfect crystal cannot be a stable equilibrium
  state. Then, the question arises as to which structure among the various
  kinds of defects is the equilibrium structure. This issue is addressed in
  \cite{Shirai24-hysteresis}.
\bibitem{note-GlassNucleat}
In the author's view, the condensation to glass means the ``nucleation" of
  glass. Usually, the term nucleation is restricted to use for crystals.
  However, it is a common understanding in the glass literature that there are
  liquid-like and solid-like parts during the glass transition, which is known
  as dynamic heterogeneity \cite{Ediger00,Berthier11}. The solid part can be
  interpreted to be the ``nucleated" glass.
\bibitem{note-INM}
In spite of this, the unorthodox usage of eigenstate (normal mode) exists for
  liquids: namely, the approach of instantaneous normal mode (INM)
  \cite{Stratt95,Keyes97,Baggioli21,Zaccone21,Stamper22}. In this approach, the
  dynamic matrix $D_{ij}$ is constructed for the instantaneous atom positions
  as $D_{ij} = (\partial^{2} E_{P}/\partial R_{i} \partial R_{j})_t$, where the
  subscript $t$ denotes that the derivative is evaluated at the instantaneous
  position $R_{i} = R_{i}(t)$. This is not the equilibrium position and hence
  the positive-definite property of the dynamic matrix is broken. Thus, the
  diagonalization of $D_{ij}$ yields imaginary frequencies. The notion of INM
  is not justified without reservation \cite{Stratt95}. To discuss the
  suitability of INM is out of the present scope.
\bibitem{note-ext-supercritical}
An interesting question is how this interpretation is extended to the limit of
  ideal gas, where $C_{V} = (3/2)R$ is expected. Normally, this continuity
  cannot be examined owing to the intervention of the liquid--vapor phase
  transition. However, recently, Trachenko has demonstrated a continuous
  convergence to $(3/2)R$ using a supercritical fluid (\cite{Trachenko23-a},
  Ch.~14).
\bibitem{note-CofWater}
The author learned from A.~Ben-Naim that the agreement with the Dulong--Petit
  law is merely accidental.
\bibitem{note-CasStateV}
Two matters complicate the analysis. The first one is that $C$ is, in more
  precisely, not the same as $\nabla U$ but $C_{X_{i}} = (\partial U/\partial
  Y_{i})_{\{X_{j}\}} (\partial Y_{i}/\partial T)_{\{X_{j}\}}$. The weighting
  factor $(\partial Y_{i}/\partial T)_{\{X_{j}\}}$ is also a state function.
  The product of two state functions makes another state function. The second
  one is that $C_{X_{i}}$ should be given more appropriately by the entropy
  change as $T (\partial S/\partial T)_{\{X_{j}\}}$, provided a reversible
  path. However, $S$ is also a state function, and hence the conclusion that
  $C_{X_{i}}$ is a state function does not change.
\bibitem{note-DiMarzio}
DiMarzio once deduced that the equality holds even when $N_{\rm op}$ is greater
  than one \cite{DiMarzio74}. However, the important matter is that $N_{\rm
  op}$ is the number of {\em independent} order parameters. If this
  independence condition is missed, then the equality would result in.
\end{thebibliography}
\providecommand{\newblock}{}

\end{document}